\documentclass[11pt]{article}
\usepackage{lscape}
\usepackage{amsthm}
\usepackage{soul}
\usepackage[utf8]{inputenc}
\usepackage{amsmath,amssymb,stmaryrd}
\usepackage{hyperref}
\usepackage[numbers]{natbib}
\usepackage{setspace}
\usepackage{subcaption}
\usepackage{multirow}
\usepackage{graphicx}
\usepackage[table]{xcolor}
\usepackage{xcolor}
\usepackage{array}

\arrayrulecolor[HTML]{DB5800}

\usepackage{bbm}
\usepackage{algorithm}
\usepackage{algorithmic}

\usepackage{pifont}
\newcommand{\bbR}{\mathbb{R}}

\newcommand{\calG}{\mathcal{G}}

\newcommand{\calM}{\mathcal{M}}
\newcommand{\calN}{\mathcal{N}}
\newcommand{\calW}{\mathcal{W}}
\newcommand{\Loss}{\text{Loss}}
\newcommand{\ot}{\leftarrow}
\newcommand{\TP}{\text{TP}}
\newcommand{\FP}{\text{FP}}
\newcommand{\TN}{\text{TN}}
\newcommand{\FN}{\text{FN}}
\DeclareMathOperator{\Ind}{\textbf{1}}
\DeclareMathOperator{\GW}{\calG\calW}
\DeclareMathOperator{\diag}{diag}
\DeclareMathOperator{\card}{card}

\floatname{algorithm}{Procedure}

\title{Optimal  transport-based   machine  learning  to   match  specific
    patterns: application to the detection of molecular regulation patterns in
    omics data} 

\author{Thi Thanh Yen Nguyen$^{1,\dagger}$, Warith Harchaoui$^{1,2}$,\\
  Lucile Mégret$^3$, Cloé Mendoza$^3$,\\
  Olivier Bouaziz$^{1,*}$, Christian Neri$^{3,*,\dagger}$, Antoine Chambaz$^{1,*}$\\~\\
  {\small $^1$ Université Paris Cité, CNRS, MAP5, F-75006 Paris, France}\\
  {\small $^2$ DERAISON.ai}\\
  {\small $^3$ Sorbonne Université, CNRS UMR 8256, Brain-C Lab, Paris,
    France}\\
  {\small $^*$ These authors contributed equally to this work.}\\
  {\small $^\dagger$ Correspondence.}
}

\begin{document}
\singlespacing
\maketitle

\begin{abstract}
  We present several algorithms designed  to learn a pattern of correspondence
  between two data sets in situations  where it is desirable to match elements
  that exhibit a  relationship belonging to a known parametric  model.  In the
  motivating  case study,  the  challenge is  to  better understand  micro-RNA
  regulation in the striatum of Huntington's disease model mice.

  The algorithms unfold  in two stages.  First, an  optimal transport plan~$P$
  and an optimal  affine transformation are learned,  using the Sinkhorn-Knopp
  algorithm and  a mini-batch  gradient descent. Second,  $P$ is  exploited to
  derive either several co-clusters or several sets of matched elements.

  A simulation study illustrates how the algorithms work and perform.  The
  real data application further illustrates  their applicability and interest. \\

\end{abstract}

%--------------------------------------------------------------------------
\doublespacing
\section{Introduction}
\label{sec:intro}

The  analysis of  numerous  omics data  is a  challenging  task in  biological
research~\cite{Benayoun19}    and    disease    research~\cite{Langfelder2016,
  Maniatis89}.  In disease research, omics data are increasingly available for
the analysis of  molecular pathology. This is notably  illustrated by research
on Huntington's Disease (HD): messenger-RNA (mRNA), micro-RNA (miRNA), protein
data collectively  quantifying several layers  of molecular regulation  in the
brain  of HD  model  knock-in  mice~\cite{Langfelder2016, Langfelder2018}  now
compose  one of  the largest  data  set available  to date  to understand  how
neurodegenerative processes  may work  on a  systems level.   The data  set is
publicly available  through the database repository  Gene Expression Omnibus
(GEO) and the \href{http://www.HDinHD.org}{HDinHD portal}.
  
Encouraged  by the  promising  findings of~\cite{megret:inserm-02512089},  our
ultimate goal  is to shed  light on the  interaction between mRNAs  and miRNAs
based on data collected in the striatum  (a brain region) of HD model knock-in
mice~\cite{Langfelder2016, Langfelder2018}.  Each data point takes the form of
multi-dimensional  profile.  The  strong biological  hypothesis is  that if  a
miRNA induces the degradation of a  target mRNA or blocks its translation into
proteins, or both, then the profile of  the former, say $y$, should be similar
to minus  the profile of  the latter, say $-x$.   We relax the  hypothesis and
consider  that $y$  is  similar to  $\theta(x)$ where  $\theta$  is an  affine
transformation in a parametric class $\Theta$ that includes minus the identity
and whose  definition translates  expert knowledge  about the  experiment that
yields the  data.  Our study  straightforwardly extends  to the case  that the
relationship is known to belong to any parametric model.  In order to identify
groups of mRNAs and miRNAs that interact, we develop a co-clustering algorithm
and a matching algorithm based on optimal transport~\cite{COT19}, spectral and
block co-clustering, and a matching procedure tailored to our needs.

Spectral       co-clustering~\cite{10.1145/502512.502550}      and       block
clustering~\cite{Keribin2014,Nadif2010}  are two  ways  among  many others  to
carry   out  co-clustering,   an   unsupervised  learning   task  to   cluster
simultaneously the rows and columns of a matrix in order to obtain homogeneous
blocks.  There  are many  efficient approaches to  solving the  problem, often
characterized as model-based or metric-based methods~\cite{PONTES2015163}.

In an enlightening  article, \citet{NK2021} review a  variety of computational
approaches to study how miRNAs ``come together to regulate the expression of a
gene or a group of genes''. They identify three different families of methods:
data-driven methods based on similarities, data-driven methods based on matrix
factorization, and hybrid methods.  Our algorithms belong to the first family.
In  view of  \cite[Section~2.5  and Fig.~2]{NK2021},  we do  not  rely on  the
standard similarity  measures (Pearson and Spearman  correlation coefficients;
cosine similarity;  mutual information) to  define our similarity  matrix but,
instead,  use  optimal transport  to  derive  it.  Moreover, as  in  canonical
correlation analysis, we do not compare  the raw mRNA and miRNA profiles $x,y$
but, instead,  we compare  a data-driven  transformation $\theta(x)$  and $y$,
where  $\theta$ is  an affine  transformation of  $x$.  Finally,  as explained
by~\citet{NK2021},   our   algorithms   cannot   discriminate   between   true
interactions and  fake interactions originating from  common hidden regulators
such  as  transcription  factors.   It  is  necessary  to  conduct  a  further
biological analysis to identify the relevant findings.

The  rest of  the  article is  organized  as follows.   Section~\ref{sec:data}
describes the data  we use.  Section~\ref{sec:elements} presents  a modicum of
optimal transport  theory.  Section~\ref{sec:main} introduces  our algorithms.
Section~\ref{sec:simulation} evaluates  the performances of the  algorithms in
various simulation settings.  Section~\ref{sec:real:data} illustrates the real
data  application.    Section~\ref{sec:discussion}  closes  the  study   on  a
discussion.

\section{Data}
\label{sec:data}
  
\subsection{Presentation}

The data  analyzed herein cover RNA-seq  data obtained in the  striatum of the
allelic series of HD knock-in mice (poly~Q lengths: Q20, Q80, Q92, Q111, Q140,
Q175) at 2-month, 6-month and 10-month of age.  For each combination of poly~Q
length  and age,  8  mice were  sacrificed  (4 females  and  4 males).   After
preprocessing~\cite[][Methods section]{megret:inserm-02512089}, the final data
set      consists      of      $M      =      13,616$      mRNA      profiles,
$X := \{x_{1},  \ldots, x_{M} \} \subset  \bbR^{d}$, and in $N  = 1,143$ miRNA
profiles, $Y := \{y_{1}, \ldots, y_{N}\} \subset \bbR^{d}$ with $d= 15$.

Informally,             we             look            for             couples
$(m, n)\in\llbracket M\rrbracket \times \llbracket N\rrbracket:=\{1,\ldots,M\}
\times \{1,\ldots,N\}$  such that the  $n$th miRNA induces the  degradation of
the  $m$th mRNA  or blocks  its translation  into proteins,  or both.   We are
guided by the strong biological hypothesis that, if that is the case, then the
profile $y_{n}$ of the  former is similar to minus the  profile $x_{m}$ of the
latter  --  then  $x_{m}$  and  $y_{n}$  exhibit  what  we  call  a  mirroring
relationship.  Of note, it is expected  that a single miRNA can target several
mRNAs.

The actual  mirroring relationships can  be more  or less acute,  for instance
because of threshold  effects, or of multiple miRNAs targeting  the same mRNA,
or of a  single miRNA targeting several mRNAs.  Therefore,  instead of rigidly
using comparisons between $-x_{m}$ and $y_{n}$, our algorithms will learn from
the data  a relevant transformation  $\theta\in\Theta$ (in a  parametric class
$\Theta$  of  transformations  that  includes  minus  the  identity)  and  use
comparisons between $\theta(x_{m})$ and $y_{n}$.

\begin{figure}
  \centering %
  \includegraphics[width=0.45\linewidth]{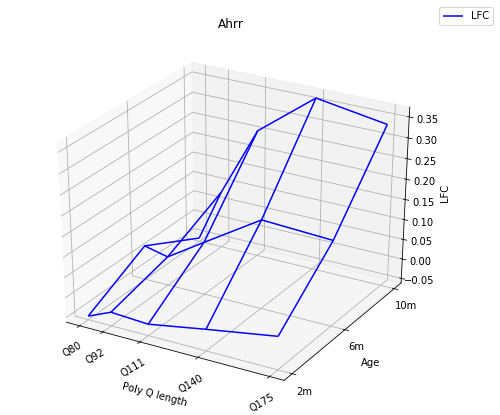}%
  \qquad%
  \includegraphics[width=0.45\linewidth]{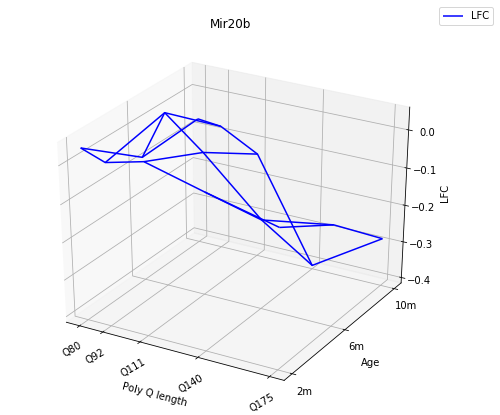}
  \caption{Left: profile $x_{m}$ of a  mRNA (Ahrr).  Right: profile $y_{n}$ of
    a miRNA (Mir20b).  It is believed that Mir20b targets Ahrr.}
  \label{fig:Mir20b}
\end{figure}

Figure~\ref{fig:Mir20b}  exhibits  two  profiles   $x_{m}$  and  $y_{n}$  that
showcase  a mirrored  similarity.  The  corresponding miRNA  and mRNA,  Mir20b
(which    may    inhibit    cerebral    ischemia-induced    inflammation    in
rats~\cite{ZWDSL19}) and  the Aryl-Hydrocarbon Receptor Repressor  (Ahrr), are
believed   to    interact   in   the    striatum   of   HD    model   knock-in
mice~\cite{megret:inserm-02512089}.

\subsection{A brief data analysis}

So as to give a sense of the distribution of the data, we propose two kinds of
visual    summaries.      The    first    one    uses     Lloyd's    $k$-means
algorithm~\citep{Lloyd1982} to build synthetic  profiles representing the real
profiles $x_{1}, \ldots, x_{M}$ on the  one hand and $y_{1}, \ldots, y_{N}$ on
the other  hand. The second one  uses kernel density estimators  of the $j$-th
component   of   $x_{1},   \ldots,   x_{M}$   on   the   one   hand   and   of
$y_{1}, \ldots, y_{N}$ on the other hand, for each $1\leq j \leq d$.
  
\subsubsection{Using  $\boldsymbol  k$-means to  cluster  the  mRNA and  miRNA
  profiles}
\label{subsec:k:means}

\begin{figure}
  \centering
  \includegraphics[width=0.45\linewidth]{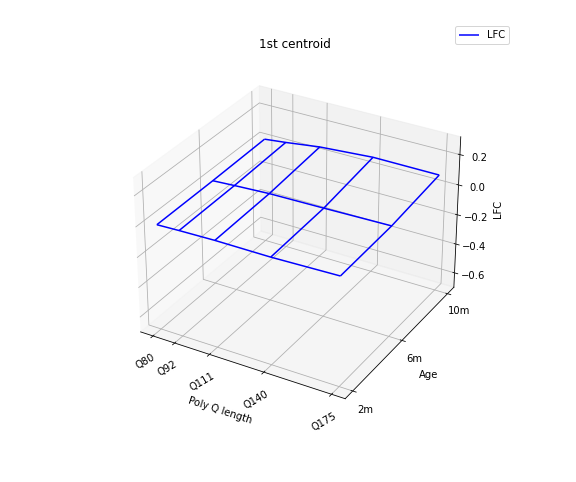}%
  \includegraphics[width=0.45\linewidth]{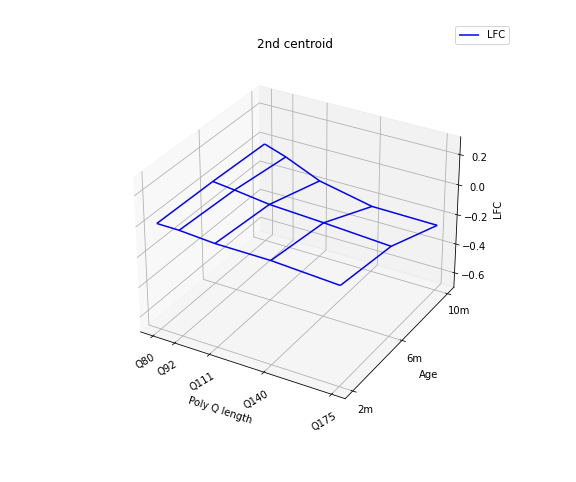}\\
  \includegraphics[width=0.45\linewidth]{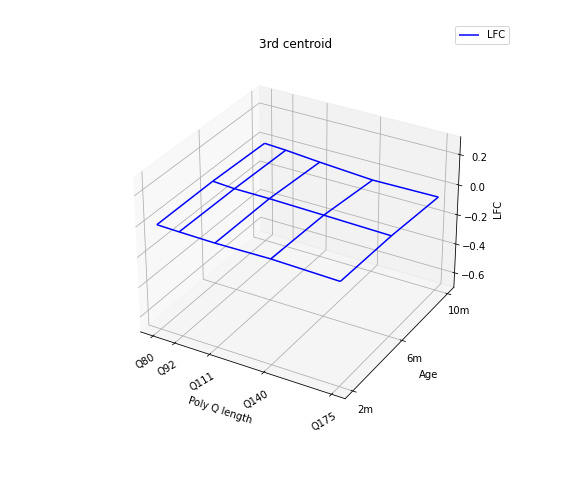}%
  \includegraphics[width=0.45\linewidth]{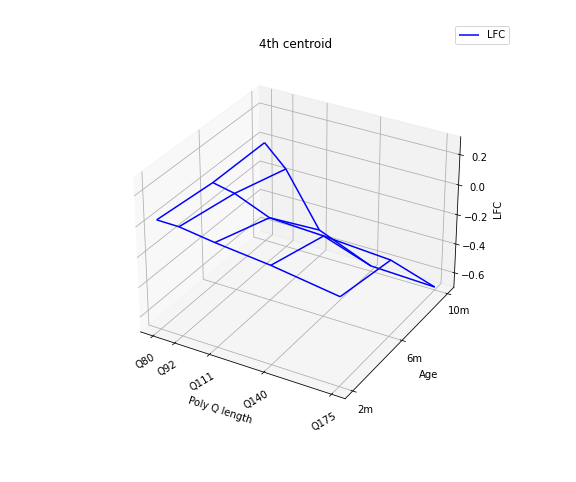}\\
  \includegraphics[width=0.45\linewidth]{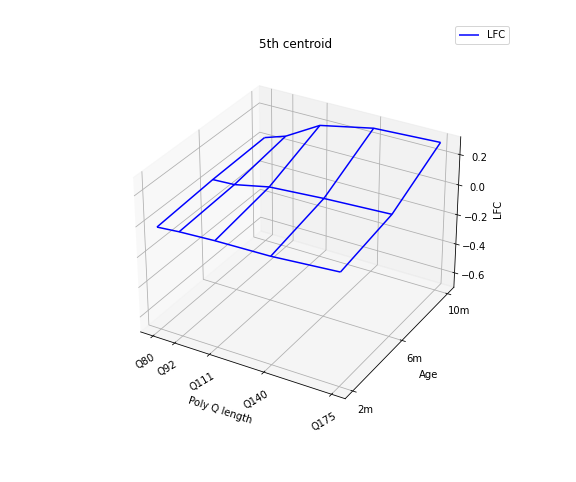}
  \caption{Profiles  $\hat{x}_{1}, \ldots,  \hat{x}_{5}$  of  the 5  centroids
    obtained   by   Lloyd's  $k$-means   algorithm   on   the  mRNA   profiles
    $x_{1}, \ldots, x_{M}$. }
  \label{fig:mRNA:centroid}  
\end{figure}

\begin{figure}
  \centering
  \includegraphics[width=0.45\linewidth]{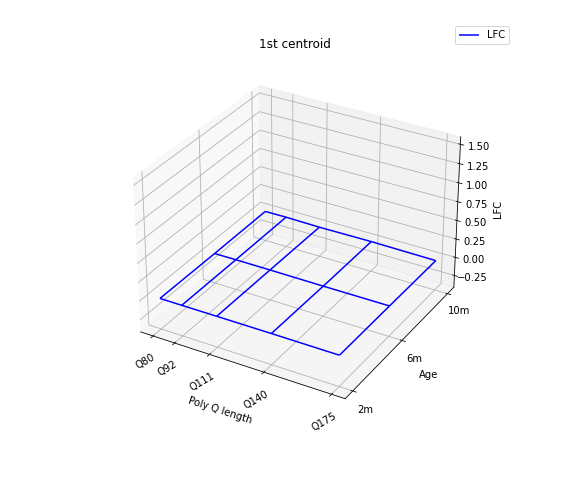}%
  \includegraphics[width=0.45\linewidth]{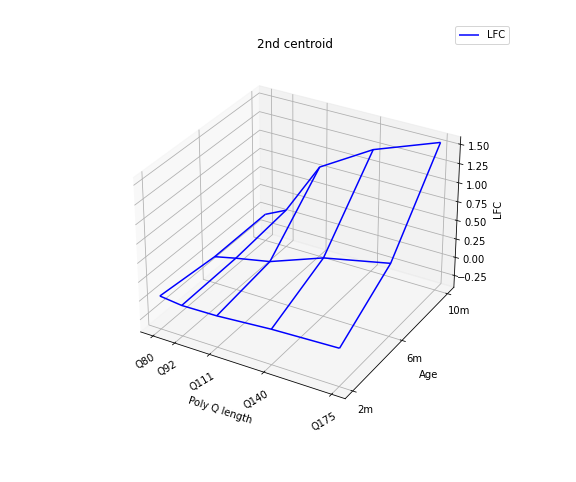}\\
  \includegraphics[width=0.45\linewidth]{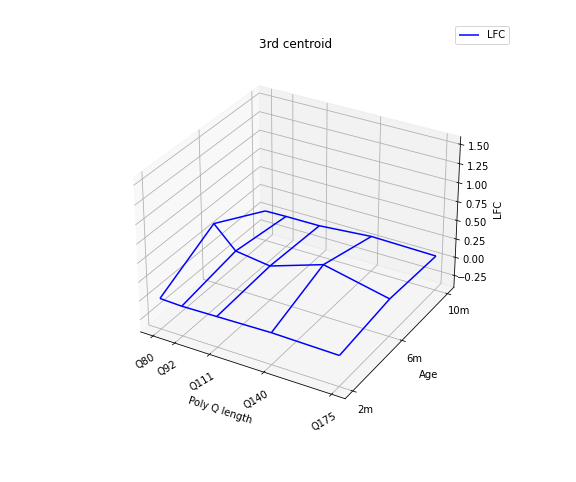}%
  \includegraphics[width=0.45\linewidth]{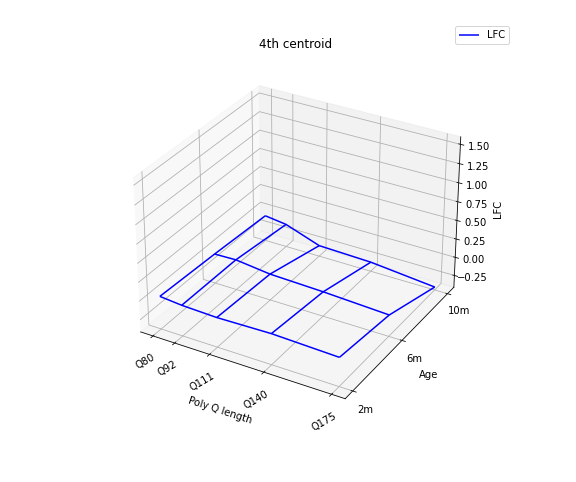}\\
  \includegraphics[width=0.45\linewidth]{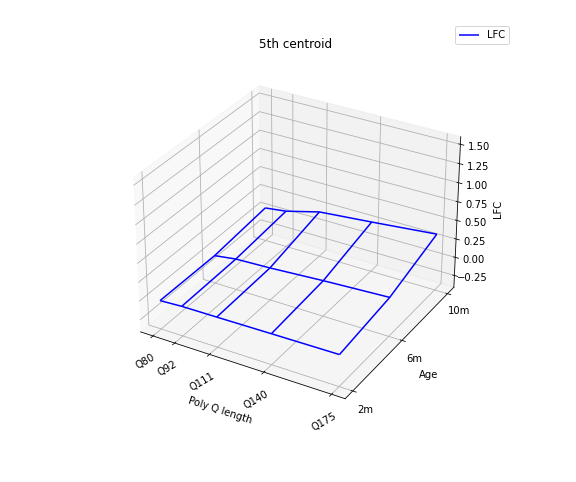}%
  \caption{Profiles  $\hat{y}_{1}, \ldots,  \hat{y}_{5}$  of  the 5  centroids
    obtained  by running  Lloyd's $k$-means  algorithm on  the miRNA  profiles
    $y_{1}, \ldots, y_{N}$. }
  \label{fig:miRNA:centroid}  
\end{figure}

In  Figure~\ref{fig:mRNA:centroid}   we  plot  the  synthetic   mRNA  profiles
$\hat{x}_{1},  \ldots, \hat{x}_{5}$  of the  5 centroids  obtained by  running
Lloyd's $k$-means algorithm  on $x_{1}, \ldots, x_{M}$  with $k=5$.  Likewise,
we  plot  in  Figure~\ref{fig:miRNA:centroid}  the  synthetic  miRNA  profiles
$\hat{y}_{1},  \ldots, \hat{y}_{5}$  of the  5 centroids  obtained by  running
Lloyd's $k$-means algorithm on $y_{1}, \ldots, y_{N}$ with $k=5$.

The 5 mRNA centroids correspond to 5319 ($\hat{x}_{1}$), 2097 ($\hat{x}_{2}$),
4688  ($\hat{x}_{3}$),  310  ($\hat{x}_{4}$)  and  1202  ($\hat{x}_{5}$)  mRNA
profiles.  The first  and third  centroids ($\hat{x}_{1}$  and $\hat{x}_{3}$),
which represent 73\%  of the real mRNA profiles, are  rather flat.  The second
and fourth  centroids ($\hat{x}_{2}$ and $\hat{x}_{4}$),  which represent 18\%
of the real mRNA profiles, are decreasing  in poly~Q length and age, in a more
pronounced  way for  the  latter  than for  the  former.   Finally, the  fifth
centroid  ($\hat{x}_{5}$), which  represents the  remaining 9\%  of real  mRNA
profiles, is increasing in poly~Q length and age.

The 5 miRNA centroids correspond to 872 ($\hat{y}_{1}$), 7 ($\hat{y}_{2}$), 80
($\hat{y}_{3}$), 81  ($\hat{y}_{4}$) and  103 ($\hat{y}_{5}$)  miRNA profiles.
The first  centroid ($\hat{y}_{1}$), which  represents 76\% of the  real miRNA
profiles,  is rather  flat. The  second and  fifth centroids  ($\hat{y}_2$ and
$\hat{y}_{5}$),  which  represent  10\%  of   the  real  miRNA  profiles,  are
increasing in poly~Q length  and age, in a more pronounced  way for the former
than for  the latter.  The  fourth centroid ($\hat{y}_{4}$),  which represents
7\%  of the  real miRNA  profiles,  is decreasing  in poly~Q  length and  age.
Finally, the third centroid ($\hat{y}_{3}$),  which represents 7\% of the real
miRNA profiles, exhibits two peaks.

In Section~\ref{sec:intro}, we stated  the following biological hypothesis: if
a miRNA  induces the degradation  of a target  mRNA or blocks  its translation
into proteins, or  both, then the profile  of the former should  be similar to
minus the  profile of the latter  (a particular form of  affine relationship).
In  view of  this hypothesis,  it is  tempting to  relate the  synthetic miRNA
profiles  $\hat{y}_{2}$  and  $\hat{y}_{5}$  to the  synthetic  mRNA  profiles
$\hat{x}_{4}$ and $\hat{x}_{2}$, respectively, and the synthetic miRNA profile
$\hat{y}_{4}$ to the synthetic mRNA profile $\hat{x}_{5}$. Our objective is to
identify groups of real mRNA and miRNA profiles that interact in this manner.

\subsubsection{Using  kernel   density  estimators   to  study   the  marginal
  distributions of the mRNA and miRNA profiles}

\begin{figure}[h]
  \centering
  \includegraphics[width=\textwidth]{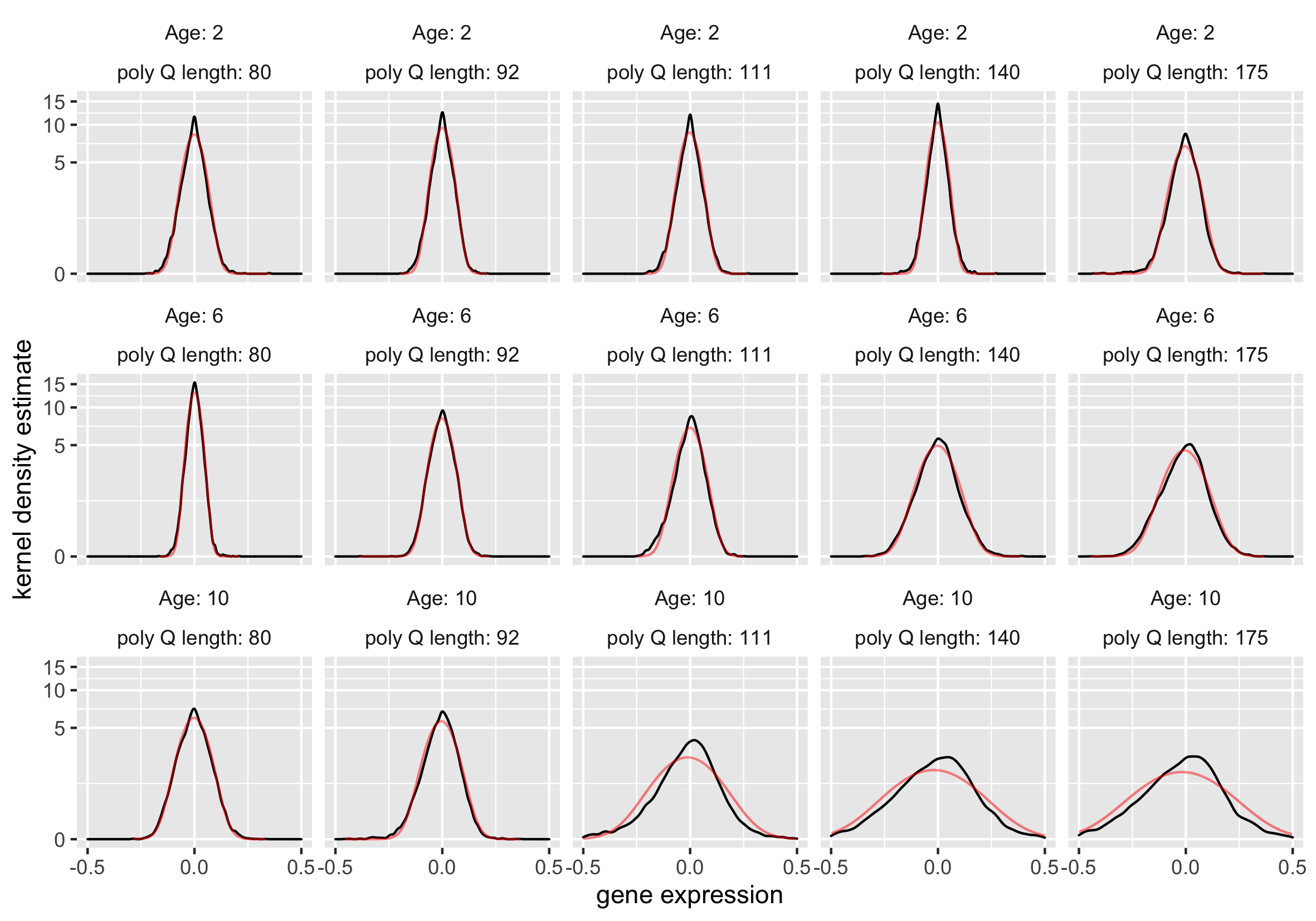}
  \caption{In black,  kernel density estimates  of the densities  of mRNA
      gene expression for  each level of poly~Q length (Q80,  Q92, Q111, Q140,
      Q175) and age  (2, 6, 10 months), zooming on  the interval $[-0.5, 0.5]$
      and using a $\log(1+\cdot)$-scale on the $y$-axis.  In red, densities of
      the Gaussian laws with a mean and a variance equal to the empirical mean
      and  variance computed  in each  stratum of  data.  Systematically,  the
      kernel density estimates  are more concentrated around  their means than
      the corresponding Gaussian densities.}
  \label{fig:KDE_mRNA}
\end{figure}

\begin{figure}[h]
  \centering
  \includegraphics[width=\textwidth]{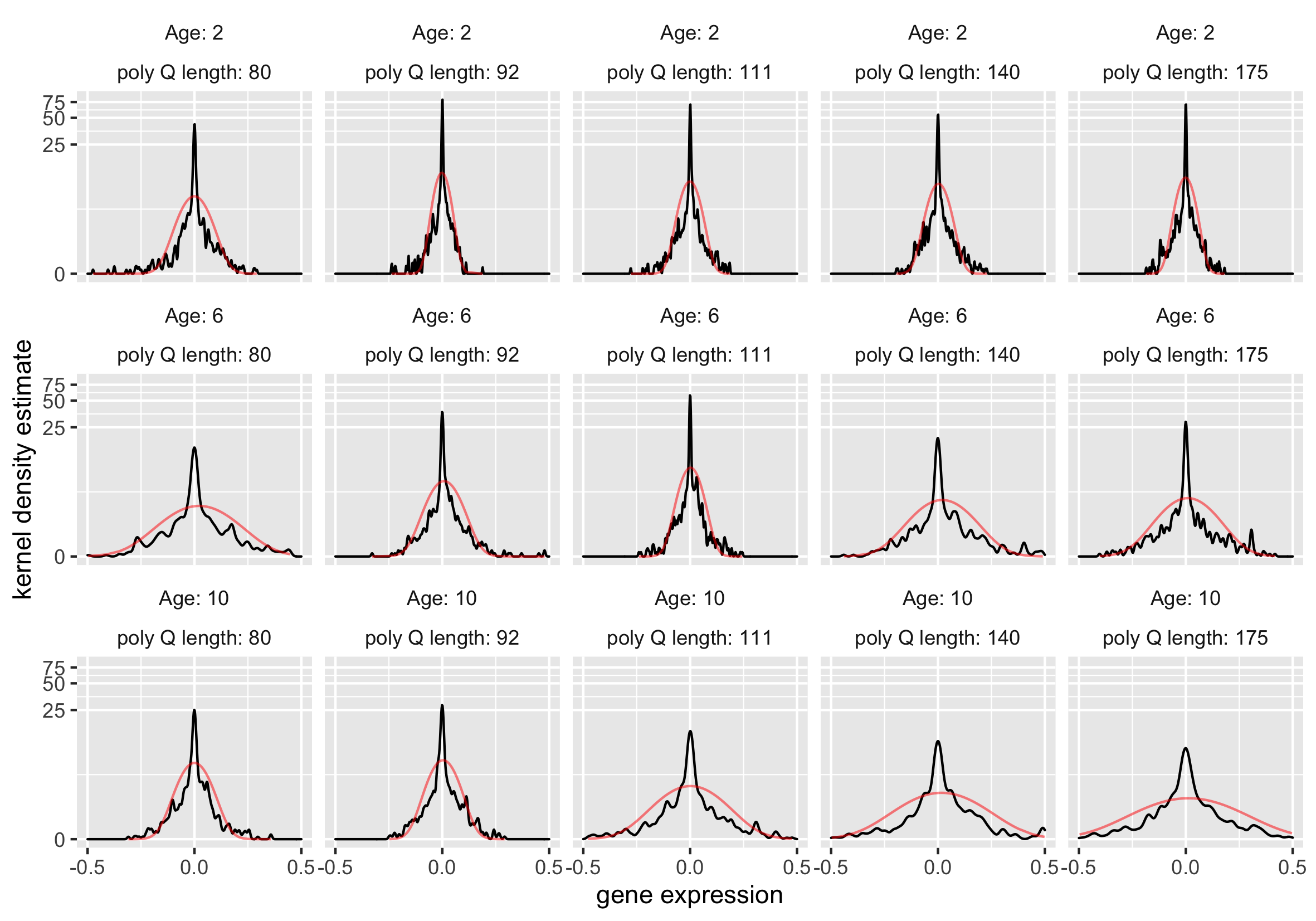}
  \caption{In black, kernel  density estimates of the densities  of miRNA gene
    expression for  each level of poly~Q  length (Q80, Q92, Q111,  Q140, Q175)
    and age (2, 6, 10 months), zooming on the interval $[-0.5, 0.5]$ and using
    a  $\log(1+\cdot)$-scale  on  the  $y$-axis.  In  red,  densities  of  the
    Gaussian laws with a  mean and a variance equal to  the empirical mean and
    variance computed  in each  stratum of  data.  Systematically,  the kernel
    density  estimates  are more  concentrated  around  their means  than  the
    corresponding Gaussian densities.}
  \label{fig:KDE_miRNA}
\end{figure}

For each $1\leq j\leq d$, we build  the kernel density estimator of the $j$-th
component of the mRNA profiles $x_{1}, \ldots, x_{M}$, using a Gaussian kernel
and  the  default  fine-tuning  of  the  \texttt{density}  function  from  the
\texttt{stats}  \texttt{R}-package~\citep{R},  see  Figure~\ref{fig:KDE_mRNA}.
We  do  the   same  for  the  miRNA  profiles  $y_{1},   \ldots,  y_{N}$,  see
Figure~\ref{fig:KDE_miRNA}.  Both  for  mRNA  and  miRNA  the  kernel  density
estimates are systematically  more concentrated around their  means (all close
to 0) than the corresponding  Gaussian densities. Moreover, the kernel density
estimates obtained  from the $M$  mRNA profiles  are much smoother  than those
obtained from $N$ miRNA profiles, a  feature that could be simply explained by
the fact that $M/N > 11$.

Table~\ref{tab:mRNA:miRNA:sd} reports,  for each level of  poly~Q length (Q80,
Q92,  Q111, Q140,  Q175) and  age (2,  6, 10  months), the  empirical standard
deviation of  mRNA (a) and miRNA  (b) gene expressions, all  normalized by the
empirical standard  deviation at poly~Q length  Q80 and 2 months  of age (that
is, by  0.0475 for mRNA  and 0.0660 for miRNA).  A clear pattern  emerges from
sub-Table~\ref{tab:mRNA:miRNA:sd}  (a):  except  for poly~Q  length  Q80,  the
poly~Q  length-specific   empirical  standard   deviation  increases   as  age
increases.   Likewise, except  for age  2 months,  the age-specific  empirical
standard deviation increases as poly~Q  length increases.  On the contrary, no
clear pattern emerges from  sub-Table~\ref{tab:mRNA:miRNA:sd} (b) but the fact
that,  except for  poly~Q  lengths  Q80 and  Q92,  the poly~Q  length-specific
empirical standard deviation increases as age increases.  We do not comment on
the  empirical  means  because  they  are  all  very  small  compared  to  the
corresponding empirical standard deviations.

\begin{table}[!ht]
  \begin{subtable}{.5\linewidth}
    \centering
    \begin{tabular}{| c | c | c | c |}
      \hline
      poly Q length & Age $2$ & Age $6$ & Age $10$ \\ 
      \hline
      Q80             & $1$         & $0.646$ & $1.39$\\
      Q92             & $0.886$  & $1.02$   & $1.48$\\
      Q111           & $0.964$   & $1.21$  & $3.08$\\
      Q140          & $0.805$   & $1.70$  & $4.11$\\
      Q175          & $1.24$     & $1.86$  & $4.32$\\
      \hline
    \end{tabular}
    \caption{mRNA}
  \end{subtable}%
  \quad
  \begin{subtable}{.5\linewidth}
    \begin{tabular}{| c | c | c | c |}
      \hline
      poly Q length & Age $2$     & Age $6$     & Age $10$ \\ 
      \hline
      Q80             & $1$        & $2.35$   & $1.03$ \\
      Q92             & $0.516$ & $1.06$   & $0.956$\\
      Q111            & $0.655$ & $0.722$ & $2.15$ \\
      Q140           & $0.698$ & $1.92$   & $2.72$ \\
      Q175           & $0.588$ & $1.80$   & $3.34$ \\
      \hline         
    \end{tabular}
    \caption{miRNA}
  \end{subtable}
  \caption{For each  level of poly~Q length  (Q80, Q92, Q111, Q140,  Q175) and
    age (2, 6, 10 months) we computed the empirical standard deviation of mRNA
    (a)  and miRNA  (b)  gene  expressions, all  normalized  by the  empirical
    standard deviation at poly~Q  length Q80 and 2 months of  age (that is, by
    0.0475 for mRNA and 0.0660 for miRNA).}
  \label{tab:mRNA:miRNA:sd}
\end{table}

\section{Elements of optimal transport}
\label{sec:elements}

Let
$\Omega :=  \{\omega \in  (\bbR_{+})^{M} | \sum_{m\in  \llbracket M\rrbracket}
\omega_{m}    =    1\}$    be     the    $(M-1)$-dimensional    simplex    and
$\bar{\omega} :=  M^{-1} \mathbf{1}_{M}$, where $\mathbf{1}_{M}  \in \bbR^{M}$
is the vector with  all its entries equal to 1.  For  any $\omega \in \Omega$,
define
\begin{equation*}
  \Pi(\omega) := \{P \in (\bbR_+)^{M\times N} | P \mathbf1_N =
  \omega, P^\top \mathbf1_M = N^{-1}\mathbf{1}_{N}\}
\end{equation*}
and                                                                        let
$\mu_X^{\omega} := \sum_{m\in\llbracket M\rrbracket} \omega_{m} \delta_{x_m}$,
$\nu_Y  := N^{-1}  \sum_{n\in \llbracket  N \rrbracket}  \delta_{y_n}$ be  the
$\omega$-weighted empirical measure attached to  $X$ and the empirical measure
attached to  $Y$. An element  $P$ of $\Pi(\omega)$  represents a joint  law on
$X \times Y$ with marginals $\mu_X^{\omega}$ and $\nu_Y$.

The celebrated  Monge-Kantorovich problem~\cite[Chapter~2]{COT19}  consists in
finding a joint law over $X  \times Y$ with marginals $\mu_X^{\bar\omega}$ and
$\nu_Y$ that  minimizes the expected  cost of  transport with respect  to some
cost function $c : \bbR^{d} \times  \bbR^{d} \rightarrow \bbR_+$.  We focus on
$c$  given by  $c(x,y)  :=  \|x-y\|_{2}^{2}$ (the  squared  Euclidean norm  in
$\bbR^{d}$).  Specifically, denoting $C_{X,Y} \in  \bbR^{M \times N}$ the cost
matrix    given   by    $(C_{X,Y})_{mn}    :=    c(x_{m},y_{n})$   for    each
$(m,n) \in \llbracket M\rrbracket  \times \llbracket N\rrbracket$, the problem
consists in solving
% \begin{equation*}
$\min_{P \in \Pi(\bar\omega)} \langle C_{X,Y}, P \rangle_F$
% \end{equation*}
where
$\langle C_{X,Y}, P \rangle_F := \sum_{(m,n) \in \llbracket M\rrbracket \times
  \llbracket N\rrbracket} (C_{X,Y})_{mn} P_{mn}$  is the $P$-specific expected
cost of transport from $X$ to $Y$.

It is well known  that it is very rewarding from  a computational viewpoint to
consider a  regularized version of the  above problem~\cite[Chapter~4]{COT19}.
The penalty term  is proportional to the discretized entropy  of $P$, that is,
to
$E(P)  :=   -  \sum_{(m,n)   \in  \llbracket  M\rrbracket   \times  \llbracket
  N\rrbracket]} P_{mn} (\log P_{mn}  -1)$.  The regularized problem (presented
here for  any $\omega  \in \Omega$  beyond the  case $\omega  = \bar{\omega}$)
consists, for  some user-supplied $\gamma  > 0$, in finding  $P_{\gamma}$ that
solves
\begin{equation}
  \label{eq:eq1}
  \calW_{\gamma}\left(\mu_X^{\omega}, \nu_Y \right) := \min_{P \in \Pi(\omega)}
  \left\{\langle C_{X,Y}, P \rangle_F - \gamma E(P)\right\}.  
\end{equation}
One  of the  advantages  of  entropic regularization  is  that  one can  solve
\eqref{eq:eq1} efficiently using the Sinkhorn-Knopp matrix scaling algorithm.

Finally, following~\cite{Genevay18}, we use  $\calW_{\gamma}$ to define the so
called Sinkhorn loss between $\mu_{X}^{\omega}$  (any $\omega \in \Omega$) and
$\nu_{Y}$ as
\begin{equation*}
  \bar{\calW}_{\gamma}\left(\mu_X^{\omega},        \nu_Y       \right)        :=       2
  \calW_{\gamma}\left(\mu_X^{\omega},         \nu_Y          \right)         -
  \calW_{\gamma}\left(\mu_X^{\omega},      \mu_X^{\omega}       \right)      -
  \calW_{\gamma}\left(\nu_Y, \nu_Y \right). 
\end{equation*}
This loss interpolates  between $\calW_{0}\left(\mu_X^{\omega}, \nu_Y \right)$
and   the   maximum  mean   discrepancy   of   $\mu_X^{\omega}$  relative   to
$\nu_Y$~\cite[Theorem~1]{Genevay18}.       Paraphrasing      the      abstract
of~\cite{Genevay18},  the  interpolation  allows  to  find  ``a  sweet  spot''
leveraging   the   geometry   of   optimal   transport   and   the   favorable
high-dimensional sample  complexity of  maximum mean discrepancy,  which comes
with unbiased gradient estimates.

\section{Optimal transport-based machine learning}
\label{sec:main}

In  this  section  we  introduce two  co-clustering  algorithms  and  one
  matching  algorithm, all  based on  the  solution of  a master  optimization
  program.      The      optimization     program     is      presented     in
  Section~\ref{subsec:method}   and   the    algorithms   are   presented   in
  Section~\ref{subsec:algos}.
  
\subsection{Stage 1:  the master  optimization program  and how  to solve
    it}
\label{subsec:method}

We  introduce  a  parametric  model $\Theta$  consisting  of  affine  mappings
$\theta      :      \bbR^{d}      \to     \bbR^{d}$      of      the      form
$x   \mapsto    \theta   (x)   =    \theta_{1}   x   +    \theta_{2}$,   where
$\theta_1\in\mathbb R^{d\times d}$ and $\theta_2\in\mathbb R^{d}$.  The formal
definition  of   $\Theta$  is  given  in   Appendix~\ref{sec:supp:mat}.   Each
$\theta \in \Theta$  is a candidate to formalize  the aforementioned mirroring
relationship.   The   set  $\Theta$   imposes  constraints  on   the  matrices
$\theta_{1}$,  in  particular  that  their  diagonals  are  made  of  negative
values. Of course, minus identity  belongs to $\Theta$. The parametrization is
identifiable,    in   the    sense   that    $\theta   =    \theta'$   implies
$(\theta_{1},\theta_{2}) = (\theta_{1}', \theta_{2}')$.  It is noteworthy that
\textit{any} identifiable, regular model $\Theta$  could be used.  We focus on
$\Theta$ as defined in  Appendix~\ref{sec:supp:mat} because of the application
that     we     consider     in    Section~\ref{sec:real:data}     (and     in
Section~\ref{sec:simulation}).

By   analogy   with   Section~\ref{sec:elements}   we   introduce,   for   any
$\theta   \in   \Theta$,    $\omega   \in   \Omega$   and    $\gamma   >   0$,
$\theta(X) :=  \{\theta(x_{1}), \ldots, \theta(x_{M})\}$  the image of  $X$ by
$\theta$;  the $\omega$-weighted  empirical measure  attached to  $\theta(X)$,
$\mu_{\theta(X)}^{\omega}  :=   \sum_{m\in\llbracket  M\rrbracket}  \omega_{m}
\delta_{\theta(x_{m})}$;   the   cost   matrix  $C_{\theta(X),Y}$   given   by
$(C_{\theta(X),Y})_{mn}     :=    c(\theta(x_{m}),     y_{n})$    for     each
$(m,n) \in \llbracket M\rrbracket \times \llbracket N\rrbracket$; and
\begin{equation}
  \label{eq:eq2}
  \calW_{\gamma}\left(\mu_{\theta(X)}^{\omega}, \nu_Y  \right) = \min_{P
    \in  \Pi(\omega)} \left\{\langle  C_{\theta(X),  Y},  P \rangle_F  -
    \gamma E(P)\right\}
\end{equation}
where
$\langle   C_{\theta(X),Y},  P   \rangle_F  :=   \sum_{(m,n)  \in   \llbracket
  M\rrbracket \times \llbracket N\rrbracket} (C_{\theta(X),Y})_{mn} P_{mn}$ is
the $P$-specific expected cost of transport from $\theta(X)$ to $Y$.

Fix arbitrarily  $\omega \in \Omega$. The  first program that we  introduce is
the $\omega$-specific program
\begin{equation}
  \label{eq:prog1}
  \min_{\theta   \in  \Theta}   \bar{\calW}_{\gamma}\left(\mu_{\theta(X)}^{\omega},
    \nu_Y\right), 
\end{equation}
where  we  are   interested  in  the  minimizer   $\hat{\theta}$  that  solves
\eqref{eq:prog1}     \textit{and}    in     the    optimal     joint    matrix
$\hat{P} \in \Pi(\omega)$ that solves
\begin{equation*}
  \min_{P \in \Pi(\omega)}  \left\{\langle  C_{\hat{\theta}(X),  Y},  P \rangle_F  -
    \gamma E(P)\right\}.
\end{equation*}

In  words,  we  look  for  an  $\omega$-specific  optimal  mirroring  function
$\hat{\theta}$ and its $\omega$-specific optimal transport plan $\hat{P}$.

How to  choose $\omega$?  We  decide to optimize  with respect to  $\omega$ as
well. This  additional optimization is  relevant because  we do not  expect to
associate  a  $y_{n}$  to  every   $x_{m}$  eventually  at  the  co-clustering
stage. So, our master program is
\begin{equation}
  \label{eq:prog2}
  \min_{\omega      \in       \Omega}      \min_{\theta       \in      \Theta}
  \bar{\calW}_{\gamma}\left(\mu_{\theta(X)}^{\omega}, \nu_Y\right),
\end{equation}
where  we  are  interested   in  the  minimizer  $(\hat{\omega},\hat{\theta})$
\textit{and} in the optimal matrix $\hat{P} \in \Pi(\hat{\omega})$ that solves
\begin{equation}
  \label{eq:secondary}
  % \calW_{\gamma}\left(\mu_{\hat{\theta}(X)}^{\hat{\omega}}, \nu_Y\right) =  
  \min_{P  \in  \Pi(\hat{\omega})}  \left\{\langle C_{\hat{\theta}(X),  Y},  P
    \rangle_F -     \gamma E(P)\right\}.
\end{equation}

We propose to solve \eqref{eq:prog2} iteratively by updating $\omega$ and then
$\theta$.  At  round $t$, given $\omega_{t}$,  we make one step  of mini-batch
gradient descent to derive $\theta_{t+1}$  from $\theta_{t}$ (here, we notably
rely on  the Sinkhorn-Knopp algorithm).  Given  $\theta_{t+1}$, $\omega_{t+1}$
is chosen proportional to the vector in $(\bbR_{+})^{M}$ whose $m$th component
equals
$h^{-1}\sum_{n     \in     \llbracket      N\rrbracket}     \varphi((y_n     -
\theta_{t+1}(x_m))/h)$ where $\varphi$ is the  standard normal density and $h$
is   the    arithmetic   mean    of   the    $c(y_{n},   y_{n'})$    for   all
$n \neq n'  \in \llbracket N\rrbracket$. Eventually, once the  final round $T$
is completed, we compute $\tilde{P} \in \Pi(\omega_{T})$ that solves
\begin{equation*}
  \min_{P \in \Pi(\omega_{T})} \left\{\langle  C_{\theta_T(X), Y}, P \rangle_F
    -     \gamma E(P)\right\}.
\end{equation*}
(again, we rely on the Sinkhorn-Knopp algorithm).

The    algorithm    to    solve     \eqref{eq:prog2}    is    summarized    in
Procedure~\ref{algo:SGD}.   We have  no  guarantee that  it converges.   Note,
however, that using the Sinkhorn-Knopp algorithm to solve \eqref{eq:secondary}
for     a     given      $(\hat{\omega},\hat{\theta})$     is     known     to
converge~\cite[Theorem~4.2]{COT19}. 

In light of \citep[Section  1.3, page~25]{MAD2019}, we inject problem-specific
knowledge onto two of the three main components of the transportation problem:
the  representation  spaces  (via  the  mapping  $\theta$)  and  the  marginal
constraints  (via  the weight  $\omega$),  leaving  aside the  cost  function.
Furthermore, we resort to mini-batch  gradient descent because the algorithmic
complexity  prevents the  direct  computation  using the  whole  data set.   A
theoretical analysis of this practice is proposed in~\citep{fatras2020}.

We can now  exploit $\tilde{P}$ so as to derive  relevant associations between
mRNAs and  miRNAs.  We  propose two  approaches.  On the  one hand,  the first
approach  outputs   \textit{bona  fide}  co-clusters.   We   expect  that  the
co-clusters  can  associate  many  mRNAs  with many  miRNAs,  thus  making  it
difficult to interpret and analyze the results.  On the other hand, the second
approach rather  \textit{matches} each mRNA with  at most $k$ miRNAs  and each
miRNA   with  at   most   $k'$   mRNAs  ($k$   and   $k'$  are   user-supplied
integers). Details follow.

\subsection{Stage 2: co-clustering or matching}
\label{subsec:algos}

\subsubsection{Co-clustering.}
\label{subsubsec:co-clust}

To carry  out the  co-clustering task  once $\tilde{P}$  has been  derived, we
propose to  rely either  on spectral  co-clustering (we  will use  the acronym
SCC)~\cite{10.1145/502512.502550}, applying it once or twice, or co-clustering
based  on   latent  block  models~\cite{Nadif2010}.   Of   course,  any  other
co-clustering algorithm could  be used as well.  Specifically,  we develop the
following  algorithms (the  acronym  WTOT stands  for weighted  transformation
optimal transport).
\begin{description}
\item[WTOT-SCC1.]   Algorithm WTOT-SCC1  applies  SCC  \textit{once} to  build
  \textit{bona  fide} co-clusters  based on  $\tilde{P}$.  It  is required  to
  provide  a number  of  clusters.  We  rely on  a  criterion involving  graph
  modularity   to    learn   from    the   data    a   relevant    number   of
  clusters~\cite[Sections~2 and 4]{AILEM2016160}.

  In  our  simulation study,  we  also  consider algorithm  WTOT-SCC1$^*$,  an
  oracular  version   of  WTOT-SCC1   that  benefits   from  relying   on  the
  \textit{true} number of clusters.  This allows to assess how relevant is the
  learned number of clusters in WTOT-SCC1.
\item[WTOT-SCC2.]   Algorithm WTOT-SCC2  applies SCC  \textit{twice} to  build
  \textit{bona fide}  co-clusters based on  $\tilde{P}$. It proceeds  in three
  successive steps.
  \begin{itemize}
  \item In  step~1, WTOT-SCC2 applies  SCC a first  time to derive  an initial
    co-clustering.   A  relevant  number  of  co-clusters  is  learned  as  in
    WTOT-SCC1.
  \item  In  step~2, WTOT-SCC2  selects  and  removes  some rows  and  columns
    corresponding  to  mRNAs  and  miRNAs that  are  deemed  irrelevant.   The
    selection is based on a numerical criterion computed from $\tilde{P}$.  In
    our simulation study (Section~\ref{sec:simulation}),  all rows and columns
    that correspond to  diagonal blocks with a variance larger  than two times
    the overall variance of $\tilde{P}$ are  selected and removed. In the real
    data  application (Section~\ref{sec:real:data}),  we implement  and use  a
    different procedure.
  \item In step~3, WTOT-SCC2 applies SCC a second time, the relevant number of
    co-clusters being learned as in WTOT-SCC1.
  \end{itemize}
  In  our  simulation study,  we  also  consider algorithm  WTOT-SCC2$^*$,  an
  oracular version of  WTOT-SCC2 that is provided the  \textit{true} number of
  clusters for  its third  step.  This  allows to assess  how relevant  is the
  sub-procedure to learn the numbers of clusters in WTOT-SCC2.
\item[WTOT-BC.]   Algorithm WTOT-BC  applies  the so  called block  clustering
  algorithm to build  \textit{bona fide} co-clusters based  on $\tilde{P}$. It
  is required to provide the row-  and column-specific numbers of clusters. We
  rely on  an integrated completed likelihood  criterion~\cite{Keribin2014} to
  learn relevant values from the data.
\end{description}
The co-clusters  obtained \textit{via} WTOT-SCC1, WTOT-SCC2  or WTOT-BC should
reveal the interplay between the (remaining, as far as WTOT-SCC2 is concerned)
mRNAs and miRNAs in HD.

\subsubsection{Matching.}
\label{subsubsec:matching}

The larger $\tilde{P}_{mn}$ is, the more we are encouraged to believe that the
profiles $x_{m}$  and $y_{n}$ reveal  a strong relationship between  the $m$th
mRNA and  the $n$th miRNA.   This simple  rule prompts the  following matching
procedure applied once $\tilde{P}$ has been derived.

\begin{description}
\item[WTOT-matching.] Fix two integers $k,k' \geq 1$ and let $\tilde{\tau}$ be
  the quantile  of order  $q$ of  all the entries  of $\tilde{P}$.   For every
  $m  \in  \llbracket M\rrbracket$  and  $n  \in \llbracket  N\rrbracket$,  we
  introduce
  \begin{align*}
    \calN_m^0
    &:=  \Big\{ n  \in  \llbracket N\rrbracket:  \tilde{P}_{mn} \in  \{  \tilde{P}_{m(1)}, \ldots,
      \tilde{P}_{m (k)} \} \textrm{ and } \tilde{P}_{mn}\geq \tilde{\tau} \Big \}, \\
    \calM_n^0
    &:=  \Big \{  m  \in  \llbracket M\rrbracket: \tilde{P}_{mn}  \in  \{ \tilde{P}_{(1)n},  \ldots,
      \tilde{P}_{(k')n} \} \textrm{ and } \tilde{P}_{mn}\geq \tilde{\tau} \Big \} 
  \end{align*}
  where  $\tilde{P}_{m(1)}, \ldots,  \tilde{P}_{m (k)}$  are  the $k$  largest
  values     among     $\tilde{P}_{m1},      \ldots,     \tilde{P}_{mN}$     and
  $\tilde{P}_{(1)n},  \ldots, \tilde{P}_{(k')m}$  are  the  $k'$ largest  values
  among $\tilde{P}_{1n}, \ldots,  \tilde{P}_{Mn}$. For instance, $\calN_{m}^{0}$
  identifies  the  miRNAs  that  are  the  $k$ more  likely  to  have  a  strong
  relationship with  the $m$th  mRNA.  However,  this does  not qualify  them as
  relevant matches yet. In order to  keep only matches that are really relevant,
  we also introduce, for each $m \in \llbracket M\rrbracket$ and $n \in \llbracket N\rrbracket$,
  \begin{align*}
    \calN_m
    &:= \calN^{0}_{m} \cap \{   n \in \llbracket N\rrbracket: m\in \calM_n^0\},\\
    \calM_n
    &:= \calM^{0}_{n} \cap \{   m \in \llbracket M\rrbracket: n\in \calN_m^0 \}.
  \end{align*}
  Algorithm WTOT-matching outputs the collections  $\{\calN_{m} : m \in \llbracket M\rrbracket\}$
  and $\{\calM_{n} : n \in \llbracket N\rrbracket\}$.
\end{description}
Now if, for  instance, $n \in \calN_{m}$  then $y_{n}$ is among  the $k$ miRNA
profiles upon which $\tilde{P}$ puts  more mass when it ``transports'' $x_{m}$
onto  $Y$ \textit{and}  $x_{m}$ is  among the  $k'$ mRNA  profiles upon  which
$\tilde{P}$ puts more mass when it ``transports'' $y_{n}$ onto $X$.

Note  that we  expect that  some $\calN_{m}$  and $\calM_{n}$  will be  empty,
depending on $k$ and  $k'$. The mRNAs and miRNAs worthy  of interest are those
for which $\calN_{m}$ and $\calM_{n}$ are not empty. The integers $k$ and $k'$
should be chosen  relatively small, to make their  interpretation and analysis
feasible, but not too small because otherwise few matchings will be made.

In the  simulation study, we  use $k=k'$ between 2  and 200, depending  on the
simulation scheme.  Moreover,  we choose $q = 50\%$ so  that $\tilde{\tau}$ is
the median of the entries of $\tilde{P}$.

\subsection{Implementation}
\label{subsec:code}

Our    code    is    written    in   \texttt{python}    and    is    available
\href{https://github.com/yen-nguyen-thi-thanh/wtot_coclust_match}{here}.    We
adapt  the  Sinkhorn  algorithm  implemented by  Aude  Genevay  and  available
\href{https://github.com/audeg/Sinkhorn-GAN/blob/master/sinkhorn.py}{here}. The
stochastic  gradient  descents  relies   on  the  machine  learning  framework
\texttt{pytorch}.   We  use  the  implementation   of  SCC  available  in  the
\href{https://scikit-learn.org/stable/modules/generated/sklearn.cluster.SpectralCoclustering.html}{\texttt{sklearn}}
\texttt{python} module.   To learn a relevant  number of clusters, we  rely on
the
\href{https://coclust.readthedocs.io/en/v0.2.1/api/evaluation.html}{\texttt{coclust}}
\texttt{python}      module.       Finally,      we      rely      on      the
\href{https://cran.r-project.org/web/packages/blockcluster/index.html}{\texttt{blockcluster}}
\texttt{R} package to carry out block clustering.

Our algorithms bear a similarity  to the one developed in~\cite{Laclau17}. The
main differences are \textit{(i)} our use of the parametric model $\Theta$ and
weights $\omega$, \textit{(ii)} the fact that we apply SCC or block clustering
to  the  approximation  of  the optimal  transport  matrix  $\tilde{P}$.   Our
algorithms  also  bear  a  similarity  to~\cite{yang2021teaser},  a  fast  and
certifiable  point  cloud  registration  algorithm.   We  plan  to  study  the
similarities and differences closely.

\section{Simulation study}
\label{sec:simulation}

To    assess   the    performances    of   the    algorithms   described    in
Section~\ref{subsec:method}, we conduct a simulation study in three parts.  As
we  go on,  the task  gets more  difficult.   In all  cases, the  laws of  the
synthetic observations are  mixtures of Gaussian laws.   Overall 12 simulation
scenarios are considered.

We think that  the first two simulation schemes produce  unrealistic data and,
on the contrary, that the  third simulation scheme produces somewhat realistic
data. The diversity of the synthetic mRNA and miRNA profiles obtained by using
Lloyd's  $k$-means  algorithm  in  order  to summarize  the  variety  of  real
profiles, see Section~\ref{subsec:k:means}, encouraged  us to rely on mixtures
in order  to simulate  data.  We  chose mixtures of  Gaussian laws  because of
their ubiquity and versatility.

In Section~\ref{subsec:simul:A}, the weights of the mixtures and parameters of
the Gaussian  laws are chosen by  us. Moreover, the two  mixtures (to simulate
$X$  and  $Y$)  share  the  same   weights  and  induce  a  perfect  mirroring
relationship  (details  below),  thus   making  the  co-clustering  task  less
difficult.  In  Section~\ref{subsec:simul:B}, the weights of  the mixtures and
parameters  of the  Gaussian laws  are randomly  generated. Moreover,  the two
mixtures do not share  the same weights and do not  induce a perfect mirroring
relationship anymore, so  that the co-clustering task is  much more difficult.
Finally, in Section~\ref{subsec:simul:C}, we use  plus or minus real, randomly
chosen miRNA profiles  \textit{and} $\textbf{0}_{d}$ as means  of the Gaussian
laws to simulate $X$ and $Y$, in such a way that there is no perfect mirroring
relationship.  We think that the  corresponding co-clustering task is the most
difficult of the three.

Section~\ref{subsec:Laclau}  briefly introduces  two  competing algorithms  to
identify  matchings~\cite{Laclau17}.   Section~\ref{subsec:competition}  lists
all   the   algorithms   that   compete    in   the   simulation   study   and
Section~\ref{subsec:criteria} presents the measure  of discrepancy between two
co-clusterings and  the matching criteria that  we rely on to  assess how well
the  algorithms perform.   Sections~\ref{subsec:simul:A}, \ref{subsec:simul:B}
and \ref{subsec:simul:C}  present in  turn the data-generating  mechanisms and
report the results in terms of co-clustering and matching performances.

\subsection{Two ``Gromov-Wasserstein co-clustering'' algorithms}
\label{subsec:Laclau}

We  compare   our  algorithms   with  two  co-clustering   algorithms  adapted
from~\cite{Laclau17}.   For self-containedness,  we summarize  here how  these
algorithms work.

The  first  step of  both  algorithms  consists  in computing  the  similarity
matrices       $K_{X}       \in        (\bbR_{+})^{M\times       M}$       and
$K_{Y} \in (\bbR_{+})^{N\times N}$ given by
\begin{align*}
  (K_{X})_{mm'}
  &  :=  \exp\left\{-\frac{\|x_{m}  -  x_{m'}\|_{2}^{2}}{2\ell_{X}^{2}}\right\}
    \quad (m, m'\in\llbracket M\rrbracket),\\
  (K_{Y})_{nn'}
  &  :=  \exp\left\{-\frac{\|y_{n}  -  y_{n'}\|_{2}^{2}}{2\ell_{Y}^{2}}\right\}
    \quad (n,n '\in\llbracket N\rrbracket)
\end{align*}
where  $\ell_{X}$  (respectively, $\ell_{Y}$)  is  the  mean of  all  pairwise
Euclidean  distances between  elements  of $X$  (respectively,  of $Y$).   The
similarity matrices $K_{X}$ and $K_{Y}$ now  represent $X$ and $Y$ through the
lens of the so called radial basis function kernel.

For any integers  $a,b \geq 1$ and  pair of matrices $A  \in \bbR^{a\times a}$
and $B \in \bbR^{b\times b}$, define
\begin{align}
  \notag
  \Pi_{a,b}
  & := \left\{P \in (\bbR_{+})^{a \times b} | P\textbf{1}_{b} =
    a^{-1}\textbf{1}_{a},       P^{\top}       \textbf{1}_{a}      =       b^{1}
    \textbf{1}_{b}\right\}, \\
  \notag
  \langle [A,B], [P,P]\rangle_{F}
  & := \sum_{i,k\in \llbracket a\rrbracket, j,\ell\in \llbracket b\rrbracket} (A_{ik} - B_{j\ell})^{2} P_{ij}P_{k\ell}
    \quad (P \in \Pi_{a,b}),\\
  \label{eq:GW}
  \GW_{\gamma} (A,B)
  &  :=  \min_{P \in  \Pi_{a,b}}  \{\langle  [A,B], [P,P]\rangle_{F}  -  \gamma
    E(P)\
\end{align}
where
$E(P) :=  -\sum_{(i,j)\in\llbracket a\rrbracket\times  \llbracket b\rrbracket}
P_{ij} (\log P_{ij}  - 1)$. The quantity $\GW_{\gamma} (A,B)$  is known in the
literature  as  an entropic  Gromov-Wasserstein  discrepancy  between $A$  and
$B$. It can be used to define an entropic Gromov-Wasserstein barycenter of $A$
and  $B$  and  its   barycenter  transport  matrices.   Specifically,  setting
$s    =    \lfloor\frac{1}{2}(a+b)\rfloor$    (one   choice    among    many),
$(\hat{\Gamma}, \hat{P}_{A},  \hat{P}_{B}) \in (\bbR_{+})^{s \times  s} \times
\Pi_{s,a} \times \Pi_{s,b}$ that solves
\begin{equation}
  \label{eq:bary}
  \min_{\Gamma,  P_{A},   P_{B}} \frac{1}{2} \left\{\Big(\langle  [\Gamma,A],
    [P_{A},P_{A}]\rangle_{F} - \gamma E(P_{A}) \Big) + \Big(\langle  [\Gamma,B],
    [P_{B},P_{B}]\rangle_{F} - \gamma E(P_{B}) \Big)\right\} 
\end{equation}
(where        $(\Gamma,        P_{A},        P_{B})$        ranges        over
$(\bbR_{+})^{s  \times   s}  \times   \Pi_{s,a}  \times  \Pi_{s,b}$)   can  be
interpreted  as a  barycenter between  $A$  and $B$  ($\hat{\Gamma}$) and  the
optimal transport matrices between  $\hat{\Gamma}$ and $A$ ($\hat{P}_{A}$) and
between $\hat{\Gamma}$ and $B$ ($\hat{P}_{B}$).

The  second step  of the  algorithms  consists either  in solving  numerically
\eqref{eq:GW}  with $(A,B)  =  (K_{X}, K_{Y})$,  yielding  $\tilde{Q}$, or  in
solving numerically \eqref{eq:bary} with $(A,B) = (K_{X}, K_{Y})$, yielding in
particular the transport matrices $\tilde{Q}_{X}$ and $\tilde{Q}_{Y}$. We call
CCOT-GWD  and  CCOT-GWB the  corresponding  algorithms.   In both  cases,  the
Sinkhorn-Knopp algorithm is used and provides solutions that decompose as
\begin{align*}
  \tilde{Q} & = \diag(\rho) \xi \diag(\rho'),\\
  \tilde{Q}_{X} & = \diag(\rho_{X}) \xi_{X} \diag(\rho_{X}'),\\
  \tilde{Q}_{Y} & = \diag(\rho_{Y}) \xi_{Y} \diag(\rho_{Y}'),
\end{align*}
for some  $\rho, \rho_{X} \in  \bbR^{M}$, $\rho', \rho_{Y}'  \in \bbR^{N}$,
$\rho_{X},          \rho_{Y}           \in          \bbR^{s}$          and
$\xi \in \bbR^{M\times N}, \xi_{X} \in  \bbR^{s \times M}, \xi_{Y} \in \bbR^{s
  \times N}$~\cite{peyre:hal-01322992}.

The   third  and   last   step   builds  upon   either   $(\rho,  \rho')$   or
$(\rho_{X}', \rho_{Y}')$  to derive  partitions of $X$  and $Y$,  by detecting
``jumps''   along  the   vectors.   The   two  partitions   finally  yield   a
co-clustering.

\subsection{Listing all competing algorithms}
\label{subsec:competition}

We  run  and  compare  algorithms WTOT-SCC1,  WTOT-SCC2  (and  their  oracular
counterparts  WTOT-SCC1$^*$,  WTOT-SCC2$^*$), WTOT-BC  on  the  one hand  (see
Sections~\ref{subsubsec:co-clust}) and CCOT-GWD and CCOT-GWB on the other hand
(see  Section~\ref{subsec:Laclau}).   In  addition,   we  also  run  algorithm
WTOT-matching (see Section~\ref{subsubsec:matching}). 

For CCOT-GWD,  we set $\gamma =  0.1$ in \eqref{eq:GW}.  For  CCOT-GWB, we set
$\gamma=0.05$ in \eqref{eq:bary}.  We tried  several values and chose the ones
that yielded the smallest errors. 

In   view  of   Procedure~\ref{algo:SGD},   we   choose  $\widetilde{M}$   and
$\widetilde{N}$   equal   approximately    $M/2$   and   $N/2$   respectively,
$(\eta, \gamma_{0})  = (1,0)$ (no  decay), $T =  500$, and an  initial mapping
$\theta_{0}$ drawn randomly (see Appendix~\ref{sec:supp:mat} for details).

We checked that  varying $\widetilde{M}$ and $\widetilde{N}$  around $M/2$ and
$N/2$ had little impact if any.   Likewise, the randomly drawn initial mapping
$\theta_{0}$ had little impact if any.  Moreover, varying $\underline{\gamma}$
in    $[\tfrac{1}{2}\times   \gamma^{*};    2    \times   \gamma^{*}]$    with
$\gamma^{*} = \text{mean}  \{\|x - x'\|_{2} :  x, x' \in X\}$  also had little
impact if any.  We did not rigorously  check the impact of the total number of
iterations  $T$, but  we  observed  that numerical  convergence  seemed to  be
reached  for fewer  iterations than  $T$. Finally,  we did  not challenge  the
choice of $h = \text{mean} \{\|y - y'\|_{2} : y, y' \in Y\}$.

\subsection{Assessing performances}
\label{subsec:criteria}

\paragraph*{A measure of discrepancy between two co-clusterings.}

In order to  assess the quality of  the co-clusterings that we  derive, and to
compare  performances, we  propose  to  rely on  a  commonly  used measure  of
discrepancy  between two  co-clusterings.  Its  definition extends  that of  a
measure of discrepancy between partitions that we first present.

Let $z$  and $z'$ be two  partitions of the set  $\llbracket M\rrbracket$ into
$K$       components,       taking       the      form       of       matrices
$z=(z_{mk})_{m\in\llbracket   M\rrbracket,k\in\llbracket   K\rrbracket}$   and
$z'=(z_{mk}')_{m\in\llbracket  M\rrbracket,k\in\llbracket  K\rrbracket}$  with
convention  $z_{mk} =  1$  (respectively, $z_{mk}'  = 1$)  if  $m$ belongs  to
component $k$ of $z$ (respectively,  $z'$) and 0 otherwise.  The corresponding
confusion matrix $C(z,z')  = (c_{k\ell})_{k,\ell\in\llbracket K\rrbracket}$ is
given by  $c_{k\ell} := \sum_{m\in \llbracket  M\rrbracket} z_{mk} z_{m\ell}'$
(every $k,\ell \in  \llbracket K\rrbracket$).  Suppose that the  labels of the
partitions $z$ and $z'$ are such that
\begin{equation*}
  \text{Tr}(C(z,z'))    =     \max_{\sigma    \in     \Sigma_{K}}    \text{Tr}
  (C(z,(z_{m\sigma(k)}')_{m\in\llbracket M\rrbracket, k \in\llbracket K\rrbracket})), 
\end{equation*}
where $\Sigma_{K}$ is  the set of permutations of the  elements of $\llbracket K\rrbracket$. Then
the proportion
\begin{equation}
  \label{eq:ezz'}
  \delta(z, z') := 1 - \frac{1}{M} \sum_{m\in\llbracket M\rrbracket,k\in\llbracket K\rrbracket} z_{mk}z'_{mk}
\end{equation}
is  a natural  measure  of  discrepancy between  $z$  and  $z'$. As  suggested
earlier, the measure can be extended to compare pairs of partitions.

Consider now  $(z, w)$ and  $(z', w')$ two pairs  of partitions, $z$  and $z'$
partitioning  $\llbracket  M\rrbracket$  into  $K$ components,  $w$  and  $w'$
partitioning  $\llbracket  N\rrbracket$  into $L$  components.   We  represent
$(z,w)$                  and                   $(z',w')$                  with
\begin{equation*}
  u       =      (u_{mnk\ell})_{m\in\llbracket       M\rrbracket,n\in\llbracket
    N\rrbracket,k\in\llbracket  K\rrbracket,\ell\in\llbracket L\rrbracket}
\end{equation*}
and
\begin{equation*}
 u'      =      (u_{mnk\ell}')_{m\in\llbracket      M\rrbracket,n\in\llbracket
  N\rrbracket,k\in\llbracket K\rrbracket,\ell\in\llbracket L\rrbracket} 
\end{equation*}
where      $u_{mnk\ell}     :=      z_{mk}      \times     w_{n\ell}$      and
$u_{mnk\ell}'     :=     z_{mk}'     \times     w_{n\ell}'$     (for     every
$m\in\llbracket    M\rrbracket,   n\in\llbracket    N\rrbracket,k\in\llbracket
K\rrbracket,\ell\in\llbracket L\rrbracket$),  supposing again that  the labels
of the partitions  $z$, $z'$ on the one  hand and $w$, $w'$ on  the other hand
maximize the traces of the confusion matrices $C(z,z')$ and $C(w,w')$ as above
(then two  pairs of partitions  define without ambiguity a  co-clustering). By
analogy with \eqref{eq:ezz'}, the proportion
\begin{equation}
  \label{eq:Delta}
  \Delta((z,w),             (z',w'))             :=            1             -
  \frac{1}{KL}\sum_{m\in\llbracket M\rrbracket,n\in\llbracket N\rrbracket,k\in\llbracket K\rrbracket,\ell\in\llbracket L\rrbracket} u_{mnk\ell}u_{mnk\ell}'  
\end{equation}
is a  measure of discrepancy  between $(z,w)$ and  $(z',w')$. It can  be shown
that 
 \begin{equation}
   \Delta((z, w),  (z', w')) = \delta(z,  z') + \delta(w, w')  - \delta(z, z')
   \times \delta(w, w').
\end{equation}

In the rest of this section  we report means and standard deviations, computed
across 30 independent  replications of each analysis, of the  above measure of
discrepancy between the derived partition/co-clustering and the true one.

\paragraph*{Matching criteria.}  Set arbitrarily $m  \in \llbracket M\rrbracket$ and suppose that
we have  derived the subset  $\calN_{m} \subset  \llbracket N\rrbracket$ that matches  $x_{m}$ to
$\{y_{n} :  n \in \calN_{m}\}$.  Suppose  moreover that in reality  $x_{m}$ is
matched    to   $\{y_{n}    :    n   \in    \calN_{m}^{\star}\}$   for    some
$\calN_{m}^{\star} \subset \llbracket N\rrbracket$.  We propose to use three real-valued criteria
to compare $\calN_{m}$ with $\calN_{m}^{\star}$.

Let     $\TP_{m}     :=      \card(\calN_{m}     \cap     \calN_{m}^{\star})$,
$\FP_{m}      :=      \card(\calN_{m}     \cap      (\calN_{m}^{\star})^{c})$,
$\TN_{m}    :=     \card((\calN_{m})^{c}    \cap    (\calN_{m}^{\star})^{c})$,
$\FN_{m} :=  \card((\calN_{m})^{c} \cap \calN_{m}^{\star})$ be  the numbers of
true  positives,   false  positives,  true  negatives   and  false  negatives,
respectively. The so called $m$-specific
\begin{itemize}
\item precision: $\TP_{m}/(\TP_{m} + \FP_{m})$,
\item sensitivity: $\TP_{m}/(\TP_{m}+\FN_{m})$,
\item specificity: $\TN_{m}/(\TN_{m}+\FP_{m})$
\end{itemize}
quantify how  similar are  $\calN_{m}$ and $\calN_{m}^{\star}$,  larger values
indicating better concordance.

In the rest of this section  we report means and standard deviations, computed
across 30  independent replications of  each analysis,  of the average  of the
$m$-specific precision, sensitivity and specificity.  We also report means and
standard deviations, computed  across the same 30  independent replications of
each analysis, of
\begin{align*}
  \tilde{k}_r &:=  \frac{\sum_{m \in  \llbracket M\rrbracket }\card(\calN_{m})}{\card(\{  m\in \llbracket M\rrbracket:
                \calN_m \neq \emptyset\})},\\ 
  \tilde{k}_c &:=  \frac{\sum_{n \in \llbracket N\rrbracket }\card(\calM_{n})}{\card(\{  n\in \llbracket N\rrbracket:
                \calM_n \neq \emptyset\})}
\end{align*}
the  row-  and column-specific  averages  of  the  cardinalities of  the  sets
$\calN_{m}$ and $\calM_{n}$ that are not empty.

\subsection{First simulation study}
\label{subsec:simul:A}

\paragraph*{Simulation scheme.}

For      four      different      choices     of      the      hyperparameters
$M    \geq    200,    N    \geq    200,    K    \geq    2,    d    \geq    2$,
$\mu_{1},   \ldots,  \mu_{K}   \in  \bbR^{d}$,   $\sigma  \in   \bbR_{+}^{*}$,
$\alpha           \in            (\bbR_{+})^{K}$           such           that
$\sum_{k\in \llbracket K \rrbracket} \alpha_{k}  = 1$, we sample independently
$x_{1}, \ldots, x_{M}$ from the mixture of Gaussian laws
\begin{equation}
  \label{eq:mixture:x}
  \sum_{k\in \llbracket K\rrbracket} \alpha_k N(\mu_k, \sigma^{2} \text{Id}_{d})
\end{equation}
and $y_{1}, \ldots, y_{N}$ from
\begin{equation}
  \label{eq:mixture:y}
  \sum_{k\in\llbracket K\rrbracket} \alpha_k N(-\mu_k, \sigma^{2} \text{Id}_{d}).
\end{equation}

One  way to  sample  $x$  from the  mixture  \eqref{eq:mixture:x} consists  in
sampling a latent  label $u$ in $\llbracket K\rrbracket$  from the multinomial
law with parameter $(1; \alpha_{1},  \ldots, \alpha_{K})$ then in sampling $x$
from  the Gaussian  law  $N(\mu_{u},  \sigma^{2} \text{Id}_{d})$.   Similarly,
sampling  $y$ from  the mixture  \eqref{eq:mixture:y}  can be  carried out  by
sampling a latent  label $v$ in $\llbracket K\rrbracket$  from the multinomial
law with parameter $(1; \alpha_{1},  \ldots, \alpha_{K})$ then by sampling $y$
from the  Gaussian law $N(-\mu_{v},  \sigma^{2} \text{Id}_{d})$.  We  think of
$x$ and $y$  as having a mirrored  relationship if $u=v$.  In  this light, the
challenge that we tackle consists in finding such relationships without having
access to the latent labels.

Table~\ref{tab:SS1}    describes    the    four   configurations    that    we
investigate. Note that configuration A2 is more difficult to deal with than A1
because \textit{(i)}  the weights in $\alpha$  are balanced in the  latter and
unbalanced in the former, and  \textit{(ii)} because the variance $\sigma^{2}$
is smaller  in A1  than in A2.   Moreover, configurations A3  and A4  are more
challenging than A2 because there is  $K=4$ components in the Gaussian mixture
under A3 and A4 and $K=3$ components under A2.

\begin{table}[!h]
  \centering
  \begin{tabular}{|c|c|c|c|c|c|}\hline
    configuration & $(M,N)$ & $K$ & $\mu_{1}, \ldots, \mu_K$ & $\sigma^{2}$ & $\alpha$\\
    \hline
    A1 &  $(200, 200)$ & 3 & $\begin{pmatrix} 
      4.0 \\ 0.5 \\ 1.5 \\
    \end{pmatrix}, \begin{pmatrix} 
      1.8 \\ 4.5 \\ 1.1 \\
    \end{pmatrix}, \begin{pmatrix} 
      1.5 \\ 1.5 \\  5.5\\
    \end{pmatrix}$ & 0.10 & $(1/3,1/3,1/3)$\\
    A2 & $(300, 300)$ & 3 & $\begin{pmatrix}
      4.0 \\ 0.5 \\ 1.5 \\
    \end{pmatrix}, \begin{pmatrix} 
      1.8 \\ 4.5 \\ 5.1 \\
    \end{pmatrix}, \begin{pmatrix} 
      3.5 \\ 1.5 \\  5.5\\
    \end{pmatrix}$ & 0.15 & $(0.2,0.3,0.5)$\\
    A3 & $(400, 300)$ & 4 & $\begin{pmatrix}
      4.0 \\ 0.5 \\
    \end{pmatrix}, \begin{pmatrix} 
      0.5 \\ 3.5
    \end{pmatrix}, \begin{pmatrix} 
      7.5 \\ 7.8
    \end{pmatrix}, \begin{pmatrix}
      0.5 \\ 0.5
    \end{pmatrix}$ & 0.20 & $(0.4,0.2,0.2,0.2)$    \\
    A4 & $(300, 300)$ & 4 & $\begin{pmatrix}
      4.0 \\ 0.5 \\
    \end{pmatrix}, \begin{pmatrix} 
      0.5 \\ 3.5
    \end{pmatrix}, \begin{pmatrix} 
      7.5 \\ 7.8
    \end{pmatrix}, \begin{pmatrix}
      0.5 \\ 0.5
    \end{pmatrix}$ & 0.10 & $(0.5,0.2,0.1,0.2)$ \\\hline
  \end{tabular}
  \caption{Four   different   configurations    for   the   first   simulation
    scheme. Configuration A1 is less challenging  than A2 which is itself less
    challenging than A3 and A4.}
  \label{tab:SS1}
\end{table}

\paragraph*{Results.}
Thirty  times, independently,  we simulated  synthetic data  sets $X$  and $Y$
under  the simulation  scheme described  above,  then we  applied the  various
algorithms as presented in  Section~\ref{subsec:competition}. We summarize the
results               in              Tables~\ref{tab:simulA:results:coclust},
\ref{tab:simulA14:results:matching},   and  \ref{tab:simulA:results:matching}.
Table~\ref{tab:simulA:results:coclust}  summarizes the  results  of the  seven
algorithms   listed   in   Section~\ref{subsec:competition}   that   rely   on
\textit{bona         fide}        co-clustering         algorithms        (see
Section~\ref{subsubsec:co-clust}), that  is, of our  algorithms WTOT-SCC1$^*$,
WTOT-SCC1, WTOT-SCC2$^*$,  WTOT-SCC2, WTOT-BC$^*$  and of  algorithms CCOT-GWD
and   CCOT-GWB.    As   for   Tables~\ref{tab:simulA14:results:matching}   and
\ref{tab:simulA:results:matching}, they summarize the results of our algorithm
that relies on matching (see Section~\ref{subsubsec:matching}).

\begin{description}
\item[Table~\ref{tab:simulA:results:coclust}.]   Except  in configuration  A1,
  where  they  perform  equally  well,  our  algorithms  WTOT-SCC1,  WTOT-SCC2
  outperform their competitors CCOT-GWD and CCOT-GWB.

  Recall that WTOT-SCC1  and WTOT-SCC2 learn the number  of co-clusters.  When
  they underestimate  it, they  pay a  high price,  partly explaining  why the
  standard deviations are rather large. In  order to assess how well they work
  relative to  their counterparts  which benefit from  knowing in  advance the
  true number of co-clusters, we can  compare their measures of performance to
  those of  algorithms WTOT-SCC1$^*$ and WTOT-SCC2$^*$.   In configurations A1
  and  A2,  algorithms   WTOT-SCC1,  WTOT-SCC2  perform  almost   as  well  as
  WTOT-SCC1$^*$ and  WTOT-SCC2$^*$, respectively.   In configuration  A3, they
  are clearly outperformed.  In configuration A4, algorithm WTOT-SCC1 performs
  better in average but not in standard deviation.

  Finally,  we  note  that  algorithm   WTOT-BC$^*$  outperforms  all  our  other
  algorithms.   Unfortunately,  its  counterpart  that learns  the  number  of
  co-clusters performs poorly (results not shown).
\item[Tables~\ref{tab:simulA14:results:matching}                           and
  \ref{tab:simulA:results:matching}.]
  Table~\ref{tab:simulA14:results:matching}   illustrates  the   influence  of
  $k=k'$ on the performances of  algorithm WTOT-matching. In configuration A1,
  specificity is not  impacted much by the value of  $k=k'$, whereas precision
  decreases  and sensitivity  increases as  $k=k'$ grows.   More specifically,
  precision does not change much when one goes from $k=k'=10$ to $k=k'=75$ but
  it drops  for larger  values of  $k=k'$.  As  for sensitivity,  it increases
  dramatically  when one  goes from  $k=k'=10$ to  $k=k'=75$ and  slightly for
  higher values  of $k=k'$.   Furthermore we note  that, in  configuration A1,
  when   $k=k'$  equal   either   65   or  75   and   are   thus  closest   to
  $N\alpha_{\ell} = M\alpha_{\ell} \approx 67$, $\tilde{k}_{r}$ is close to 67
  and  precision,  sensitivity  and  specificity  are  quite  satisfying.   In
  configuration A4 (as in configuration  A1), specificity is not impacted much
  by the value of $k=k'$; on the contrary, precision decreases and sensitivity
  increases steadily as $k=k'$ grows.   The best performances are achieved for
  $k=k'=95$   and   $k=k'=150$,  that   is,   when   $k=k'$  get   closer   to
  $M\max_{i\leq  4}\{\alpha_{i}\}  =   N  \max_{i\leq  4}\{\alpha_{i}\}$.   As
  emphasized  earlier,  deriving  relevant  matchings  is  more  difficult  in
  configuration  A4 than  in configuration  A1  because the  weights given  in
  parameter $\alpha$ are unbalanced in the former and balanced in the latter.

  Table~\ref{tab:simulA:results:matching}    summarizes    the   results    of
  WTOT-matching in all configurations for a specific choice of $k=k'$ in terms
  of the  row- and  column-specific averages $\tilde{k}_r$  and $\tilde{k}_c$,
  precision, sensitivity and specificity.  In each configuration, we chose the
  value of $k=k'$ among many  retrospectively, so that the overall performance
  (in  terms  of  precision,  sensitivity   and  specificity)  is  good.   The
  left-hand-side ($m$-specific)  and right-hand-side ($n$-specific)  tables in
  Table~\ref{tab:simulA:results:matching}  are very  similar.   This does  not
  come as a surprise because the first simulation scheme imposes symmetry.
\end{description}

\subsection{Second simulation study}
\label{subsec:simul:B}

\paragraph*{Simulation scheme.}

The second simulation scheme also relies on mixtures of Gaussian laws, but the
means and weights  are generated randomly from a  Gaussian determinantal point
process (DPP) for  the former and from  a Dirichlet law for  the latter.  More
specifically,              given              the              hyperparameters
$M \geq 200, N \geq 200, K \geq L\geq 3$, $\sigma \in \bbR_{+}^{*}$,
\begin{enumerate}
\item we sample $\mu_{1}, \ldots, \mu_{K}$  from a Gaussian DPP on $[0,1]^{2}$
  with  a   kernel  proportional  to  $x   \mapsto  \exp(-\|x/0.05\|_{2}^{2})$
  conditionally on obtaining exactly $K$ points~\cite{Lavancier2015,spatstat};
\item   independently,    we   sample   $\alpha   \in    (\bbR_{+})^{K}$   and
  $\beta  \in   (\bbR_{+})^{L}$  from  the  Dirichlet   laws  with  parameters
  $7 \Ind_{K}$ and $7 \Ind_{L}$;
\item  we sample  independently $x_{1},  \ldots,  x_{M}$ from  the mixture  of
  Gaussian laws
  \begin{equation*}
    \sum_{k\in \llbracket K\rrbracket} \alpha_k N(\mu_k, \sigma^{2} \text{Id}_{2})
  \end{equation*}
  and $y_{1}, \ldots, y_{N}$ from
  \begin{equation*}
    \sum_{k\in\llbracket L\rrbracket} \beta_k N(-\mu_k, \sigma^{2} \text{Id}_{2}).
  \end{equation*}
\end{enumerate}
We use  a DPP to  generate $\mu_{1}, \ldots,  \mu_{K}$ to avoid  the arbitrary
choice  of  the  mean parameters  in  such  a  way  that the  randomly  picked
$\mu_{1}, \ldots, \mu_{K}$ are dispersed in  $[0,1]^{2}$ (because the DPP is a
repulsive point process).

Table~\ref{tab:SS2} describes the four configurations that we investigate. The
larger $L$  is the more  challenging the configuration is.   In configurations
B2, B3,  B4, it holds  that $K =  L+1$, hence the  data points from  the $K$th
cluster should  not be  matched. Moreover,  for given  $(K,L)$ and  $(M,N)$, a
configuration gets  more challenging as its  $\sigma^{2}$ parameter increases.
It   is  noteworthy   that  the   values  of   $\sigma^{2}$  as   reported  in
Table~\ref{tab:SS2} cannot be compared  straightforwardly to those reported in
Table~\ref{tab:SS1}, because $\mu_{1}, \ldots, \mu_{K}$ live in $[0,1]^{2}$ in
the present  simulation study whereas they  do not in the  simulation study of
Section~\ref{subsec:simul:A}.

\begin{table}[!ht]
\center
\begin{tabular}{|c | c|c |c |c|c} \hline  configuration & $ (M,N)$ & $(K,L)$ &
  $\sigma^{2}$ \\ \hline
  B1 &$(200, 200)$& $(3, 3)$ & $5\times 10^{-4}$\\
  B2  &$(300, 300)$& $(7,6)$ & $10^{-4}$ \\
  B3  &$(300, 300)$& $(16,15)$ & $10^{-5}$ \\
  B4 &$(300, 300)$& $(16,15)$ & $10^{-4}$ \\\hline
\end{tabular}
\caption{Four different  configurations for the second  simulation scheme. The
  larger $\ell \in \llbracket 4\rrbracket$ is the more challenging configuration B$\ell$
  is.}
 \label{tab:SS2}
\end{table}

\paragraph*{Results.}
Thirty  times, independently,  we simulated  synthetic data  sets $X$  and $Y$
under  the simulation  scheme described  above,  then we  applied the  various
algorithms      as     presented      in     Section~\ref{subsec:competition}.
Table~\ref{tab:simulB:results:coclust}  summarizes the  results  of the  seven
algorithms   listed   in   Section~\ref{subsec:competition}   that   rely   on
\textit{bona         fide}        co-clustering         algorithms        (see
Section~\ref{subsubsec:co-clust}).  Tables~\ref{tab:simulB14:results:matching}
and \ref{tab:simulB:results:matching}  summarize the results of  our algorithm
that relies on matching (see Section~\ref{subsubsec:matching}).
\begin{description}
\item[Table~\ref{tab:simulB:results:coclust}.]  We first  note that WTOT-SCC1,
  WTOT-SCC2 and CCOT-GWD  perform similarly in configurations B1  and B2, much
  better  than   CCOT-GWB,  but  less   well  than  the   oracular  algorithms
  WTOT-SCC1$^*$,  WTOT-SCC2$^*$  and   WTOT-BC$^*$.   More  generally,  across
  configurations B1,  B2, B3,  B4, the  oracular algorithms  WTOT-SCC1$^*$ and
  WTOT-SCC2$^*$ perform much better than the other algorithms (and WTOT-BC$^*$
  fails to  find a partition  with the given number  of co-clusters in  B3 and
  B4).  Moreover, WTOT-SCC1 and WTOT-SCC2 perform poorly in configurations B2,
  B3 and B4 though not as poorly as CCOT-GWD and CCOT-GWB in configurations B3
  and B4.  It seems that WTOT-SCC1 and WTOT-SCC2 fail to learn a ``practical''
  number  of co-clusters  from $\tilde{P}$,  in  part because  of those  among
  $x_{1},   \ldots,   x_{M}$   that   are  drawn   from   the   Gaussian   law
  $N(\mu_{K},  \sigma^{2}  \text{Id}_{2})$  when $K=L+1$  (these  data  points
  should  not be  matched  at all).   The fact  that  WTOT-SCC1 and  WTOT-SCC2
  perform  similarly  in  configurations   B3  and  B4  although  $\sigma^{2}$
  is  10  times  larger in  B4  than  in  B3  gives credit  to  the  previous
  interpretation. 
\item[Tables~\ref{tab:simulB14:results:matching}                           and
  \ref{tab:simulB:results:matching}.]
  Table~\ref{tab:simulB14:results:matching}   illustrates  the   influence  of
  $k=k'$ on the  performances of algorithm WTOT-matching  in configurations B1
  and  B4. In  each configuration,  the  values of  $k=k'$ are  chosen in  the
  vicinity of  $M/K$ (67  in configuration  B1, 11  in configuration  B4).  We
  observe the same  patterns in configurations B1 and  B4: precision decreases
  (gradually)  and specificity  decreases  (slightly) as  $k=k'$ grows,  while
  sensitivity increases (strongly in B1 and dramatically in B4).

  Table~\ref{tab:simulB:results:matching}    summarizes    the   results    of
  WTOT-matching in  configurations B1,  B2, B3,  B4 for  a specific  choice of
  $k=k'$ in terms  of the row- and column-specific  averages $\tilde{k}_r$ and
  $\tilde{k}_c$,   precision,   sensitivity    and   specificity.    In   each
  configuration, we chose  the value of $k= k'$ among  many retrospectively so
  that  the  overall  performance  (in terms  of  precision,  sensitivity  and
  specificity) is good.  The left-hand-side ($m$-specific) and right-hand-side
  ($n$-specific)  tables in  Table~\ref{tab:simulB:results:matching} are  very
  similar although $K > L$ in configuration B3 and B4. Interestingly, the fact
  that $\sigma^{2}$ is 10 times larger in configuration B4 than in B3 does not
  affect much the performance of the matching algorithm.
\end{description}

\subsection{Third simulation study}
\label{subsec:simul:C}

\paragraph*{Simulation scheme.}

The third  simulation scheme aspires to  generate synthetic data sets  $X$ and
$Y$ that are  more similar to the  real data sets than those  generated in the
two first  simulation studies.  Once  again, we  rely on mixtures  of Gaussian
laws. This  time, however,  the various means  are neither  chosen arbitrarily
(unlike  in the  first simulation  study) nor  drawn randomly  (unlike in  the
second  simulation  study)   but  are  sampled  in  the   real  collection  of
miRNAs. Moreover, the weights of the mixtures are random.

Specifically,      given     the      hyperparameters     $K      \geq     3$,
$\lambda_{x},  \lambda_{x}' \geq  0$, $\lambda_{y},  \lambda_{y}' \geq  0$ and
$\sigma, \sigma' \in \bbR_{+}^{*}$ (with $\sigma'$ much larger than $\sigma$),
\begin{enumerate}
\item we sample $\mu_1, \ldots,  \mu_K$ uniformly without replacement from the
  collection     of    observed     miRNA     profiles    conditionally     on
  $\min_{k\neq k'} \|\mu_{k} - \mu_{k'}\|_{2} \geq 2$;
\item independently, we sample independently  $(m_{1} - 1), \ldots, (m_{K}-1)$
  from     the      Poisson     law     with      parameter     $\lambda_{x}$,
  $(n_{1}  -  1), \ldots,  (n_{K}-1)$  from  the  Poisson law  with  parameter
  $\lambda_{y}$, $(m_{K+1}  - 1)$ and  $(n_{K+1} -  1)$ from the  Poisson laws
  with parameter $\lambda_{x}'$ and $\lambda_{y}'$;
\item    for    each    $1\leq    k\leq   K$,    we    sample    independently
  $x_{k,    1},    \ldots,    x_{k,    m_k}$    from    the    Gaussian    law
  $N(\mu_k, \sigma^{2} \text{Id}_{18})$ and $y_{k,1}, \ldots, y_{k, n_k}$ from
  the Gaussian law $N(-\mu_k,  \sigma^{2} \text{Id}_{18})$.  Moreover, we also
  sample   independently   $x_{K+1,   1},  \ldots,   x_{K+1,   m_{K+1}}$   and
  $y_{K+1,   1},   \ldots,   y_{K+1,   n_{K+1}}$   from   the   Gaussian   law
  $N(\textbf{0}_{18}, (\sigma')^2 \text{Id}_{18})$.
\end{enumerate}

Here, we  think of  $x$ and  $y$ as  having a  mirrored relationship  if there
exists  $k\in\llbracket K\rrbracket$   such  that   $x$  and   $y$  are   drawn  from   the  laws
$N(\mu_{k},             \sigma^{2}            \text{Id}_{18})$             and
$N(-\mu_{k}, \sigma^{2}  \text{Id}_{18})$.  Furthermore,  we view $x$  and $y$
drawn  from  the  law $N(\textbf{0}_{18},  (\sigma')^{2}  \text{Id}_{18})$  as
noise.

Table~\ref{tab:SS3} describes the four configurations that we investigate. The
larger $K$ is the more challenging the configuration is.

\begin{table}[!h]
  \center
  \begin{tabular}{|c|c|c|c|c|}\hline
    configuration & $(\lambda_x, \lambda_y)$ & $(\lambda_x',\lambda_y')$ & $K$ & $(\sigma, \sigma')$\\ \hline
    C1 & $(50, 50)$ & $(50, 10)$ & $3 $&$(0.1, 5)$ \\
    C2&$ (15,15)$ &$(0,0)$&$15$&$(0.01,5)$\\
    C3 & $ (15, 15)$ & $(30, 30)$ &$ 15$& $(0.01, 5)$\\
    C4& $ (15, 15)$ & $(30, 30)$ &$ 15$&$(0.1, 5)$ \\
    \hline
  \end{tabular}
  \caption{Four different configurations for  the third simulation scheme. The
    larger  $\ell   \in  \llbracket  4\rrbracket$  is   the  more  challenging
    configuration C$\ell$ is.}
  \label{tab:SS3}
\end{table}

\paragraph*{Results.}
Thirty  times, independently,  we simulated  synthetic data  sets $X$  and $Y$
under  the simulation  scheme described  above,  then we  applied the  various
algorithms      as     presented      in     Section~\ref{subsec:competition}.
Table~\ref{tab:simulC:results:coclust}  summarizes the  results  of the  seven
algorithms   listed   in   Section~\ref{subsec:competition}   that   rely   on
\textit{bona         fide}        co-clustering         algorithms        (see
Section~\ref{subsubsec:co-clust}).  Tables~\ref{tab:simulC34:results:matching}
and \ref{tab:simulC:results:matching}  summarize the results of  our algorithm
that relies on matching (see Section~\ref{subsubsec:matching}).

\begin{description}
\item[Table~\ref{tab:simulC:results:coclust}.]     We     first    focus    on
  configuration C1.  We  note that WTOT-SCC1 and  WTOT-SCC2 perform similarly,
  much better than  CCOT-GWD and CCOT-GWB, better than  the oracular algorithm
  WTOT-BC$^*$, but  not as well  as the oracular algorithms  WTOT-SCC1$^*$ and
  WTOT-SCC2$^*$.

  We  now turn  to configurations  C2,  C3 and  C4. Configuration  C3 is  more
  challenging than configuration C2 because it shares the same hyperparameters
  as C2 except  for $(\lambda_{x}',\lambda_{y}')$ (which drives  the number of
  noisy data points), set to $(0,0)$ in  C2 and to $(30,30)$ in C3. Similarly,
  configuration C4 is more challenging than configuration C3 because it shares
  the same hyperparameters  as C3 except for $\sigma$  (the standard deviation
  of the  Gaussian variations around the  mean profiles), set to  $0.01$ in C3
  and to $0.1$ in C4. The  comparisons will not concern algorithms WTOT-BC$^*$
  (which never converges  in these simulations), CCOT-GWD  and CCOT-GWB (which
  perform very poorly).

  In  configuration  C2,  in  the  absence of  noisy  data  points,  algorithm
  WTOT-SCC1 performs slightly  better than WTOT-SCC2, as well  as the oracular
  algorithm  WTOT-SCC2$^*$,  and almost  as  well  as the  oracular  algorithm
  WTOT-SCC1$^*$ (in average).   In configurations C3 and  C4, the introduction
  of noisy data  points then the increase in variability  strongly degrade the
  performances of WTOT-SCC1,  WTOT-SCC1$^*$ and, to a lesser  extent, those of
  WTOT-SCC2 and WTOT-SCC2$^*$.  Algorithm  WTOT-SCC2 outperforms WTOT-SCC1 and
  the oracular algorithm WTOT-SCC1$^*$ too.
  
\item[Tables~\ref{tab:simulC34:results:matching}                           and
  \ref{tab:simulC:results:matching}.]
  Table~\ref{tab:simulC34:results:matching}   illustrates  the   influence  of
  $k=k'$ on the  performances of algorithm WTOT-matching  in configurations C1
  and C4. In each configuration, the  values $k=k'$ are chosen in the vicinity
  of $\lambda_x$ or  $\lambda_y$ (50 in configuration C1,  15 in configuration
  C4).   For specificity  and sensitivity,  we  observe the  same patterns  in
  configurations C1 and  C4: specificity is not impacted much  as $k=k'$ grows
  whereas  sensitivity  increases  dramatically.  Precision  remains  high  in
  configuration C1 for  all choices of $k=k'$. In  configuration C4, precision
  remains high  for $k=k'$ ranging  between 5 and  20, then it  decreases when
  $k=k'$ grows from 25 to 30.
  
  Table~\ref{tab:simulC:results:matching}    summarizes    the   results    of
  WTOT-matching in  configurations C1,  C2, C3,  C4 for  a specific  choice of
  $k=k'$ in terms  of the row- and column-specific  averages $\tilde{k}_r$ and
  $\tilde{k}_c$,   precision,   sensitivity    and   specificity.    In   each
  configuration, we chose  the value of $k=k'$ among  many retrospectively, so
  that  the  overall  performance  (in terms  of  precision,  sensitivity  and
  specificity) is good.  The left-hand-side ($m$-specific) and right-hand-side
  ($n$-specific)  tables in  Table~\ref{tab:simulC:results:matching} are  very
  similar.   In  configurations C1  and  C2,  all precision,  sensitivity  and
  specificity are quite satisfying.  In configurations C3, C4, sensitivity and
  specificity are quite satisfying as well while precision falls bellow 0.86.
\end{description}

\section{Illustration on  real data: matching  mRNA and miRNA  in Huntington's
  disease mice}
\label{sec:real:data}

Next, we  apply algorithms  WTOT-SCC2 and  WTOT-matching to  discover patterns
hidden  in  RNA-seq data  obtained  in  the striatum  of  HD  model mice.   As
explained   in  Section~\ref{sec:intro},   multidimensional  mRNA   and  miRNA
sequencing    data    were    obtained    in    the    striatum    of    these
mice~\cite{Langfelder2016, Langfelder2018}  and an  earlier analysis  of these
data   using   shape   analysis   concepts~\cite{megret:inserm-02512089}   has
demonstrated their value.

\subsection{Tuning}
\label{subsec:tuning}

Specifically,    in    view    of    Procedure~\ref{algo:SGD},    we    choose
$\widetilde{M} = 1,024$,  $\widetilde{N}=512$, $T = 500$.  The  entries of the
$3\times                              5$                              matrices
$\tilde{\theta}_{1}^{a},  \tilde{\theta}_{1}^{b}, \tilde{\theta}_{1}^{c}$  are
constrained to  take their values  in $]-10,0[$  (for WTOT-SCC2) or  $]-2, 0[$
(for  WTOT-matching), $]-0.2,0.2[$  and  $]-0.2,0.2[$  respectively.  We  also
choose  $(\eta,  \gamma_{0})  =  (0.95, 3)$.   Finally,  the  initial  mapping
$\theta_{0}$ is drawn randomly.

Furthermore,  regarding step~2  of  algorithm WTOT-SCC2,  we  remove rows  and
columns based on  the following loop: 100 times  successively, \textit{(i)} we
compute the  Kullback-Leibler divergence  between each row  (renormalized) and
the  uniform  distribution  then  remove   the  100  rows  with  the  smallest
divergences,  then \textit{(ii)}  we compute  the Kullback-Leibler  divergence
between each  column (renormalized) and  the uniform distribution  then remove
the 5 columns with the smallest divergences.  By doing so, we successively get
rid of  the rows and columns  which, viewed as distributions,  are too uniform
and therefore  deemed irrelevant. Finally,  we remove  all rows for  which the
(columnwise) sum of  the remaining entries of $\tilde{P}$ is  smaller than one
tenth of the maximal (columnwise) sum, and all columns for which the (rowwise)
sum of the remaining  entries of $\tilde{P}$ is smaller than  one tenth of the
maximal (rowwise) sum.

\subsection{Results}
\label{sec:results}

\paragraph{Co-clustering.}

\begin{figure}[ht]
  \centering \includegraphics[width=0.4\linewidth]{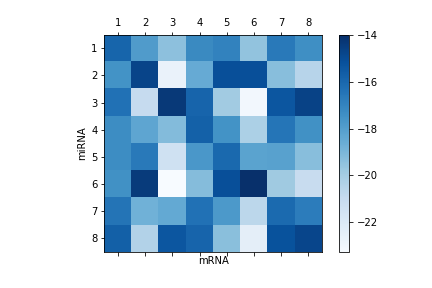}
  \caption{Logarithms  of the  averages, computed  across all  blocks, of  the
    entries  of  the   matrix  derived  from  the   optimal  transport  matrix
    $\tilde{P}$   during  step~2   of  algorithm   WTOT-SCC2  and   after  its
    rearrangement. }
    \label{fig:res:real:co-clusters}
\end{figure}

The selection procedure (step 2 of WTOT-SCC2) keeps 3,409 mRNA profiles (among
the 13,616 available in the data set) and 602 miRNA (among the 1,143 available
in the data  set). Eventually, algorithm WTOT-SCC2 outputs  8 co-clusters. The
co-clusters's sizes  (numbers of mRNA  and miRNA gathered in  each co-cluster)
are  $(321,86)$, $(333,30)$,  $(261,6)$, $(498,125)$,  $(127,5)$, $(708,203)$,
$(703,119)$, $(458,28)$.  Figure~\ref{fig:res:real:co-clusters} represents the
averages, computed  across all blocks,  of the  entries of the  matrix derived
from  the optimal  transport  matrix $\tilde{P}$  during  step~2 of  algorithm
WTOT-SCC2 and after  its rearrangement.  Squares located on  the diagonal tend
to be slightly darker than the other squares. This reveals that, in average, a
pair $(x_{m},y_{n})$ of mRNA and miRNA gathered in a diagonal co-cluster tends
to exhibit a mirrored relationship that is slightly stronger than those of the
form $(x_{m},  y_{n'})$ or  $(x_{m'}, y_{n})$  which do not  fall in  the same
co-cluster.  However, few of the off-diagonal averages are small in comparison
to the on-diagonal averages, a disappointing  observation that comes on top of
the fact  that the  co-clusters' sizes are  so large that  it is  difficult to
interpret the results. This makes it  even more relevant to focus on algorithm
WTOT-matching.

\paragraph{Matching.}

We run  the WTOT-matching  algorithm with  $k=k'=10$ and $q  = 90\%$.  For the
anecdote,  we observe  $(\tilde{k}_{r}, \tilde{k}_{c})  \approx (1.82,  6.04)$
(recall that  $\tilde{k}_{r},\tilde{k}_{c}$ are  the row-  and column-specific
averages of the cardinalities of the sets $\calN_{m}$ and $\calM_{n}$ that are
not  empty).   We   report  the  parameters  that   characterize  the  mapping
$\widehat{\theta}$ in Appendix~\ref{sec:supp:mat}.

\begin{figure}[ht]
  \centering%
  \includegraphics[width=0.45\linewidth]{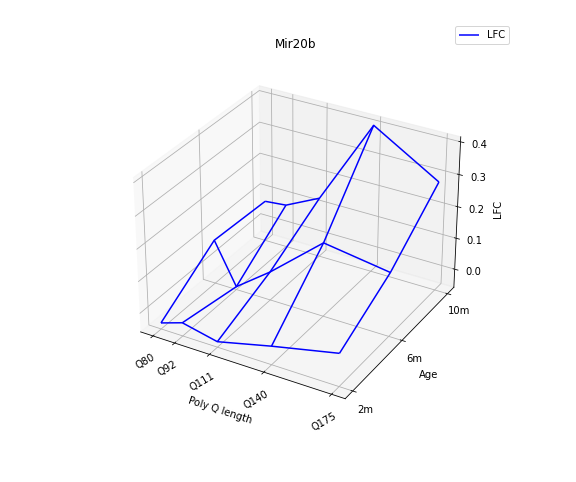}
  \includegraphics[width=0.45\linewidth]{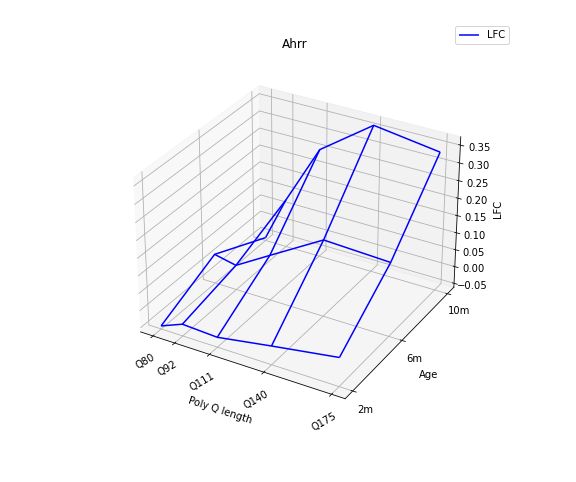}\\
  \includegraphics[width=0.45\linewidth]{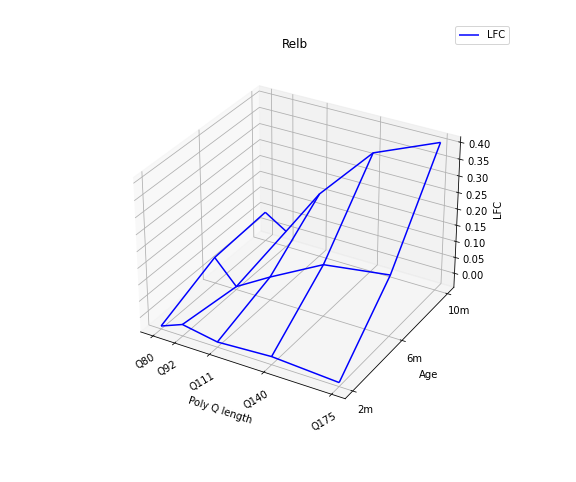}
  \includegraphics[width=0.45\linewidth]{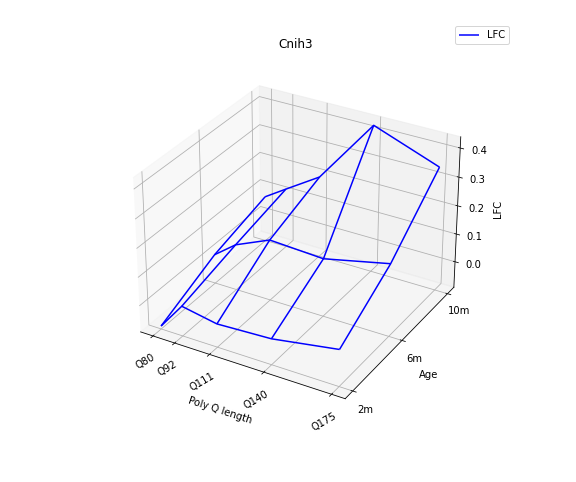}
  \caption{Minus  the profile  $-y_{n}$ of  the Mir20b  miRNA (top  left), and
    profiles  $x_{m}$ of  its matched  mRNAs, Ahrr  (top right),  Relb (bottom
    left) and Cnih3 (bottom right).}
  \label{fig:res:real:matching}
\end{figure}

As an illustration, the mirrored profile  (the opposite value of $y_n$) of the
Mir20b miRNA is displayed in Figure~\ref{fig:res:real:matching} along with its
three  matched mRNAs  (Ahrr, Cnih3  and  Relb) obtained  by running  algorithm
WTOT-matching algorithm with  $k=k'=10$.  Recall that the  original profile of
Mir20b can be found in Figure~\ref{fig:Mir20b}.

\subsection{Biological analysis of the results}
\label{subsec:biological:analysis}

In  an effort  to guarantee  biological relevance  to the  matchings, we  only
retain those showing evidence for binding  sites as indicated in the databases
TargetScan~\cite{LEWIS2005},           MicroCosm~\cite{Betel2010}          and
miRDB~\cite{Ding2016}.  Specifically, a  pair $(x,y)$ is retained  if and only
if the mRNA whose  profile is $x$ and the miRNA whose profile  is $y$ are both
among the 27,355 mRNAs and 1,478 miRNAs appearing in the TargetScan, MicroCosm
and miRDB databases.  The 1,247 matchings retained out of  the 7,521 output by
the      WTOT-matching      algorithm      are      all      presented      on
\href{http://www.broca.inserm.fr/WTOT/results.php}{this page} of the companion
website.  We  stress that  we would  have obtained fewer  matchings if  we had
excluded from the collections $X$ and $Y$  the profiles of mRNA or miRNA which
do not appear in the databases.

Furthermore, we  build upon two previous  analyses of miRNA regulation  in the
striatum  of HD  knock-in-mice~\cite{Langfelder2018,megret:inserm-02512089} to
comment on  the biological relevance and  novelty of our findings.   The first
analysis~\cite{Langfelder2018} relies on the  WGCNA algorithm, a weighted gene
co-expression network analysis which yields clusters of genes whose expression
profiles  are correlated.   The second  analysis~\cite{megret:inserm-02512089}
relies on  the MiRAMINT  algorithm. MiRAMINT  is a  pipeline whose  main steps
consist in  \textit{(a)} carrying  out a  weighted gene  co-expression network
analysis, \textit{(b)} using random forests to select candidate matchings, and
\textit{(c)}  using  Spearman's  correlation   test  and  a  multiple  testing
procedure to  identify the more  reliable matchings.  We highlight  that WGCNA
outputs 1,583 mRNA-miRNA  matchings showing evidence for binding  sites in the
databases TargetScan,  MicroCosm and  miRDB, which  involve only  46 different
miRNAs.   As for  MiRAMINT, it  only  outputs 31  matchings of  which 20  show
evidence for binding  sites in the databases TargetScan,  MicroCosm and miRDB,
involving 14 different miRNAs.  The 31 mRNA-miRNA matchings output by MiRAMINT
are                    all                     presented                    on
\href{http://www.broca.inserm.fr/MiRAMINT/results.php}{this webpage}.

\paragraph*{Analyzing the overlaps.}

Three mRNA-miRNA  matchings are retained  both by the WTOT-matching  and WGCNA
algorithms:  Mir186-Chl1, Mir132-Fam196b,  Mir212-Fam196b.   No matchings  are
retained both by  the WTOT-matching and the MiRAMINT algorithms.   One pair is
retained both by the MiRAMINT and WGCNA algorithms: Mir132-Pafah121.

Figure~\ref{fig:venn:all}  in  Appendix~\ref{sec:supp:mat} presents  two  Venn
diagrams summarizing  the overlaps between  the sets of  miRNAs (respectively,
mRNAs) which belong to a pair  output by the WGCNA, MiRAMINT and WTOT-matching
algorithms.   On the  one  hand, focusing  on  miRNAs, $13/14$  (respectively,
$29/46$)   miRNAs  involved   in  a   mRNA-miRNA  pair   output  by   MiRAMINT
(respectively,  WGCNA) are  among the  miRNAs  involved in  a mRNA-miRNA  pair
output  by  WTOT-matching.  On  the  other  hand,  focusing on  mRNAs,  $1/20$
(respectively, $100/1,583$)  miRNAs involved  in a  mRNA-miRNA pair  output by
MiRAMINT (respectively, WGCNA)  are among the miRNAs involved  in a mRNA-miRNA
pair output by WTOT-matching.  We carry  out one-sided Fisher's exact tests to
quantify  to  what  extent  the  overlaps reflect  an  agreement  between  two
algorithms  (using  the  1,478  miRNAs  and  27,355  mRNAs  appearing  in  the
TargetScan,  MicroCosm  and miRDB  databases  as  reference populations).  The
$p$-value of  the test comparing  WTOT-matching and MiRAMINT equals  0.45. The
other $p$-values are smaller than~$10^{-6}$. 

It is desirable to identify miRNAs that are particularly susceptible to play a
distinct role in HD in mice.  To do so, we evaluate two simple criteria on the
mRNAs associated to each miRNA (the miRNAs with no matched mRNAs are obviously
less interesting  in our  study). The  criteria assess to  what extent  a mRNA
profile is ``monotonic" and, on the  contrary, to what extent it is ``peaked",
accounting for  the amplitude  of log-fold  change.  Formally,  rewriting each
profile        $x        \in        \bbR^{15}$       as        a        matrix
$(\tilde{x}_{tq})_{t   \in   \llbracket    3\rrbracket,   q   \in   \llbracket
  5\rrbracket}$, the first criterion is the  minimum (relative to time $t$) of
the  absolute  values of  the  slopes  of the  regression  lines  of the  sets
$\{(q,  \tilde{x}_{tq})  : q  \in  \llbracket  5\rrbracket\}$ and  the  second
criterion                                                                   is
$\max_{q\in\llbracket 5  \rrbracket} (\tilde{x}_{1q} -  \tilde{x}_{2q}) \times
(\tilde{x}_{2q} - \tilde{x}_{3q})$.  By convention, a miRNA profile is labeled
monotonic  (respectively, peaked)  if  at  least one  of  its associated  mRNA
profiles is  such that  its first (respectively,  second) criterion  is larger
than 95\% (respectively, smaller than 99\%) of the similar criteria. Moreover,
all mRNA  profiles $x$ appearing  in a pair $(x,y)$  are labeled like  $y$. We
stress that no mRNA labeling conflicts occur.

Below, we reproduce the same analysis  as above focusing in turn on mRNA-miRNA
matchings labeled as peaked, monotonic, and neither peaked nor monotonic.

\begin{description}
\item[Peaked        profiles.]          Figure~\ref{fig:venn:peaked}        in
  Appendix~\ref{sec:supp:mat}  presents  two  Venn  diagrams  summarizing  the
  overlaps between the sets of miRNAs  (respectively, mRNAs) which belong to a
  pair   output  by   the  WGCNA,   MiRAMINT  and   WTOT-matching  algorithms,
  \textit{looking at the WTOT-matching matchings  labeled as peaked}.  None of
  the 17 miRNAs and none of the  12 mRNAs involved in a mRNA-miRNA pair output
  by WTOT-matching  are involved in a  mRNA-miRNA pair output by  the WGCNA or
  MiRAMINT algorithms.

  The  take-home  message  is   that  the  WTOT-algorithm  retains  mRNA-miRNA
  matchings that we label as peaked whereas neither the WGCNA nor the MiRAMINT
  algorithms do.
\item[Monotonic      profiles.]        Figure~\ref{fig:venn:monotonic}      in
  Appendix~\ref{sec:supp:mat}  presents  two  Venn  diagrams  summarizing  the
  overlaps between the sets of miRNAs  (respectively, mRNAs) which belong to a
  pair   output  by   the  WGCNA,   MiRAMINT  and   WTOT-matching  algorithms,
  \textit{looking at  the WTOT-matching  matchings labeled as  monotonic}.  On
  the  one hand,  focusing  on miRNAs,  $8/14$  (respectively, $9/46$)  miRNAs
  involved in a  mRNA-miRNA pair output by MiRAMINT  (respectively, WGCNA) are
  among the miRNAs involved in a  mRNA-miRNA pair output by WTOT-matching.  On
  the other hand, focusing on  mRNAs, $0/20$ (respectively, $14/1,583$) miRNAs
  involved in a  mRNA-miRNA pair output by MiRAMINT  (respectively, WGCNA) are
  among the miRNAs involved in a  mRNA-miRNA pair output by WTOT-matching.  We
  carry out  one-sided Fisher's  exact tests  to quantify  to what  extent the
  overlaps reflect an agreement between two algorithms (using the 1,478 miRNAs
  and 27,355 mRNAs appearing in  the TargetScan, MicroCosm and miRDB databases
  as  reference populations),  excluding the  comparison of  the MiRAMINT  and
  WTOT-matching  algorithms in  mRNAs  (due to  an  empty intersection).   The
  $p$-values are smaller than~$10^{-5}$. 

  The take-home message is that, in  matchings that we label as monotonic, the
  agreement between the WTOT-matching and WGCNA algorithms is better than that
  between the WTOT-matching and MiRAMINT algorithms.
\item[Neither      peaked     nor      monotonic     profiles.]       Finally,
  Figure~\ref{fig:venn:others}  in  Appendix~\ref{sec:supp:mat}  presents  two
  Venn  diagrams  summarizing   the  overlaps  between  the   sets  of  miRNAs
  (respectively, mRNAs) which  belong to a pair output by  the WGCNA, MiRAMINT
  and WTOT-matching  algorithms and labeled  neither as peaked  nor monotonic.
  On the one hand, focusing  on miRNAs, $12/14$ (respectively, $28/46$) miRNAs
  involved in a  mRNA-miRNA pair output by MiRAMINT  (respectively, WGCNA) are
  among the miRNAs involved in a  mRNA-miRNA pair output by WTOT-matching.  On
  the other hand, focusing on  mRNAs, $1/20$ (respectively, $86/1,583$) miRNAs
  involved in a  mRNA-miRNA pair output by MiRAMINT  (respectively, WGCNA) are
  among the miRNAs involved in a  mRNA-miRNA pair output by WTOT-matching.  We
  carry out  one-sided Fisher's  exact tests  to quantify  to what  extent the
  overlaps reflect an agreement between two algorithms (using the 1,478 miRNAs
  and 27,355 mRNAs appearing in  the TargetScan, MicroCosm and miRDB databases
  as  reference populations),  excluding the  comparison of  the MiRAMINT  and
  WTOT-matching  algorithms in  mRNAs (due  to  an intersection  reduced to  a
  singleton).  The $p$-value are smaller than~$10^{-5}$.

  The take-home message is that, in  matchings that we label as neither peaked
  nor monotonic, the agreement between  the WTOT-matching and WGCNA algorithms
  is better than that between the WTOT-matching and MiRAMINT algorithms.
\end{description}

\paragraph*{Enrichment analysis.}

Next, we assess and compare the  biological significance of the mRNAs retained
by the WGCNA, MiRAMINT and WTOT-matching algorithms.  To do so we carry out an
enrichment           analysis            using           the           EnrichR
tools~\cite{chen2013enrichr,kuleshov2016enrichr,xie2021357}.  We consider only
top annotations  (balancing a small $p$-value  and a large number  of hits) as
provided  by \href{http://geneontology.org/}{Gene  Ontology data}  (biological
process,                 cellular                 content)                 and
\href{https://www.genome.jp/kegg/pathway.html}{KEGG  data}.   When  necessary,
only the top 40  hits are considered so as to guarantee  a sufficient level of
biological precision.  Pubmed searches are  also used to assess the biological
significance of predicted miRNA regulation.

Figures~\ref{fig:network:peaked},          \ref{fig:network:monot}         and
\ref{fig:network:others} in Appendix~\ref{sec:supp:mat} present the mRNA-miRNA
networks  based  on  the  mRNA-miRNA matchings  output  by  the  WTOT-matching
algorithm, focusing  on the matchings  which are labeled as  peaked, monotonic
and  neither peaked  nor  monotonic (in  that order).   The  mRNAs and  miRNAs
retained by  the WGCNA  and MiRAMINT algorithms  are colored.   The enrichment
analysis reveals
  \begin{itemize}
  \item  that the  mRNA-miRNA  matchings  output by  the  WGCNA algorithm  are
    primarily annotated for \textit{axonogenesis}\footnote{GO:0007409, de novo
      generation  of  a long  process  of  a  neuron, including  the  terminal
      branched region.   Refers to the  morphogenesis or creation of  shape or
      form of  the developing axon,  which carries efferent  (outgoing) action
      potential from  the cell body  towards target cells.}, which  relates to
    cytoskeleton dynamics and cell morphology;
  \item  that the  matchings output  by the  MiRAMINT algorithm  are primarily
    annotated  for   \textit{regulation  of  defense  response   to  virus  by
      host}\footnote{GO:0050691,   any  host   process   that  modulates   the
      frequency, rate  or extent of the  antiviral response of a  host cell or
      organism.}, which relates to stress response and innate immunity;
  \item that the matchings output by the WTOT-matching algorithm are primarily
    annotated for \textit{extracellular matrix organization} (which relates to
    cell identity)\footnote{GO:0030198, a  process that is carried  out at the
      cellular level which results in the assembly, arrangement of constituent
      parts,  or  disassembly of  an  extracellular  matrix.   }, due  to  the
    matchings  labeled  as  neither  peaked  nor  monotonic,  and  secondarily
    annotated    for   \textit{mitigation    of    host   antiviral    defense
      response}\footnote{GO:0050690,   evasion  by   virus   of  host   immune
      response.},  due  to  the  matchings   labeled  as  monotonic,  and  for
    \textit{conventional motile  cilium}\footnote{GO:0097729, a  motile cilium
      where the  axoneme has a  ring of 9  outer microtubules doublets  plus 2
      central micro tubules.}, due to the matchings labeled as peaked.
  \end{itemize}
  Although the  numbers of hits in  some of these annotations  are small, they
  suggest that the WTOT-matching algorithm is  able to uncover a role of miRNA
  regulation in responding  to mutant huntingtin that was not  detected by the
  WGCNA and MiRAMINT algorithms (despite the large number of mRNAs retained by
  the former).

  We now interpret the above results from a biological viewpoint.  Recall that
  the peaked  and monotonic profiles  are especially interesting  because they
  are more susceptible to correspond to  mRNAs and miRNAs that play a distinct
  role  in  HD  in  mice.   Extracellular  matrix  organization  (the  primary
  annotation of the matchings output by the WTOT-matching algorithm, driven by
  the mRNA-miRNA matchings  labeled as neither peaked nor  monotonic) is known
  to be regulated by miRNAs~\cite{PMID23159731}  and HD mutations are known to
  strongly affect neuronal identity via down-regulating a large number of cell
  identity  genes~\cite{PMID25784504}. Mitigation  of  host antiviral  defense
  response  (the first  secondary annotation  of the  matchings output  by the
  WTOT-matching algorithm, due to  the mRNA-miRNA matchings labeled monotonic)
  is similar to the primary annotation of the matchings output by the MiRAMINT
  algorithm.   Finally,  conventional  motile  cilium  (the  second  secondary
  annotation of  the matchings output  by the WTOT-matching algorithm,  due to
  the mRNA-miRNA matchings labeled peaked) is a new finding. 

  Additionally, although  miRNA levels  and regulation  in response  to mutant
  huntingtin is anticipated  to be dependent on cellular context  and could be
  differentially influenced across murine models  of HD, it is noticeable that
  the analysis of  miRNA regulation in the striatum of  HD knock-in mice based
  on the WTOT-matching  algorithm retained several miRNAs that  are altered in
  the striatum of other types of  HD mice such as BACHD~\cite{PMID33906482} or
  altered  in the  human HD  caudate nucleus~\cite{PMID35869509}  such as  for
  example Mir100, Mir127, Mir132, Mir 212 and Mir133, supporting the relevance
  of our  findings for the  study of molecular  regulation in mouse  and human
  HD.

  We believe  that these facts  substantiate our claim that  the WTOT-matching
  algorithm strikes a good balance between the low and high selectivity of the
  WGCNA and  MiRAMINT algorithms. Moreover,  our findings related  to striatal
  alterations in HD mice lead  to reconsidering the formerly-expressed view on
  a limited role of  miRNA regulation in the striatum of HD  mice on a systems
  level~\cite{megret:inserm-02512089}.

\section{Discussion}
\label{sec:discussion}

We have developed two co-clustering  algorithms (WTOT-SCC1 and WTOT-SCC2) and
a matching algorithm (WTOT-matching) for  the purpose of identifying groups of
mRNAs and miRNAs that interact. The algorithms apply in any situation where it
is of interest  to cluster or match the  elements of two data sets  based on a
parametric model  $\Theta$ expressing what  it means  to interact for  any two
pair  of elements  from the  two data  sets.  The  algorithms rely  on optimal
transport,  spectral  co-clustering and  a  matching  procedure. In  light  of
\citep[Section 1.3, page~25]{MAD2019},  problem-specific knowledge is injected
onto  two of  the three  main components  of the  transportation problem:  the
representation  spaces (via  $\Theta$) and  the marginal  constraints, leaving
aside the cost function.

During the first  stage, an optimal optimal transport plan~$P$  and mapping in
$\Theta$ are  learned from the data  using the Sinkhorn-Knopp algorithm  and a
mini-batch  gradient descent.  During the  second stage,  $P$ is  exploited to
derive either co-clusters or several sets of matched elements. 

As in~\cite{megret:inserm-02512089},  the motivation of  our study is  to shed
light on the  interaction between mRNAs and miRNAs based  on data collected in
the striatum of HD  model knock-in mice~\cite{Langfelder2016, Langfelder2018}.
Each  data point  takes the  form  of multi-dimensional  profile.  The  strong
biological hypothesis is  that if a miRNA induces the  degradation of a target
mRNA or blocks its translation into proteins, or both, then the profile of the
former  should  be  similar to  minus  the  profile  of  the latter  ---  this
particular form  of affine relationship  drives the formulation of  a loosened
hypothesis  and definition  of model  $\Theta$.  The  fact that  the algorithm
learns from the data a best element in $\Theta$ provides more flexibility than
in~\cite{megret:inserm-02512089}.

The simulation  study reveals  on the  one hand  that WTOT-SCC2  works overall
better than WTOT-SCC1, but that the co-clustering task can be very challenging
in  the presence  of many  irrelevant  data points  (data points  that do  not
interact). On the other hand, it  shows that the performances of WTOT-matching
are satisfying.

An illustration on  real data is given. The results  are biologically relevant
and illustrate how our algorithm strikes a good balance between two moderately
and highly selective, competing algorithms. Our findings lead to reconsidering
the  formerly-expressed view  on a  limited role  of miRNA  regulation in  the
striatum of HD mice on a systems level~~\cite{megret:inserm-02512089}.

In conclusion, there  are several directions for future work.   First, we will
develop  a  similar  study  to  better  understand  miRNA  regulation  in  the
\textit{cortex} of HD model mice  (ongoing project).  Second, we will evaluate
the performances of our algorithms by simulation studies based on a simulation
scheme \textit{learned}  from the real  data so as  to better mimic  their law
(ongoing project).  Third, we will put our algorithms into the general context
of co-clustering and  matching of datasets and carry out  more benchmark tests
and comparisons.

\section*{Declarations}

\begin{itemize}
\item \textbf{Funding:} T.   T. Y.  Nguyen is funded by  Université Paris Cité
  thanks to a Ph.D. fellowship granted  by Domaine d'Intérêt Majeur Math Innov
  (Région  Île-de-France and  Fondation Sciences  Mathématiques de  Paris). O.
  Bouaziz and A.   Chambaz are funded by Université Paris  Cité, W.  Harchaoui
  by \href{http://deraison.ai}{Déraison.ai}.   C.  Mendoza, L.  Mégret  and C.
  Neri  are  funded by  the  CHDI  Foundation  (grant no.  A-14814),  Sorbonne
  Université, CNRS and INSERM.
\item \textbf{Conflicts of interest/Competing interests:} None.
\item \textbf{Availability of data and material:}  The omics data used in this
  study  are  publically  available   through  the  database  repository  Gene
  Expression   Omnibus  (GEO)   and  the   \href{http://www.HDinHD.org}{HDinHD
    portal}.   Overlaps   between  the   results  obtained  by   applying  the
  WTOT-matching algorithm and  results previously obtained based  on two other
  algorithms, and  full display  of biological  annotations, are  available on
  \href{http://www.broca.inserm.fr/WTOT/results.php}{this    page}   of    the
  companion website.
\item\textbf{Code      availability:}      The     code      is      available
  \href{http://www.broca.inserm.fr/WTOT/}{here}                            and
  \href{https://github.com/yen-nguyen-thi-thanh/wtot_coclust_match}{here}.
\item  \textbf{Authors' contributions:}  C. Neri,  O. Bouaziz  and A.  Chambaz
  conceived  the study.   T.   T.  Y.   Nguyen, O.   Bouaziz  and A.   Chambaz
  developed the  methodology, formally and computationally,  and performed the
  data analysis based  on insights from L.   Mégret and C. Neri  on how mutant
  huntingtin may significantly influence expression patterns across CAG repeat
  alleles and age points in the brain of HD mice.  L.  Mégret, C.  Mendoza and
  C.  Neri performed the biological analysis of the results, the comparison to
  other algorithms,  and the data base  construction.  T.  T.  Y.   Nguyen, O.
  Bouaziz  and A.   Chambaz  wrote the  first draft  of  the manuscript.   All
  authors commented on subsequent versions.  All authors read and approved the
  final manuscript.
\item \textbf{Ethics approval:} Not applicable.
\item \textbf{Consent to participate:} Not applicable.
\item \textbf{Consent for publication:} Not applicable.
\end{itemize}

\bibliographystyle{plainnat}
\bibliography{OT4H}

\appendix

\newpage
\section{Supplementary material}
\label{sec:supp:mat}

\paragraph*{Parametric model $\Theta$.}

Introduced  in  Section~\ref{subsec:method},  the  parametric  model  $\Theta$
consists  of affine  mappings $\theta  : \bbR^{d}  \to \bbR^{d}$  of the  form
$x \mapsto \theta_{1} x + \theta_{2}$,  where $\theta_{1}$ takes its values in
a subset  $T_{1}$ of $\bbR^{d\times d}$  and $\theta_{2}$ takes its  values in
$\bbR^{d}$ (without  any constraint).   It is  easier to  describe the  set of
linear mappings  $\{x \mapsto \theta_{1} x  : \theta_{1} \in T_{1}\}$  after a
reparametrization.

In the  rest of  this section  only, we  rewrite the  mRNA and  miRNA profiles
$x,y  \in  \bbR^{d}$   under  the  form  of  $d_{1}   \times  d_{2}$  matrices
$\tilde{x}  =  (\tilde{x}_{tq})_{t  \in   \llbracket  d_{1}\rrbracket,  q  \in
  \llbracket                       d_{2}\rrbracket}$                       and
$\tilde{y}  =  (\tilde{y}_{tq})_{t  \in   \llbracket  d_{1}\rrbracket,  q  \in
  \llbracket d_{2}\rrbracket}$.   For each $t \in  \llbracket d_{1}\rrbracket$
and   $q   \in   \llbracket   d_{2}\rrbracket$,   $\tilde{x}_{t\bullet}$   and
$\tilde{x}_{\bullet q}$  are the  $t$th row and  $q$th column  of $\tilde{x}$.
Here, indices $t$  and $q$ correspond to  the age and CAG lengths  of the mice
whose   RNA   sequencing   yielded  $\tilde{x}_{tq}$   and   $\tilde{y}_{tq}$.

The  definition  of  $T_{1}$  should  formalize  what  we  consider  to  be  a
(plausible) mirroring  relationship.  The  simplest mirroring  relationship is
$y  = -x$  or, equivalently,  $\tilde{y} =  -\tilde{x}$.  The  equality is  of
course too  stringent/rigid, and the  definition of  $T_{1}$ is driven  by our
wish to relax it.

Biological  arguments encourage  us to  consider that  $y$ and  $x$ exhibit  a
(plausible)     mirroring     relationship     if,    for     each     $(t,q)$
($t  \in \llbracket  d_{1}\rrbracket$,  $q  \in \llbracket  d_{2}\rrbracket$),
$\tilde{y}_{tq}$  is  strongly  negatively correlated  with  $\tilde{x}_{tq}$,
mainly, and  (positively or  negatively) correlated  with $\tilde{x}_{(t-1)q}$
(if $t>1$) and/or with $\tilde{x}_{t(q-1)}$  (if $q>1$), secondarily.  We thus
formalize $\{x  \mapsto \theta_{1} x :  \theta_{1} \in T_{1}\}$ as  the set of
all linear mappings of the form
\begin{equation*}
  x  \mapsto \tilde{\theta}_{1}^{a}  \odot \tilde{x}  + \tilde{\theta}_{1}^{b}
  \odot \left(%
    \begin{smallmatrix}
      \textbf{0}_{d_{2}}^{\top}\\\tilde{x}_{1\bullet}\\\vdots\\\tilde{x}_{(d_{1}-1)\bullet}
    \end{smallmatrix}\right) %
  + \tilde{\theta}_{1}^{c}
  \odot \left(%
    \textbf{0}_{d_{1}}
    \tilde{x}_{\bullet 1} \; \cdots \tilde{x}_{\bullet (d_{2}-1)} \right) 
\end{equation*}
where                       $\tilde{\theta}_{1}^{a}$                       and
$\tilde{\theta}_{1}^{b},   \tilde{\theta}_{1}^{c}$  are   $d_{1}\times  d_{2}$
matrices (here, $\odot$  is the componentwise multiplication).  The entries of
$\tilde{\theta}_{1}^{a}$  correspond to  comparisons between  $\tilde{x}_{tq}$
and $\tilde{y}_{tq}$  (same poly~Q  length $q$  and age  $t$). The  entries of
$\tilde{\theta}_{1}^{b}$  (whose  first  row  consists of  0s)  correspond  to
comparisons  between $\tilde{x}_{(t-1)q}$  and  $\tilde{y}_{tq}$ (same  poly~Q
length  $q$,  different age  $t$).   The  entries of  $\tilde{\theta}_{1}^{c}$
(whose  first  column  consists  of  0s)  correspond  to  comparisons  between
$\tilde{x}_{t(q-1)}$ and  $\tilde{y}_{tq}$ (different poly~Q length  $q$, same
age $t$).

In the simulation study presented in Section~\ref{sec:simulation}, the entries
of  $\tilde{\theta}_{1}^{a}$  are constrained  to  take  their values  in  the
interval             $]-5,0[$            while             those            of
$\tilde{\theta}_{1}^{b}, \tilde{\theta}_{1}^{c}$ are constrained to take their
values in  $]-1/2, 1/2[$.  The initial  mapping is drawn randomly  by sampling
the  entries  of  $\tilde{\theta}_{1}^{a}$   independently  and  uniformly  in
$]-5,0[$    and,     independently,    by    sampling    the     entries    of
$\tilde{\theta}_{1}^{b}$   and   $\tilde{\theta}_{1}^{c}$  independently   and
uniformly in $]-1/2,1/2[$.

In   the   illustration   of   the  WTOT-matching   algorithm   presented   in
Section~\ref{sec:results}, the  mapping $\widehat{\theta}$ is  parametrized by
$\tilde{\theta}$ given by
\begin{gather*}
  \tilde{\theta}_{1}^{a} = \left(%
    \begin{smallmatrix}
      -0.88  &-1.47&  -0.73 \\
      -0.59& -0.90&  -0.89\\
      -0.62  &-0.70&   -1.17 \\
      -0.97  &-1.30   &-0.95\\
       -0.56 &-1.16& -1.24 \\
           \end{smallmatrix}\right), \quad
         \tilde{\theta}_{1}^{b} = \left(%
    \begin{smallmatrix}
      0 & 0 & 0  \\
      0.13 & -0.19 &  0.13\\
      0.17 &  0.09 &  0.13\\
      0.19 & 0.09 &-0.00  \\
      0.18 &  0.15 &  0.08\\
           \end{smallmatrix}\right), \\
         \tilde{\theta}_{1}^{c} = \left(%
    \begin{smallmatrix}
      0 &0.18& -0.18\\
      0 &0.19 & 0.17\\
      0 & 0.04  &0.15\\
      0 & 0.05&  0.11\\
      0 & 0.18  &0.14\\
           \end{smallmatrix}\right),\quad
         \theta_{2} = \left(%
    \begin{smallmatrix}
-0.01  &0.01& -0.00 \\ 
  0.00   & 0.01 & 0.01\\
  0.00    &0.01&  0.00  \\
  0.01 & 0.01 & 0.01\\
 -0.01 & 0.01&  0.01\\
           \end{smallmatrix}\right)
\end{gather*}
(the numbers are rounded to two decimal places). We note that:
  \begin{itemize}
  \item  On  the  one  hand,   the  entries  of  $\tilde{\theta}_{1}^{a}$  are
    distributed around -1. On the other  hand, the entries of $\theta_{2}$ are
    small.  This  is in  line with  the \textit{strong}  biological hypothesis
    (that is, if  a miRNA induces the  degradation of a target  mRNA or blocks
    its translation  into proteins, or  both, then  the profile of  the former
    should be similar to minus the profile of the latter).
  \item The  entries of $\tilde{\theta}_{1}^{b}$  and $\tilde{\theta}_{1}^{c}$
    are small.
  \end{itemize}

\newpage

\begin{algorithm}[h]
  \caption{\textit{Master optimal transport algorithm.}
    % In our  experiments, we choose $m=64$, $\alpha=0.005$.
  }
  \label{algo:SGD}
  \begin{algorithmic}
    \REQUIRE  $X, Y$,  minibatch sizes  $\widetilde{M}, \widetilde{N}$,  decay
    rate  $\eta \in  ]0,1]$,  initial  regularization parameter  $\gamma_{0}$,
    initial mapping $\theta_{0} \in \Theta$, maximal number of iterations $T$
    \ENSURE  Transport coupling  $\tilde{P}_{T}  \in (\bbR_{+})^{M\times  N}$,
    mapping $\theta_{T}\in\Theta$, weight $\omega_{T}$
    \STATE Compute: 
    \begin{itemize}
    \item $\underline{\gamma} = \text{mean} \{\|x -  x'\|_{2} : x, x' \in X\}$
      \COMMENT{for entropy regularization}
    \item $h  = \text{mean}  \{\|y -  y'\|_{2} : y,  y' \in  Y\}$ \COMMENT{for
        window calibration}
    \end{itemize}
    \STATE Set $t \ot 0$
    \STATE Set stop $\ot$ FALSE
    \WHILE{$\neg$ stop or $t < T$}
    \STATE
    $\gamma_{t} \ot \max(\gamma_{0} \times \eta^{t}, \underline{\gamma})$
    \STATE  Sample  uniformly  a  minibatch  of  $\widetilde{M}$  observations
    $\tilde{x}_{1:\widetilde{M}}       :=        (\tilde{x}_{1},       \ldots,
    \tilde{x}_{\widetilde{M}})$ from $X$
    \STATE  Sample  uniformly  a  minibatch  of  $\widetilde{N}$  observations
    $\tilde{y}_{1:\widetilde{N}}       :=        (\tilde{y}_{1},       \ldots,
    \tilde{y}_{\widetilde{N}})$ from $Y$
    \STATE                 Define                  and                 compute
    $\theta_{t}(\tilde{x}_{1:\widetilde{M}})  := \big(\theta_{t}(\tilde{x}_1),
    \ldots, \theta_{t}(\tilde{x}_{\widetilde{M}})\big)$
    \STATE Define and compute $\omega_{t} \in (\bbR_{+})^{\widetilde{M}}$ such
    that $\sum_{m \in \llbracket\widetilde{M}\rrbracket} (\omega_{t})_{m} = 1$
    by setting
    \begin{equation*}
      (\omega_{t})_{m}        \propto       \sum_{n        \in       \llbracket\widetilde{N}\rrbracket}
      \varphi\left(\frac{\tilde{y}_{n}    -   \theta_{t}(\tilde{x}_{m})}{h}\right)
      \quad (\text{all\;} m\in \llbracket\widetilde{M}\rrbracket) 
    \end{equation*}
    where $\varphi$ is the standard normal density
    \STATE                                                              Define
    $\mu_{\theta_{t}(\tilde{x}_{1:\widetilde{M}})}^{\omega_{t}}$,          the
    $\omega_{t}$-weighted       empirical       measure      attached       to
    $\theta_{t}(\tilde{x}_{1:\widetilde{M}})$,                             and
    $\nu_{\tilde{y}_{1:\widetilde{N}}}$,  the  empirical measure  attached  to
    $\tilde{y}_{1:\widetilde{N}}$
    \STATE                                                             Compute
    $\Loss_{t}                                                               =
    \bar{\calW}_{\gamma_{t}}\left(\mu^{\omega_{t}}_{\theta_{t}(\tilde{x}_{1:\widetilde{M}})},
      \nu_{\tilde{y}_{1:\widetilde{N}}} \right)$  and $\nabla  \Loss_{t}$, the
    gradient of  $\Loss_{t}$ relative  to the parameter  defining $\theta_{t}$
    \COMMENT{relies on the Sinkhorn-Knopp algorithm}
    \STATE Update the  parameter defining $\theta_{t}$ by  performing one step
    of stochastic gradient descent, yielding $\theta_{t+1}$
    \STATE Check stopping criterion and update stop variable accordingly
    \STATE $t \ot t+1$
    \ENDWHILE
    \STATE Set $\theta_{T} \ot \theta_{t-1}$
    \STATE Set $\gamma_{T} \ot \gamma_{t-1}$
    \STATE  Define  and  compute  $\omega_{T} \in  (\bbR_{+})^{M}$  such  that
    $\sum_{m \in \llbracket M\rrbracket} (\omega_{T})_{m} = 1$ by setting
    \begin{equation*}
      (\omega_{T})_{m} \propto \sum_{n \in \llbracket N\rrbracket}
      \varphi\left(\frac{y_{n}    -   \theta_{T}(x_{m})}{h}\right)
      \quad (\text{all\;} m\in \llbracket M\rrbracket) 
    \end{equation*}
    \STATE    Compute    $\tilde{P}_{T}     \in    \Pi(\omega_{T})$    solving
    $\min_{P                        \in                       \Pi(\omega_{T})}
    \calW_{\gamma_{T}}\left(\mu^{\omega_{T}}_{\theta_{T}(X)}, \nu_{Y} \right)$
\end{algorithmic}
\end{algorithm}

\begin{landscape}
  \begin{table}[h]
    \small{ \center
      \begin{tabular}{c|c|c|c|c|c|c |c}
        \hline
        &   \multicolumn{5}{   |c|    }{the   WTOT(\ldots{})   algorithms}   &
                                                                  \multicolumn{2}{
                                                                               |c}{the
                                                                               CCOT(\ldots{})
                                                                               algorithms}
        \\
        \hline

        & WTOT-SCC1$^*$ & WTOT-SCC1 & WTOT-SCC2$^*$  &  WTOT-SCC2 & WTOT-BC$^*$ &  CCOT-GWD  &CCOT-GWB \\
        \hline
        A1 & $0$ & $0.068\pm 0.126$ &   $0$ & $0.068\pm 0.126$ &$0$& $0.054\pm 0.14$ & 
                                                                                       $0.092\pm 0.15$\\
        A2 & $0\pm 0.001$& $0.014\pm 0.029$ & $0\pm 0.001$ &$0.016\pm 0.035$ &$0.033\pm 0.125$ & $0.105\pm 0.13$& $0.121\pm 0.146$ \\
        A3 & $0.005\pm0.005$ & $0.189\pm 0.175$  &   $0.0182\pm 0.033$ & $0.233\pm 0.179$ &$0.029\pm 0.087$&  $0.612\pm 0.03$ & $0.532\pm 0.068$ \\
        A4 & $0.326\pm0.064$ & $0.282\pm 0.232$  &   $0.257\pm 0.256$ & $0.393\pm 0.164$ &$0.05\pm 0.093$&$0.507\pm 0.123$   &$0.522\pm 0.116$   \\
        \noalign{\smallskip} \hline \noalign{\smallskip}
      \end{tabular}
      \caption{Mean  ($\pm$   standard  deviation)  computed  across   the  30
        independent replications of the co-clustering discrepancy obtained for
        configurations A1, A2, A3, A4.    \label{tab:simulA:results:coclust}   }}
  \end{table}

  \begin{table}[!h]
    \tiny{
      \begin{subtable}{.5\linewidth}
        \centering    
        \begin{tabular}{c|c|c|c|c|c}
	 & $k=k'$ & $\tilde{k}_r$ & precision & sensitivity & specificity\\
	\hline
A1 &$ 10 $&$ 7.825 \pm 0.091 $&$ 1.0 \pm 0.0 $&$ 0.118 \pm 0.001 $&$ 1.0 \pm 0.0 $\\
A1  &$ 35 $&$ 29.373 \pm 0.261 $&$ 1.0 \pm 0.0 $&$ 0.442 \pm 0.003 $&$ 1.0 \pm 0.0 $\\
A1 &$ 65 $&$ 60.649 \pm 0.998 $&$ 0.999 \pm 0.002 $&$ 0.913 \pm 0.014 $&$ 1.0 \pm 0.0 $\\
A1  &$ 75 $&$  67.418 \pm 0.9 $&$ 0.981 \pm 0.006 $&$ 0.991 \pm 0.013 $&$ 0.994 \pm 0.002 $\\
A1 &$ 95 $&$ 76.335 \pm 1.282 $&$ 0.888 \pm 0.014 $&$ 1.0 \pm 0.0 $&$ 0.957 \pm 0.005 $\\
A1 &$ 150 $&$ 97.049 \pm 1.182 $&$ 0.727 \pm 0.012 $&$ 1.0 \pm 0.0 $&$ 0.879 \pm 0.005 $\\	\hline
		\end{tabular}     
      \end{subtable}%
      \begin{subtable}{.5\linewidth}
        \centering
\begin{tabular}{c|c|c|c|c|c}
 & $k=k'$ & $\tilde{k}_r$ & precision & sensitivity & specificity\\
\hline
A4 &$ 10 $&$ 6.964 \pm 0.161 $&$ 0.998 \pm 0.003 $&$ 0.089 \pm 0.003 $&$ 1.0 \pm 0.0 $\\
A4  &$ 35 $&$ 28.632 \pm 0.668 $&$ 0.995 \pm 0.009 $&$ 0.374 \pm 0.01 $&$ 1.0 \pm 0.0 $\\
 A4  &$ 65 $&$ 54.653 \pm 0.927 $&$ 0.986 \pm 0.011 $&$ 0.668 \pm 0.018 $&$ 0.998 \pm 0.002 $\\
A4 &$ 75 $&$ 61.193 \pm 0.724 $&$ 0.963 \pm 0.016 $&$ 0.709 \pm 0.022 $&$ 0.993 \pm 0.003 $\\
A4 &$ 95 $&$  75.856 \pm 0.749 $&$ 0.893 \pm 0.017 $&$ 0.768 \pm 0.022 $&$ 0.975 \pm 0.003 $\\
A4 &$ 150 $&$ 121.273 \pm 3.63 $&$ 0.783 \pm 0.025 $&$ 0.976 \pm 0.023 $&$ 0.936 \pm 0.011 $\\
\hline
\end{tabular}
\end{subtable} 
\caption{Mean ($\pm$  standard deviation)  computed across the  30 independent
  replications of  $\tilde{k}_{r}$, precision, sensitivity and  specificity of
  the $m$-specific  matchings averaged across  all mRNAs for  configuration A1
  (left) and A4 (right).  \label{tab:simulA14:results:matching}}}
\end{table}

\begin{table}[!h]
    \tiny{
      \begin{subtable}{.5\linewidth}
        \centering
      \begin{tabular}{c|c|c|c|c|c}
 & $k=k'$ & $\tilde{k}_r$ & precision & sensitivity & specificity\\
 \hline
A1  &$ 75 $&$ 67.418 \pm 0.9 $&$ 0.981 \pm 0.006 $&$ 0.991 \pm 0.013 $&$ 0.994 \pm 0.002 $\\
A2 &$ 130 $&$ 100.217 \pm 2.127 $&$ 0.976 \pm 0.017 $&$ 0.894 \pm 0.027 $&$ 0.995 \pm 0.004 $\\
A3 &$ 120 $&$ 82.764 \pm 1.105 $&$ 0.881 \pm 0.015 $&$ 0.902 \pm 0.025 $&$ 0.968 \pm 0.004 $\\
A4 &$ 120 $&$ 97.561 \pm 1.836 $&$ 0.821 \pm 0.015 $&$ 0.853 \pm 0.025 $&$ 0.95 \pm 0.005 $\\
\hline
\end{tabular}
      \end{subtable}%
      \begin{subtable}{.5\linewidth}
        \centering
      \begin{tabular}{c|c|c|c|c|c}
 & $k=k'$ & $\tilde{k}_c$ & precision & sensitivity & specificity\\
\hline
A1 &$ 75 $&$ 67.418 \pm 0.9 $&$ 0.982 \pm 0.006 $&$ 0.991 \pm 0.015 $&$ 0.994 \pm 0.002 $\\
A2 &$ 130 $&$ 100.217 \pm 2.127 $&$ 0.984 \pm 0.012 $&$ 0.894 \pm 0.028 $&$ 0.995 \pm 0.004 $\\
A3 &$ 120 $&$ 110.352 \pm 1.473 $&$ 0.878 \pm 0.017 $&$ 0.9 \pm 0.024 $&$ 0.967 \pm 0.004 $\\
A4 &$ 120 $&$ 97.561 \pm 1.836 $&$ 0.84 \pm 0.018 $&$ 0.853 \pm 0.026 $&$ 0.951 \pm 0.004 $\\
\hline
\end{tabular}
\end{subtable}
\caption{Mean  ($\pm$   standard  deviation)  computed  across   the  30
  independent  replications   of  $\tilde{k}_{r}$   or  $\tilde{k}_{c}$,
  precision, sensitivity  and specificity of the  $m$-specific matchings
  (left) and  $n$-specific matchings  (right) averaged across  all mRNAs
  (left) and all miRNAs (right).  \label{tab:simulA:results:matching} }}
\end{table}
\end{landscape}

\begin{landscape}
  \begin{table}[!ht]
    \small{%
      \centering
      \begin{tabular}{c|c|c|c|c|c|c|c |}
        % \noalign{\smallskip} \hline \hline \noalign{\smallskip}
        \hline
        &   \multicolumn{5}{   |c|    }{the   WTOT(\ldots{})   algorithms}   &
                                                                               \multicolumn{2}{
                                                                               |c}{the
                                                                               CCOT(\ldots{})
                                                                               algorithms}
        \\
        \hline
        & WTOT-SCC1$^*$ & WTOT-SCC1 & WTOT-SCC2$^*$  &  WTOT-SCC2 & WTOT-BC$^*$ &  CCOT-GWD  &CCOT-GWB \\
        \hline
        B1&$ 0.062 \pm 0.151 $&$ 0.204 \pm 0.221 $&$ 0.082 \pm 0.161 $&$ 0.204 \pm 0.221 $& $0.049 \pm 0.125$ &
                                                                                                                $ 0.276 \pm 0.204 $&$ 0.53 \pm 0.168 $  \\
        B2&$ 0.114 \pm 0.108 $&$ 0.418 \pm 0.265 $&$ 0.178 \pm 0.207 $&$ 0.455 \pm 0.258 $& $0.382 \pm 0.121$ &
                                                                                                                $ 0.477 \pm 0.14 $&$ 0.523 \pm 0.115 $ \\
        B3&$ 0.175 \pm 0.086 $&$ 0.724 \pm 0.236 $&$ 0.163 \pm 0.082 $&$ 0.775 \pm 0.176 $& $- $ &
                                                                                                   $ 0.858 \pm 0.042 $&$ 0.867 \pm 0.044 $ \\
        B4&$ 0.174 \pm 0.092 $&$ 0.747 \pm 0.196 $&$ 0.171 \pm 0.112 $&$ 0.782 \pm 0.159 $& $- $ &
                                                                                                       $ 0.882 \pm 0.041 $&$ 0.883 \pm 0.04 $ \\
        \noalign{\smallskip} \hline \noalign{\smallskip}
      \end{tabular}
    }
    \caption{Mean  ($\pm$   standard  deviation)  computed  across   the  30
      independent replications of the co-clustering discrepancy obtained for
      configurations B1, B2, B3, B4. }
    \label{tab:simulB:results:coclust}
    % \label{tab:SS2}        
  \end{table}
  
  \begin{table}[!htb]
    \tiny{%
    \begin{subtable}{.5\linewidth}
      \centering
      \begin{tabular}{c|c|c|c|c|c}
        & $k=k'$ & $\tilde{k}_r$ & precision & sensitivity & specificity\\
        \hline
        B1&$ 60 $&$ 48.578 \pm 5.201 $&$ 0.885 \pm 0.209 $&$ 0.658 \pm 0.191 $&$ 0.985 \pm 0.025 $\\
        B1&$ 80 $&$ 63.96 \pm 6.126 $&$ 0.851 \pm 0.199 $&$ 0.816 \pm 0.222 $&$ 0.968 \pm 0.03 $\\
        B1&$ 85 $&$ 67.537 \pm 6.193 $&$ 0.837 \pm 0.193 $&$ 0.842 \pm 0.222 $&$ 0.961 \pm 0.031 $\\
        B1&$ 90 $&$ 71.214 \pm 6.208 $&$ 0.823 \pm 0.186 $&$ 0.864 \pm 0.219 $&$ 0.953 \pm 0.031 $\\
        B1&$ 110 $&$ 85.833 \pm 6.358 $&$ 0.753 \pm 0.156 $&$ 0.918 \pm 0.202 $&$ 0.913 \pm 0.029 $\\
        \hline
      \end{tabular}
    \end{subtable}%
    \begin{subtable}{.5\linewidth}
      \centering
      \begin{tabular}{c|c|c|c|c|c}
        & $k=k'$ & $\tilde{k}_r$ & precision & sensitivity & specificity\\
        \hline
        B4&$ 10 $&$ 6.78 \pm 0.259 $&$ 0.926 \pm 0.102 $&$ 0.321 \pm 0.046 $&$ 0.999 \pm 0.001 $\\
        B4&$ 20 $&$ 15.163 \pm 0.619 $&$ 0.873 \pm 0.091 $&$ 0.72 \pm 0.087 $&$ 0.996 \pm 0.003 $\\
        B4&$ 25 $&$ 19.033 \pm 0.784 $&$ 0.817 \pm 0.084 $&$ 0.837 \pm 0.084 $&$ 0.991 \pm 0.004 $\\
        B4&$ 30 $&$ 22.889 \pm 0.997 $&$ 0.754 \pm 0.076 $&$ 0.907 \pm 0.077 $&$ 0.984 \pm 0.005 $\\
        B4&$ 40 $&$ 31.118 \pm 1.086 $&$ 0.618 \pm 0.053 $&$ 0.969 \pm 0.049 $&$ 0.963 \pm 0.005 $\\
        \hline
      \end{tabular}
    \end{subtable}
  }
  \caption{Mean   ($\pm$  standard   deviation)  computed   across  the   30
    independent replications of  $\tilde{k}_{r}$, precision, sensitivity and
    specificity of the $m$-specific matchings  averaged across all mRNAs for
    configurations           B1            (left)           and           B4
    (right).}
  \label{tab:simulB14:results:matching}
\end{table}

\begin{table}[!htb]
  \tiny{%
    \begin{subtable}{.5\linewidth}
      \centering
      \begin{tabular}{c|c|c|c|c|c}
        & $k=k'$ & $\tilde{k}_r$ & precision & sensitivity & specificity\\
        \hline
        B1&$ 85 $&$ 67.537 \pm 6.193 $&$ 0.837 \pm 0.193 $&$ 0.842 \pm 0.222 $&$ 0.961 \pm 0.031 $\\
        B2&$ 60 $&$ 48.282 \pm 3.449 $&$ 0.751 \pm 0.194 $&$ 0.838 \pm 0.2 $&$ 0.979 \pm 0.022 $\\
        B3&$ 25 $&$ 19.546 \pm 1.151 $&$ 0.833 \pm 0.136 $&$ 0.837 \pm 0.152 $&$ 0.992 \pm 0.006 $\\
        B4&$ 25 $&$ 19.033 \pm 0.784 $&$ 0.817 \pm 0.084 $&$ 0.837 \pm 0.084 $&$ 0.991 \pm 0.004 $\\
        \hline
      \end{tabular}
    \end{subtable}%
    \begin{subtable}{.5\linewidth}
      \centering
      \begin{tabular}{c|c|c|c|c|c}
        & $k=k'$ & $\tilde{k}_c$ & precision & sensitivity & specificity\\
        \hline
        B1&$ 85 $&$ 63.732 \pm 8.642 $&$ 0.844 \pm 0.175 $&$ 0.836 \pm 0.229 $&$ 0.96 \pm 0.033 $\\
        B2&$ 60 $&$ 44.349 \pm 2.495 $&$ 0.792 \pm 0.218 $&$ 0.819 \pm 0.227 $&$ 0.971 \pm 0.024 $\\
        B3&$ 25 $&$ 18.766 \pm 0.97 $&$ 0.847 \pm 0.125 $&$ 0.833 \pm 0.152 $&$ 0.991 \pm 0.005 $\\
        B4&$ 25 $&$ 18.833 \pm 0.793 $&$ 0.834 \pm 0.087 $&$ 0.827 \pm 0.099 $&$ 0.99 \pm 0.005 $\\
        \hline
      \end{tabular}
    \end{subtable}
  }
  \caption{Mean ($\pm$  standard deviation)
    computed  across  the  30  independent replications  of  $\tilde{k}_{r}$  or
    $\tilde{k}_{c}$, precision, sensitivity and  specificity of the $m$-specific
    matchings  (left) and  $n$-specific  matchings (right)  averaged across  all
    mRNAs (left) and all miRNAs (right).}
  \label{tab:simulB:results:matching}  
\end{table}
\end{landscape}

\begin{landscape}
  \begin{table}[!ht]
    \small{%
      \center
      \begin{tabular}{c|c|c|c|c|c|c|c|c|c }
        \hline
        & \multicolumn{5}{ |c| }{the  WTOT(\ldots{}) algorithms} & \multicolumn{2}{
                                                                   |c}{the
                                                                   CCOT-(\ldots{}) algorithms} \\
        \hline
        % & SpecC  & SpecC  & Adapted SpecC  & Adapted SpecC & BlockC & CCOT  & CCOT-GW \\
        & WTOT-SCC1$^*$ & WTOT-SCC1 & WTOT-SCC2$^*$  &  WTOT-SCC2 & WTOT-BC$^*$ &  CCOT-GWD  &CCOT-GWB \\
        \hline
        C1&  $0.106  \pm 0.1 $&  $0.203 \pm 0.135$ & $0.101 \pm 0.056$ & $0.194 \pm 0.116$ & $ 0.265 \pm 0.255 $ &$0.496\pm 0.16 $  &  $0.902\pm 0.007 $  \\
        C2&$ 0.209 \pm 0.131 $&$ 0.252 \pm 0.182 $&$ 0.262 \pm 0.141 $&$ 0.345 \pm 0.205 $& $- $ &$ 0.938 \pm 0.023 $&$ 0.971 \pm 0.026$\\
        C3&$ 0.609 \pm 0.113 $&$ 0.693 \pm 0.154 $&$ 0.385 \pm 0.151 $&$ 0.521 \pm 0.198 $& $- $ &
                                                                                                   $0.926 \pm 0.027 $ & $0.987\pm0.002 $  \\
        C4&$ 0.63 \pm 0.141 $&$ 0.751 \pm 0.145 $&$ 0.435 \pm 0.197 $&$ 0.6 \pm 0.233 $& $- $ &
                                                                                                    $0.939\pm0.027 $  & $0.987\pm 0.002$   \\
        \noalign{\smallskip} \hline \noalign{\smallskip}
      \end{tabular}
    }
    \caption{Mean  ($\pm$   standard  deviation)  computed  across   the  30
      independent replications of the co-clustering discrepancy obtained for
      configurations C1, C2, C3, C4.}
    \label{tab:simulC:results:coclust} 
  \end{table}

  \begin{table}[!ht]
  \tiny{%
    \begin{subtable}{.5\linewidth}
      \centering
      \begin{tabular}{c|c|c|c|c|c}
        & $k=k'$ & $\tilde{k}_r$ & precision & sensitivity & specificity\\
        \hline
        C1 &$ 10 $&$ 7.748 \pm 0.446 $&$ 0.973 \pm 0.03 $&$ 0.156 \pm 0.01 $&$ 1.0 \pm 0.0 $\\
        C1 &$ 30 $&$ 25.888 \pm 1.418 $&$ 0.972 \pm 0.029 $&$ 0.526 \pm 0.032 $&$ 1.0 \pm 0.0 $\\
        C1 &$ 50 $&$ 45.521 \pm 2.441 $&$ 0.944 \pm 0.025 $&$ 0.916 \pm 0.04 $&$ 1.0 \pm 0.001 $\\
        C1 &$ 55 $&$ 49.108 \pm 3.018 $&$ 0.93 \pm 0.021 $&$ 0.972 \pm 0.025 $&$ 0.999 \pm 0.002 $\\
        C1 &$ 60 $&$ 51.365 \pm 3.335 $&$ 0.919 \pm 0.024 $&$ 0.993 \pm 0.011 $&$ 0.997 \pm 0.004 $\\
        C1 &$ 70  $&$ 55.296 \pm 3.312 $&$  0.881 \pm 0.034 $&$ 1.0 \pm  0.0 $&$ 0.985
                                                                                \pm  0.01
                                                                                $\\
        \hline
      \end{tabular}
    \end{subtable}%
    \begin{subtable}{.5\linewidth}
      \centering
      \begin{tabular}{c|c|c|c|c|c}
        & $k=k'$ & $\tilde{k}_r$ & precision & sensitivity & specificity\\
        \hline
        C4&$ 5 $&$ 3.293 \pm 0.096 $&$ 0.895 \pm 0.023 $&$ 0.185 \pm 0.012 $&$ 1.0 \pm 0.0 $\\
        C4&$ 10 $&$ 7.278 \pm 0.303 $&$ 0.899 \pm 0.022 $&$ 0.474 \pm 0.029 $&$ 1.0 \pm 0.0 $\\
        C4&$ 15 $&$ 11.982 \pm 0.578 $&$ 0.888 \pm 0.02 $&$ 0.787 \pm 0.04 $&$ 1.0 \pm 0.0 $\\
        C4&$ 20 $&$ 15.935 \pm 0.864 $&$ 0.843 \pm 0.022 $&$ 0.96 \pm 0.023 $&$ 0.997 \pm 0.001 $\\
        C4&$ 25 $&$ 19.138 \pm 0.89 $&$ 0.762 \pm 0.032 $&$ 0.997 \pm 0.005 $&$ 0.989 \pm 0.003 $\\
        C4&$ 30 $&$ 22.578 \pm 1.11 $&$ 0.671 \pm 0.04 $&$ 1.0 \pm 0.0 $&$ 0.978 \pm 0.004 $\\
        \hline
      \end{tabular}
    \end{subtable}
  }
  \caption{Mean ($\pm$ standard deviation)  computed across the 30 independent
    replications of $\tilde{k}_{r}$, precision, sensitivity and specificity of
    the $m$-specific matchings averaged across all mRNAs for configurations C1
    (left) and C4 (right).}
  \label{tab:simulC34:results:matching}
\end{table}  

\begin{table}[!ht]
  \tiny{%
    \begin{subtable}{.5\linewidth}
      \centering
      \begin{tabular}{c|c|c|c|c|c}
        & $k=k'$ & $\tilde{k}_r$ & precision & sensitivity & specificity\\
        \hline
        C1 &$ 55 $&$ 49.108 \pm 3.018 $&$ 0.93 \pm 0.021 $&$ 0.972 \pm 0.025 $&$ 0.999 \pm 0.002 $\\
        C2&$ 20 $&$ 16.203 \pm 0.956 $&$ 0.955 \pm 0.016 $&$ 0.965 \pm 0.021 $&$ 0.997 \pm 0.001 $\\
        C3&$ 20 $&$ 15.552 \pm 0.877 $&$ 0.854 \pm 0.024 $&$ 0.968 \pm 0.019 $&$ 0.997 \pm 0.001 $\\
        C4&$ 20 $&$ 15.935 \pm 0.864 $&$ 0.843 \pm 0.022 $&$ 0.96 \pm 0.023 $&$ 0.997 \pm 0.001 $\\
        \hline
      \end{tabular}
    \end{subtable}%
    \begin{subtable}{.5\linewidth}
      \centering
      \begin{tabular}{c|c|c|c|c|c}
        & $k=k'$ & $\tilde{k}_c$ & precision & sensitivity & specificity\\
        \hline
        C1 &$ 55 $&$ 49.056 \pm 3.461 $&$ 0.898 \pm 0.06 $&$ 0.971 \pm 0.026 $&$ 0.981 \pm 0.009 $\\
        C2&$ 20 $&$ 16.371 \pm 0.812 $&$ 0.953 \pm 0.018 $&$ 0.963 \pm 0.023 $&$ 0.997 \pm 0.001 $\\
        C3&$ 20 $&$ 15.879 \pm 0.691 $&$ 0.804 \pm 0.025 $&$ 0.969 \pm 0.018 $&$ 0.993 \pm 0.001 $\\
        C4&$ 20 $&$ 15.867 \pm 0.635 $&$ 0.812 \pm 0.032 $&$ 0.961 \pm 0.021 $&$ 0.993 \pm 0.002 $\\
      \end{tabular}
    \end{subtable}
  }
  \caption{Mean   ($\pm$  standard   deviation)  computed   across  the   30
    independent   replications   of  $\tilde{k}_{r}$   or   $\tilde{k}_{c}$,
    precision,  sensitivity and  specificity of  the $m$-specific  matchings
    (left)  and $n$-specific  matchings  (right) averaged  across all  mRNAs
    (left) and all miRNAs (right). }
  \label{tab:simulC:results:matching}
\end{table}
\end{landscape}

\begin{figure}
  \centering
  \includegraphics[width=.45\textwidth]{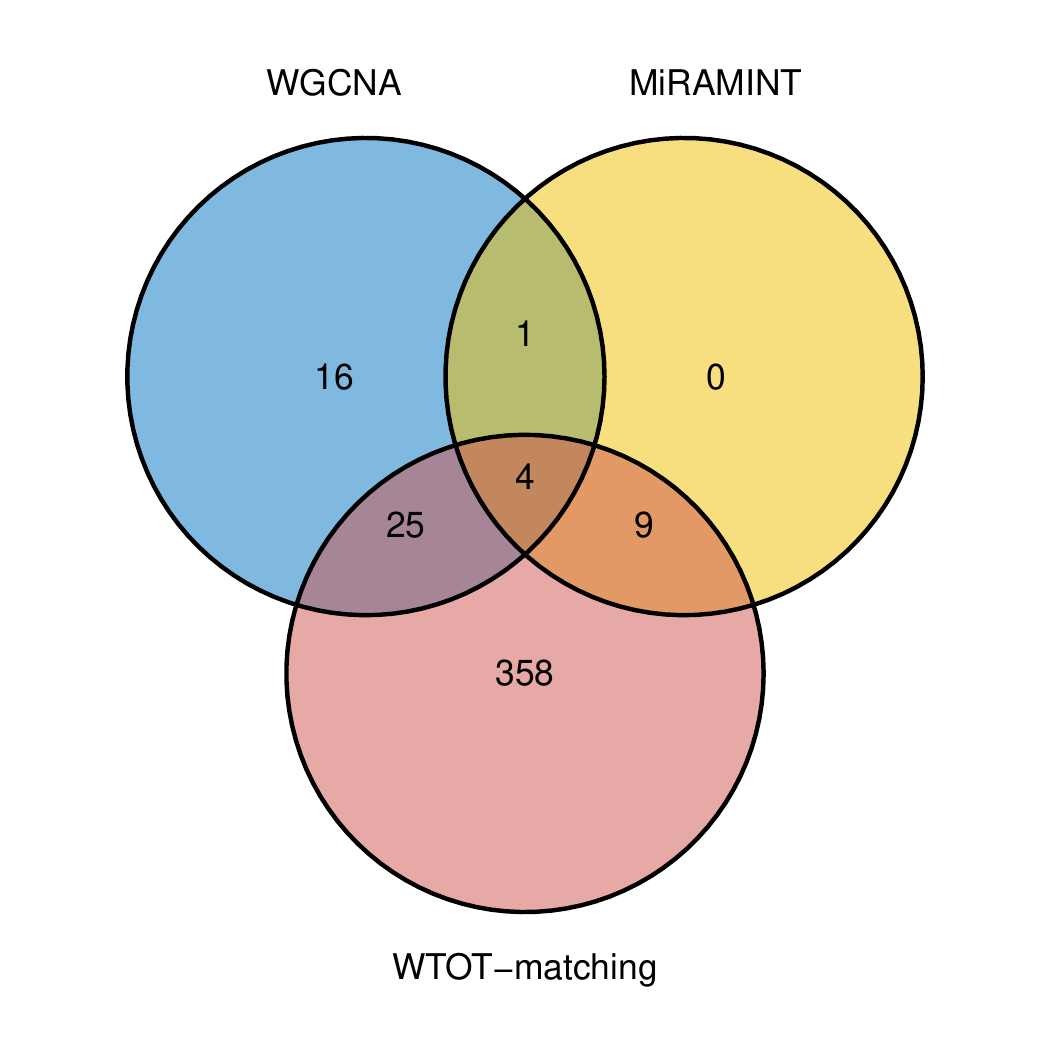}
  \includegraphics[width=.45\textwidth]{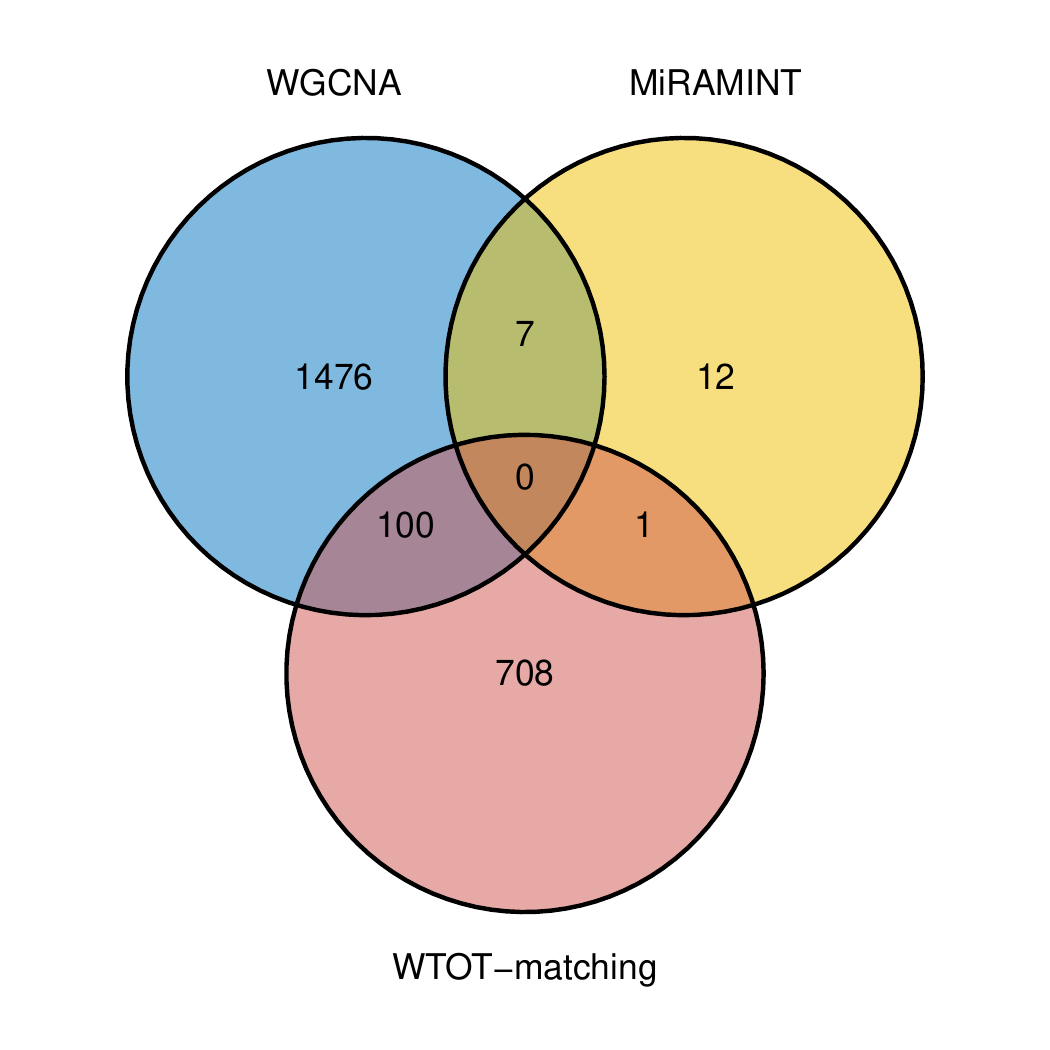}
  \caption{Venn diagrams summarizing  the overlaps between the  sets of miRNAs
    (left)  and mRNAs  (right) which  belong to  a pair  output by  the WGCNA,
    MiRAMINT and WTOT-matching algorithms.}
  \label{fig:venn:all}
\end{figure}

\begin{figure}
  \centering
  \includegraphics[width=.45\textwidth]{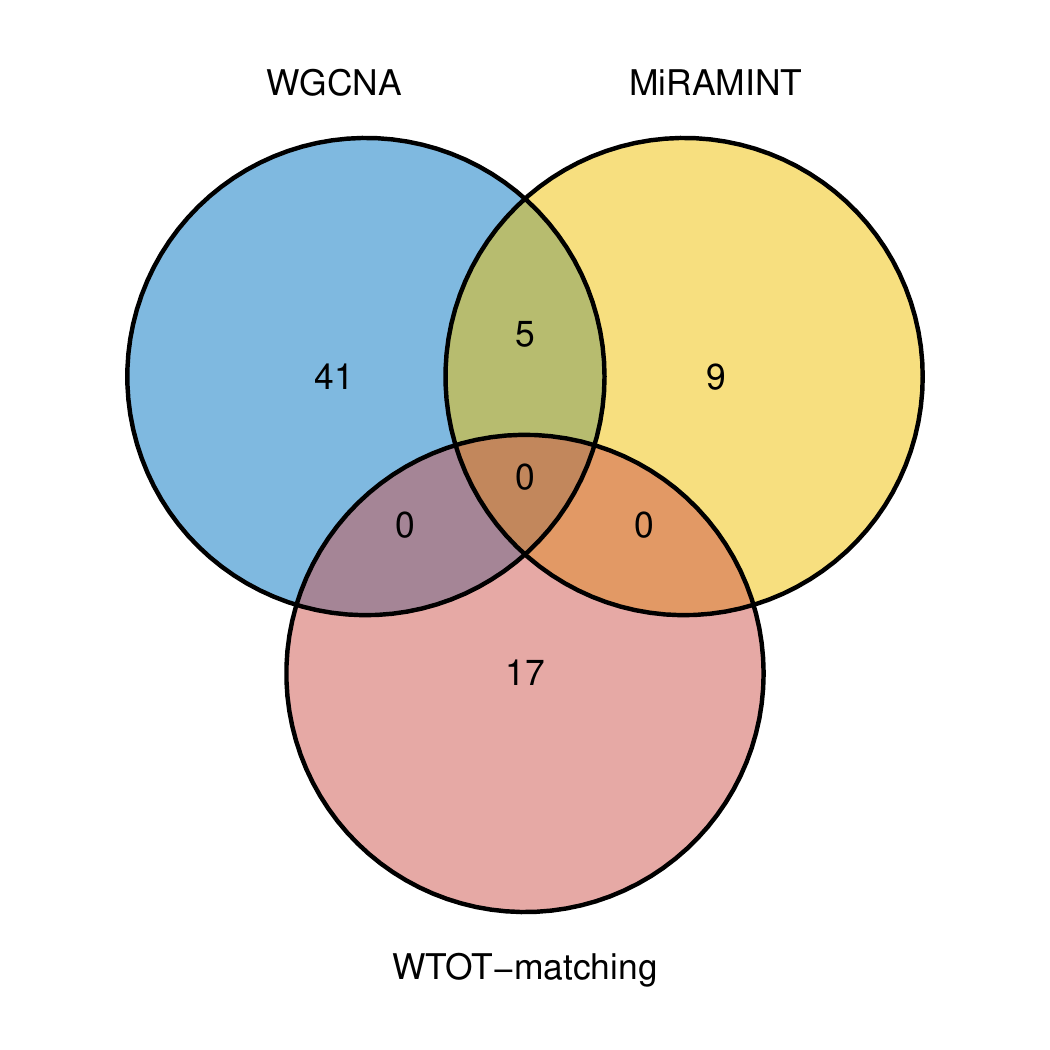}
  \includegraphics[width=.45\textwidth]{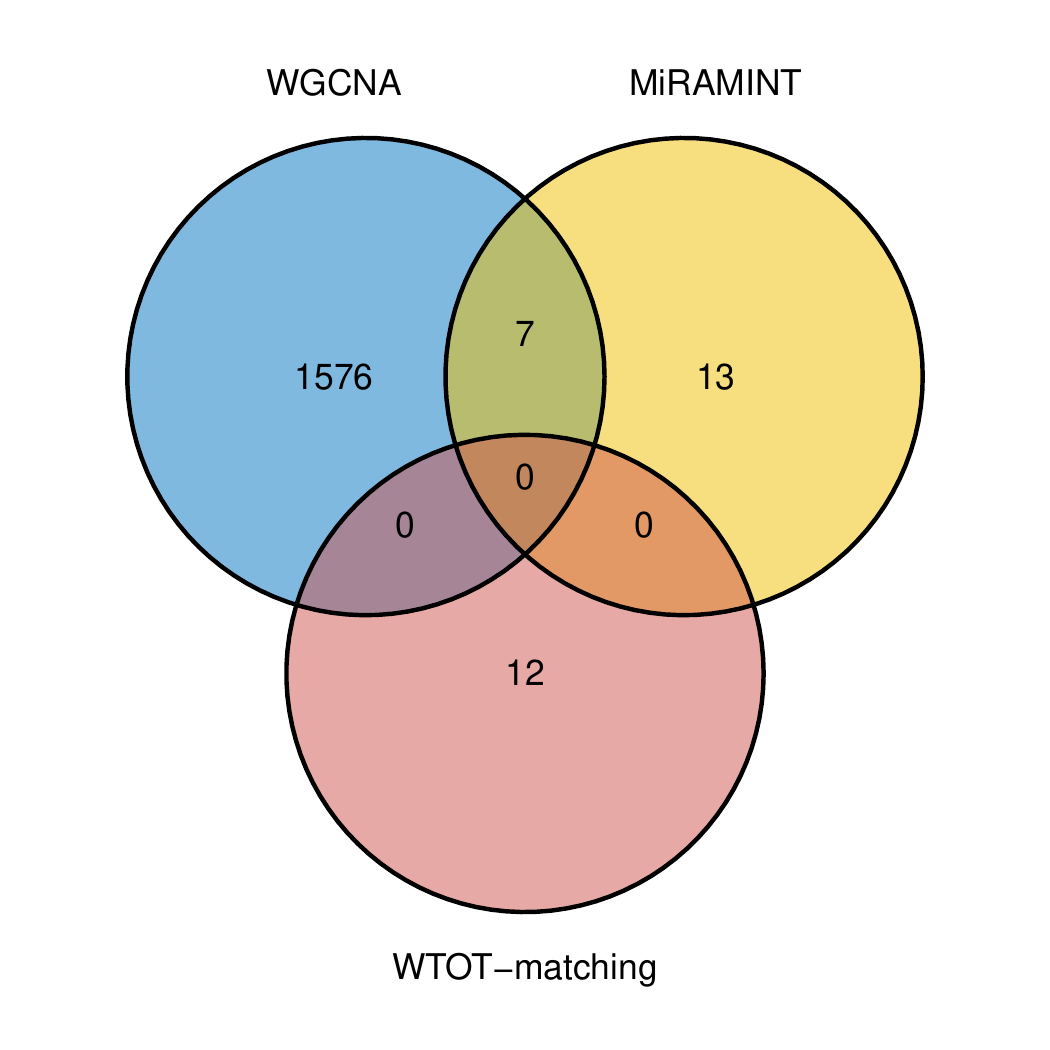}
  \caption{Venn diagrams summarizing  the overlaps between the  sets of miRNAs
    (left)  and mRNAs  (right) which  belong to  a pair  output by  the WGCNA,
    MiRAMINT   and   WTOT-matching   algorithms,   \textit{focusing   on   the
      WTOT-matching matchings labeled as peaked}.}
  \label{fig:venn:peaked}
\end{figure}

\begin{figure}
  \centering
  \includegraphics[width=.45\textwidth]{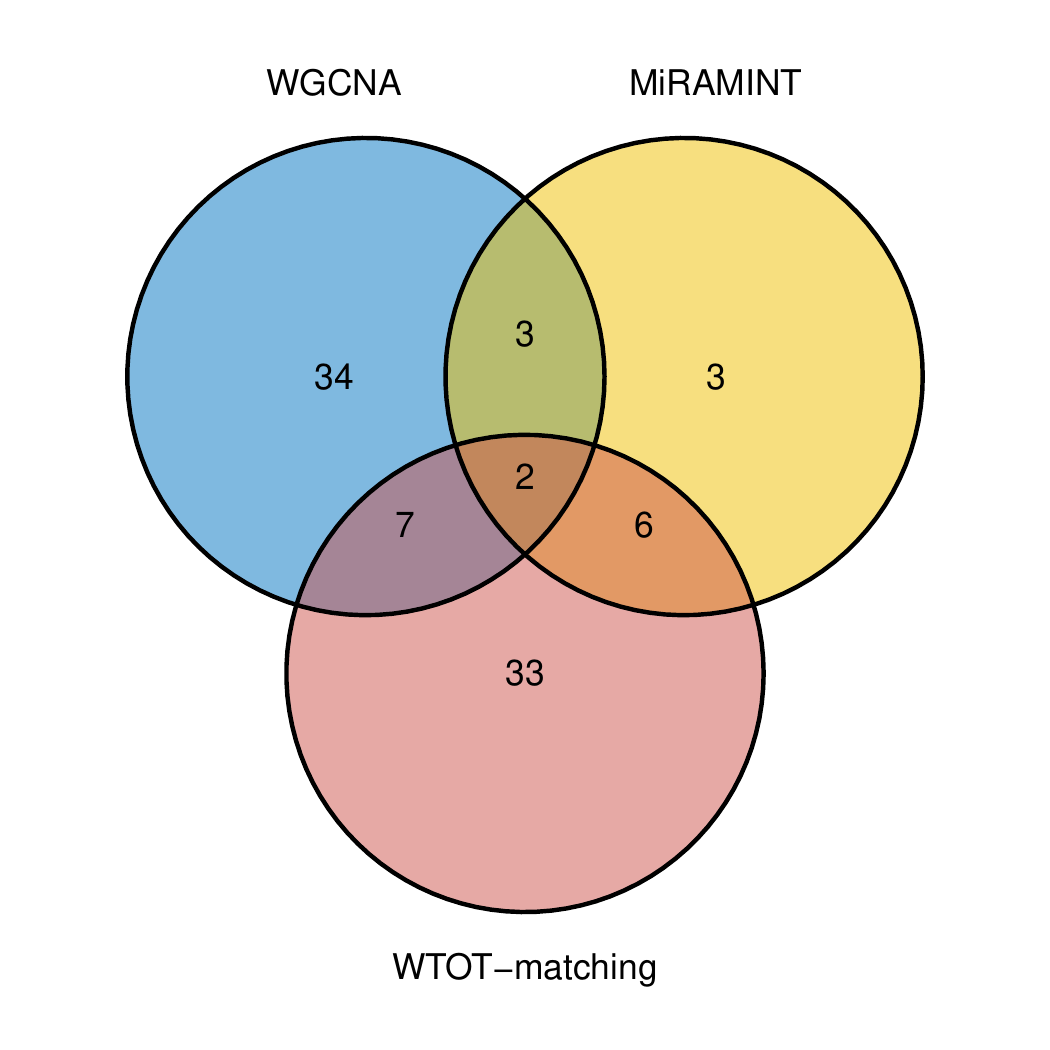}
  \includegraphics[width=.45\textwidth]{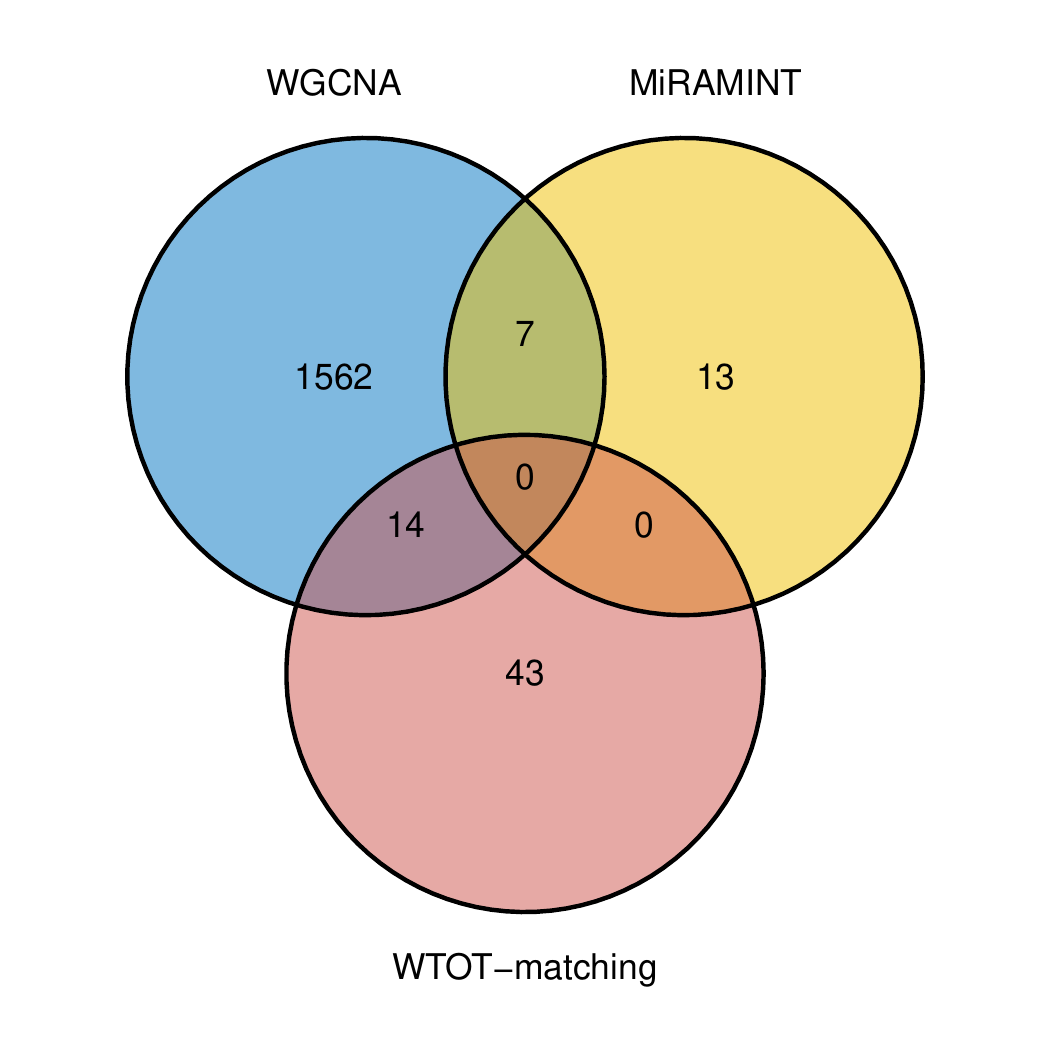}
  \caption{Venn diagrams summarizing  the overlaps between the  sets of miRNAs
    (left)  and mRNAs  (right) which  belong to  a pair  output by  the WGCNA,
    MiRAMINT   and   WTOT-matching   algorithms,   \textit{focusing   on   the
      WTOT-matching matchings labeled as monotonic}.}
  \label{fig:venn:monotonic}
\end{figure}

\begin{figure}
  \centering
  \includegraphics[width=.45\textwidth]{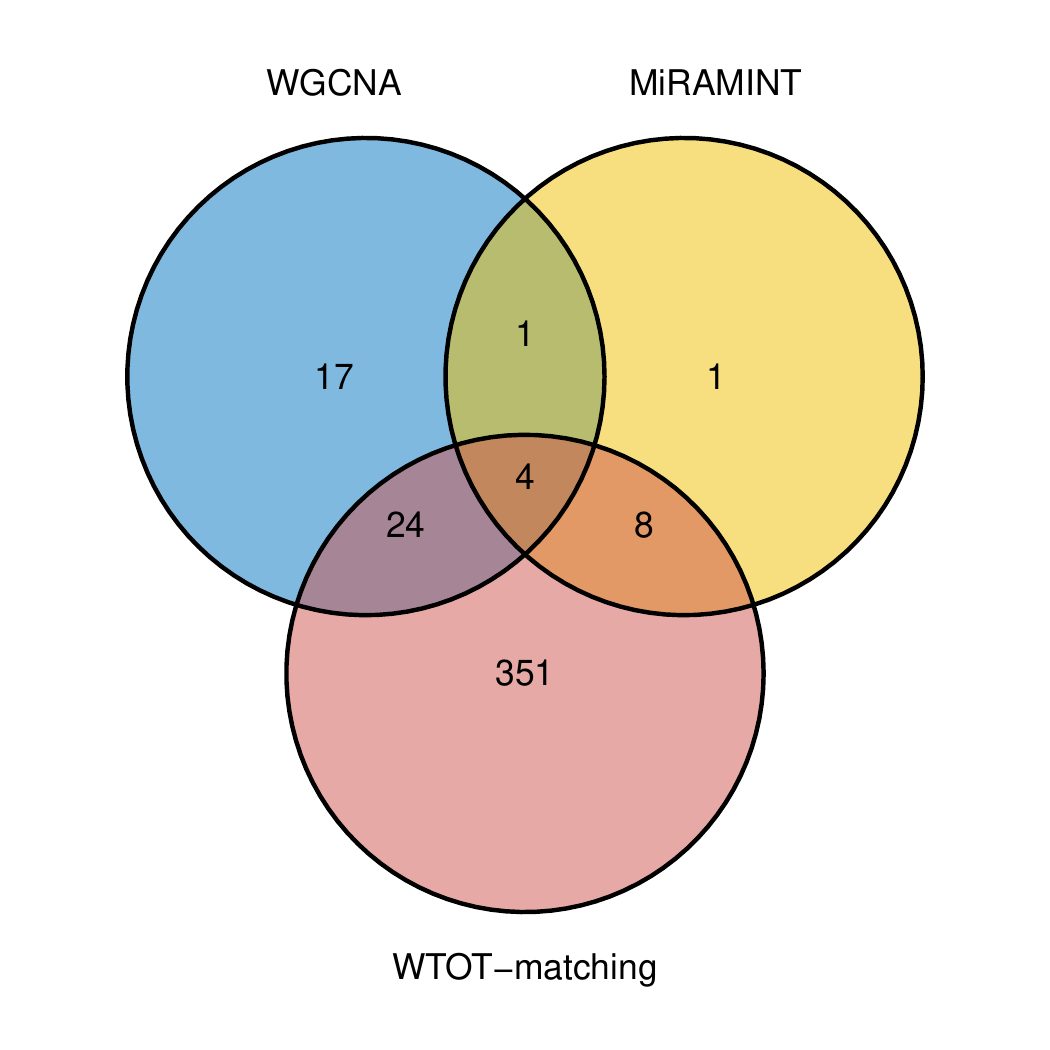}
  \includegraphics[width=.45\textwidth]{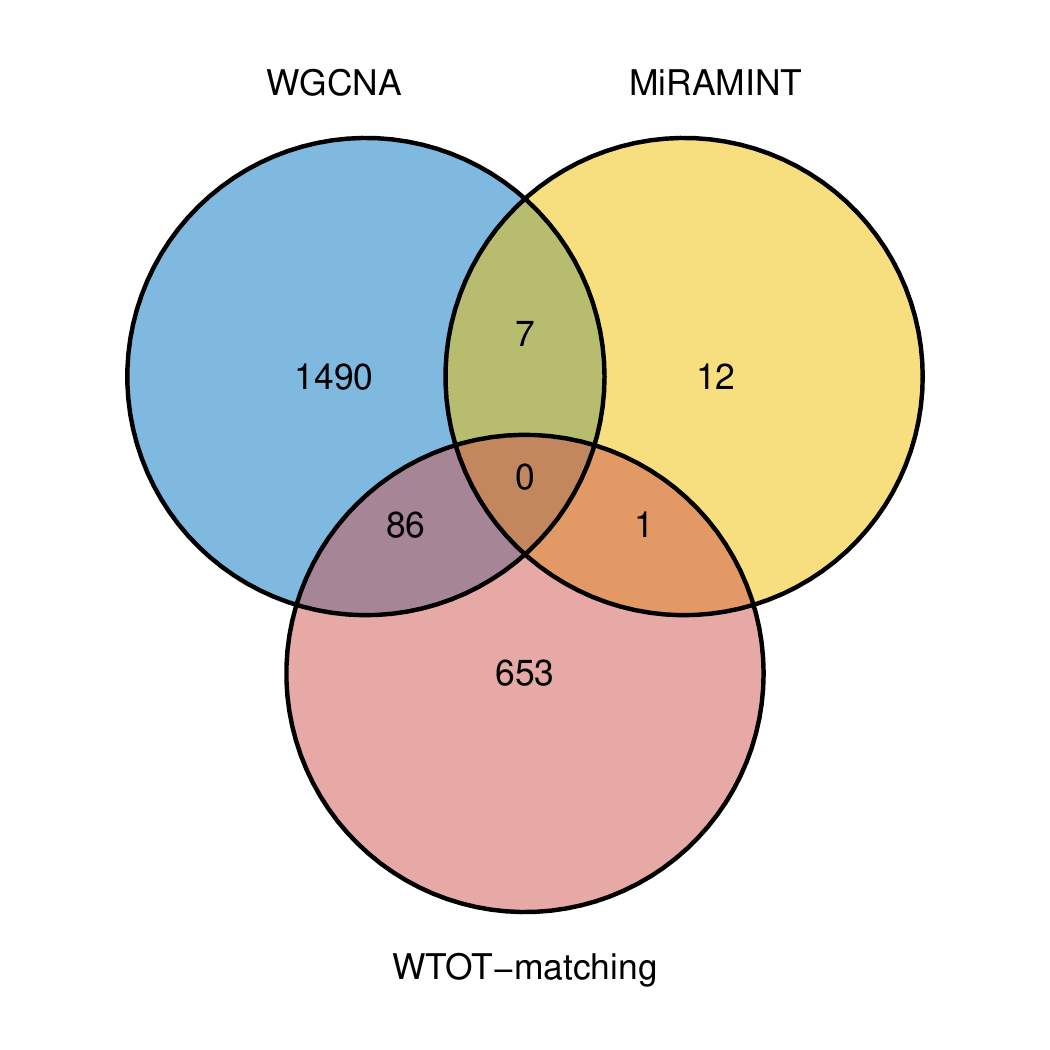}
  \caption{Venn diagrams summarizing  the overlaps between the  sets of miRNAs
    (left)  and mRNAs  (right) which  belong to  a pair  output by  the WGCNA,
    MiRAMINT   and   WTOT-matching   algorithms,   \textit{focusing   on   the
      WTOT-matching  matchings  which  are   labeled  as  neither  peaked  nor
      monotonic}.}
  \label{fig:venn:others}
\end{figure}

\begin{figure}
  \centering
  \includegraphics[width=.9\textwidth]{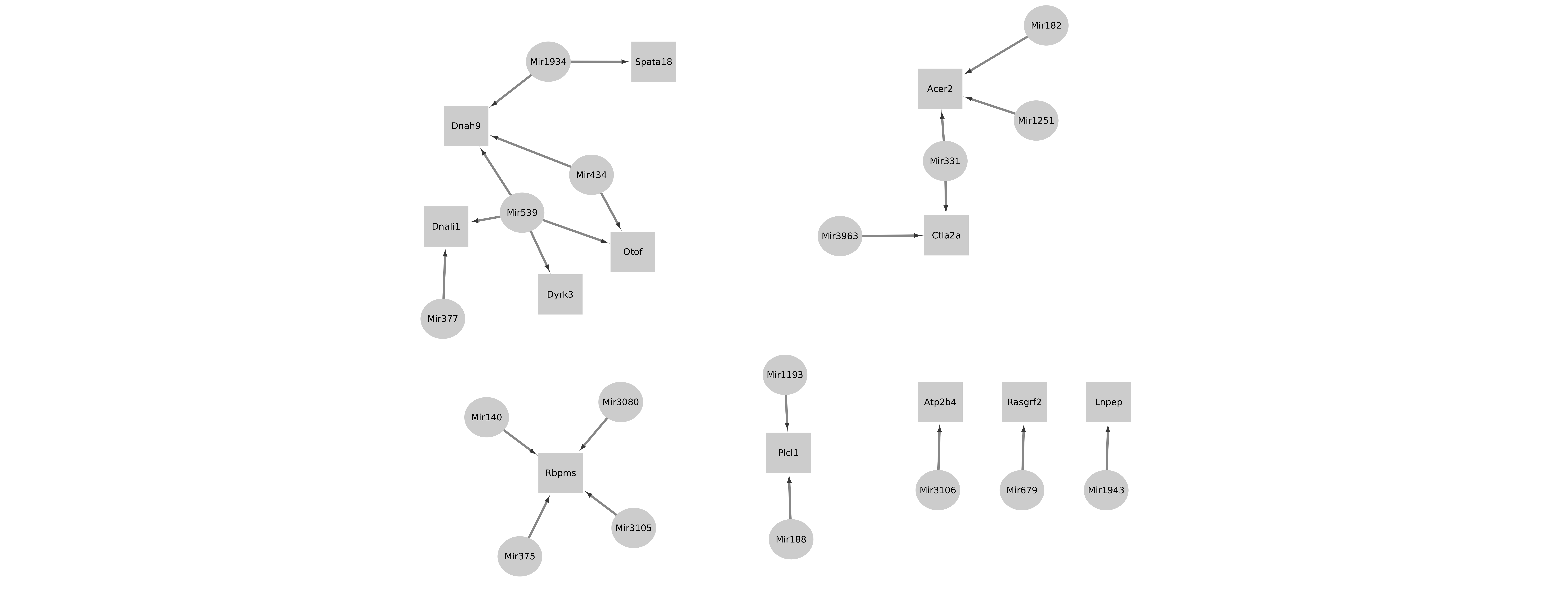}
  \caption{The mRNA-miRNA networks based on the mRNA-miRNA matchings output by
    the WTOT-matching  algorithm, \textit{focusing on the  matchings which are
      labeled as  peaked}.  Disks correspond  to miRNAs and squares  to mRNAs.
    The top
    \label{fig:network:peaked}
    annotation is \textit{conventional motile cilium} (GO:0097729, 3 hits).}
\end{figure}

\begin{figure}
  \centering
  \includegraphics[width=.9\textwidth]{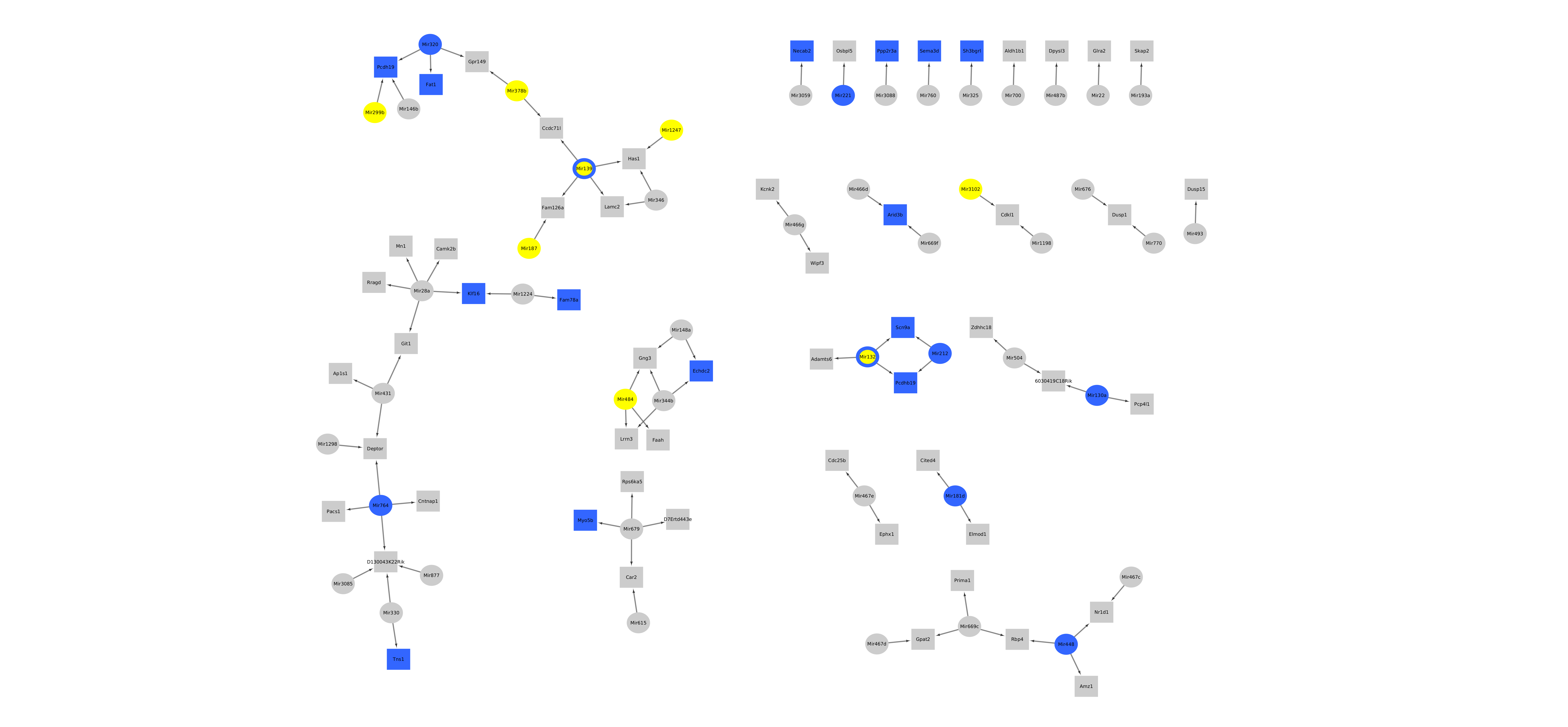}
  \caption{The mRNA-miRNA networks based on the mRNA-miRNA matchings output by
    the WTOT-matching  algorithm, \textit{focusing on the  matchings which are
      labeled as monotonic}.  Disks correspond to miRNAs and squares to mRNAs.
    Elements also retained by the  WGCNA algorithm (respectively, the MiRAMINT
    algorithm)  are  indicated  in   blue  (respectively,  yellow).   The  top
    annotation  is  \textit{mitigation  of host  antiviral  defense  response}
    (GO:0050690, 2 hits).}
  \label{fig:network:monot}
\end{figure}

\begin{landscape}
  \begin{figure}
    \centering
    \includegraphics[width=.9\textwidth]{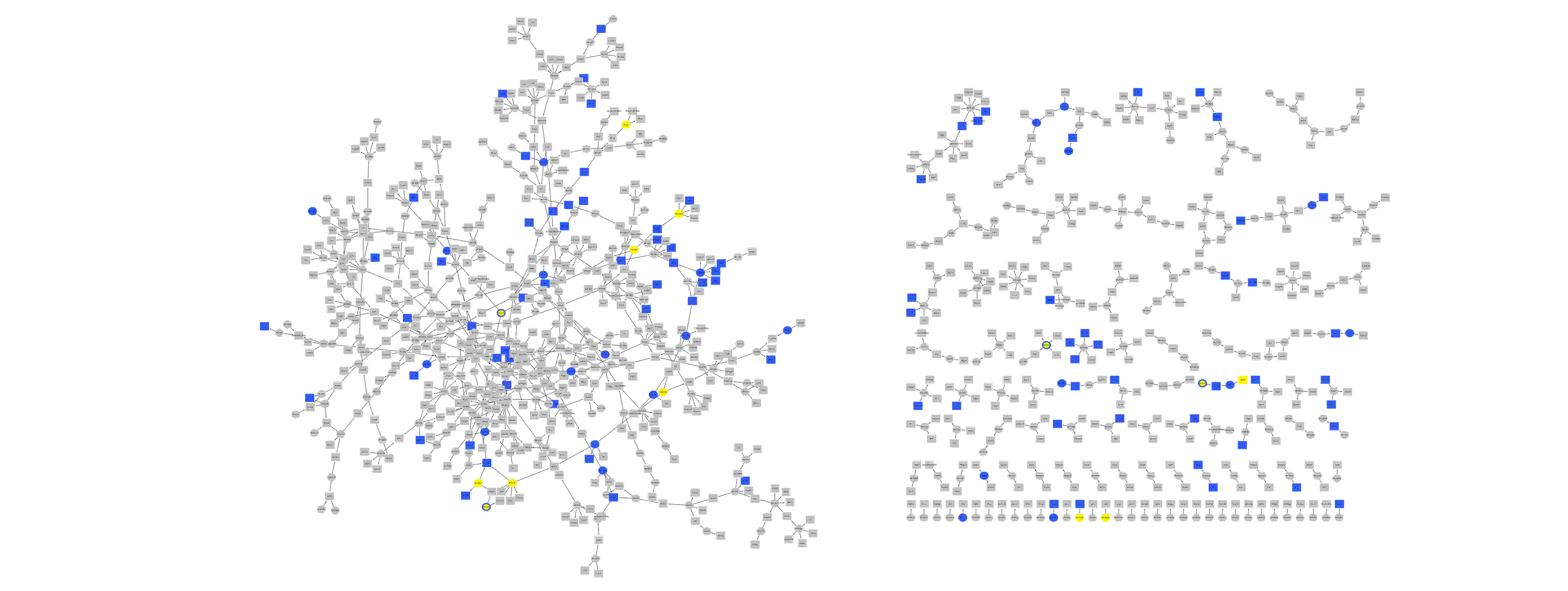}
    \caption{The mRNA-miRNA networks based  on the mRNA-miRNA matchings output
      by the WTOT-matching algorithm,  \textit{focusing on the matchings which
        are labeled  as neither  peaked nor  monotonic}.  Disks  correspond to
      miRNAs  and squares  to  mRNAs.   Elements also  retained  by the  WGCNA
      algorithm (respectively,  the MiRAMINT algorithm) are  indicated in blue
      (respectively,  yellow).  The  top  annotation is  \textit{extracellular
        matrix organization} (GO:0030198, 22 hits).}
    \label{fig:network:others}
  \end{figure}  
\end{landscape}

\end{document}